\documentclass[12pt,a4paper]{JHEP3}
\preprint{ZU-TH 25/11\\ DESY 11-232\\LPN11-71\\MCNET-11-26}

\usepackage{snapshot}
\usepackage{cite}
\usepackage{amssymb}
\usepackage{amsmath}
\usepackage{epsfig}
\usepackage{subfigure}
\usepackage{color}
\let\normalcolor\relax
\usepackage{graphicx}
\usepackage{multirow}
\usepackage{inputenc}
\usepackage{xspace}
\usepackage{slashed}
\input{latin1.def}
\usepackage{epstopdf}
\usepackage{axodraw4j}

\def\gev{~{\rm GeV}}

\def\lsim{\mathrel{\raise.3ex\hbox{$<$\kern-.75em\lower1ex\hbox{$\sim$}}}}
\def\gsim{\mathrel{\raise.3ex\hbox{$>$\kern-.75em\lower1ex\hbox{$\sim$}}}}
\def\ifmath#1{\relax\ifmmode #1\else $#1$\fi}

\newcommand{\beq}{\begin{equation}}
\newcommand{\eeq}{\end{equation}}
\def\beqn{\begin{eqnarray}}
\def\eeqn{\end{eqnarray}}

\newcommand{\bea}{\begin{eqnarray}}
\newcommand{\eea}{\end{eqnarray}}

\newcommand{\Herwigpp}{H\protect\scalebox{0.8}{ERWIG++}\xspace}

\title{Determining the helicity structure of third generation resonances}

\author{Andreas Papaefstathiou$^{1}$, Kazuki Sakurai$^{2}$\\
$^1$Institut f\"ur Theoretische Physik, Universit\"at Z\"urich, Wintherturerstrasse 190, CH-8057 Z\"urich, Switzerland\\
$^2$Deutsches Elektronen Synchrotron DESY, Notkestrasse 85, D-22607
Hamburg, Germany\\
Email: \email{andreasp@physik.uzh.ch, kazuki.sakurai@desy.de}}

\abstract{We examine methods that have been proposed for determining the
  helicity structure of decays of new resonances to third
  generation quarks and/or leptons. We present 
analytical and
  semi-analytical predictions and assess
  the applicability of the relevant variables in realistic reconstruction scenarios using
  Monte Carlo-generated events, including the effects of QCD radiation
  and multiple parton interactions,
  combinatoric ambiguities and fast
  detector simulation.}

\keywords{Hadronic Colliders, Beyond Standard Model}

\begin{document}
\section {Introduction}
The third generation is thought to be intimately connected to the
mechanism responsible for electroweak symmetry breaking due to
the comparatively large top, tau and bottom Yukawa couplings. New, third generation resonances
are further favoured in comparison to those of the first and second generation by flavour constraints, coming from experiments
investigating flavour-changing processes~\cite{Gripaios:2009dq}. 

Collider searches for new resonances decaying to third generation quarks or
leptons (or both in the case of leptoquarks) are particularly 
challenging. 
If the new resonance decays leptonically to either a $\tau$ or $\nu_{\tau}$, the final state necessarily possesses missing energy, which prevents us from 
directly reconstructing the events.
If it decays to $b$ quarks, the signal has to compete with the high QCD cross sections, which
can imitate the topology.
In the case that the new resonance decays to top quarks, these may
subsequently decay to several jets (for the hadronic decay), 
or to a jet plus lepton and missing energy (for the semi-leptonic decay).
Both top quark decay modes require further reconstruction, where
combinatorial ambiguities, QCD backgrounds and the unknown neutrino
momenta can be problematic.

Despite the aforementioned difficulties, third generation resonances
provide us with an interesting opportunity. The decays of the resulting third generation fermions (apart from $\nu_{\tau}$) may allow us
to determine the chirality structure of their couplings to the new resonance.
To fully determine the underlying theory and reconstruct the Lagrangian terms,
determination of the coupling structure, as well as the spin of the new resonance, is necessary.
The determination of the helicity of top quarks and $\tau$ leptons is
made possible by the fact that the angular distributions of their daughter particles are highly
correlated to the helicity of the parents. 
Determination of the helicity of top quarks has been investigated
theoretically in detail~\cite{Jezabek:1988ja, Czarnecki:1990pe,
  Shelton:2008nq}. Several variables have been
proposed and QCD corrections to these have been
calculated. Similar variables have also been proposed for $\tau$
leptons~\cite{DTP-91-42,KEK-TH-332,Anderson:1992jz}.
The $b$ quark hadronizes before decay and produces $B$ mesons which have
relatively long lifetimes and thus produce displaced vertices. This allows for tagging those jets that
originate from $b$ quarks but washes away the effect of its helicity
from the angular distributions of the associated
daughter particles. Thus, the helicity couplings of resonances containing a $b$ quark
has to be inferred by first determining the spin of the parent and
sister particles.

In this paper we investigate the feasibility of helicity measurement 
at the Large Hadron Collider (LHC) for top quarks and $\tau$
leptons resulting from the decays of new third generation resonances. 
We first review the variables that have been previously defined for top and $\tau$ decays
and reproduce the relevant distributions,
comparing them to results from a general-purpose Monte Carlo event
generator, \Herwigpp~\cite{Gieseke:2011na} (section~\ref{sec:vardef}).\footnote{In
  appendix~\ref{app:angular} we also present a set of angular
  variables that complement the energy fraction variables given in the
  literature thus far.} In the case of hadronic
top quark decays, the variables have thus far mostly been considered in the
highly-boosted case. We relax this approximation and attempt to
determine their usefulness in more realistic reconstruction situations. 
For simplicity, in sections~\ref{sec:zprime} and~\ref{sec:ztautau} respectively, we examine the applicability of these methods to
a model containing a new heavy vector boson, a $Z'$, which possesses decays to the third
generation: either to a top and a light quark (specifically
$t\bar{u}$, $\bar{t}u$)\footnote{Note that such models have been proposed as
  possible explanations to the Tevatron $t\bar{t}$
  asymmetry~\cite{Abazov:2011rq, Aaltonen:2011kc, Duraisamy:2011pt, Buckley:2011vc}.}
or a pair of $\tau$ leptons. We consider reconstruction of these
topologies using mass-shell constraints which lead to polynomial
equations. In the latter case we also employ the $\tau$ lepton decay
vertex information. Subsequently, in section~\ref{sec:leptoquark}, we
consider the more challenging case of a third-generation leptoquark model, focusing on pair-production of
these followed by decays to a top quark and a $\tau$ lepton. The
reconstruction technique we employ is related to those presented in
Ref.~\cite{Gripaios:2010hv}. Finally, we present our conclusions in
section~\ref{sec:conclusions}. 

\section{Variable definitions}\label{sec:vardef}
In the decays of top quarks and $\tau$ leptons, the distribution of
the cosine of the angle between the spin axis of
the parent particle $k$ and the daughter particle $i$, $\cos
\theta_i$, is given by: 
\beq\label{eq:costhetabdist}
P(\cos \theta_i) = \frac{1}{2} ( 1 + P_k \kappa_i \cos \theta_i )\;\;,
\eeq
where $P_k = \pm 1$ corresponds to the spin-up or spin-down states of the parent
respectively and $\kappa_i$ is known as the `resolving power' of $i$. The
decay product $i$ can also be taken to be a jet, i.e. containing more
than one particle. The resolving power has been calculated in higher
orders and for various decay products~\cite{Brandenburg:2002xr,
  Jezabek:1994qs, Bernreuther:2008ju, Godbole:2010kr}.
\subsection{Daughter-to-parent energy ratios, $x_{p,i}$}
If tops and taus are produced as a result of the decay of a heavy
resonance, their own decay products may be collimated 
due to their large velocities.
In this case, it is useful to define the energy fraction $x_{k,i} = \mathcal{E}_{k,i}/\mathcal{E}_k$ between
the parent particle $k$ $(=t, \tau)$ and the one of its daughter particles, $i$, rather than 
the angle between them.
These are
formed in the laboratory frame for both the tau leptons and top quarks as:
\begin{eqnarray} 
&& x_{\tau,\rm jet} = \mathcal{E}_{\mathrm{jet}} /
\mathcal{E}_{\tau} \nonumber\;,\\
&& x_{\mathrm {top}, \rm b} = \mathcal{E}_{b}
/\mathcal{E}_{\mathrm{top}}\;,
\label{eq:x}
\end{eqnarray}
respectively. For brevity, we will write $x_{\rm top}$ and
$x_\tau$ to denote the preceding variables. 

\begin{figure}[!t]
  \centering 
  \vspace{5.0cm}
  \includegraphics[scale=0.6]{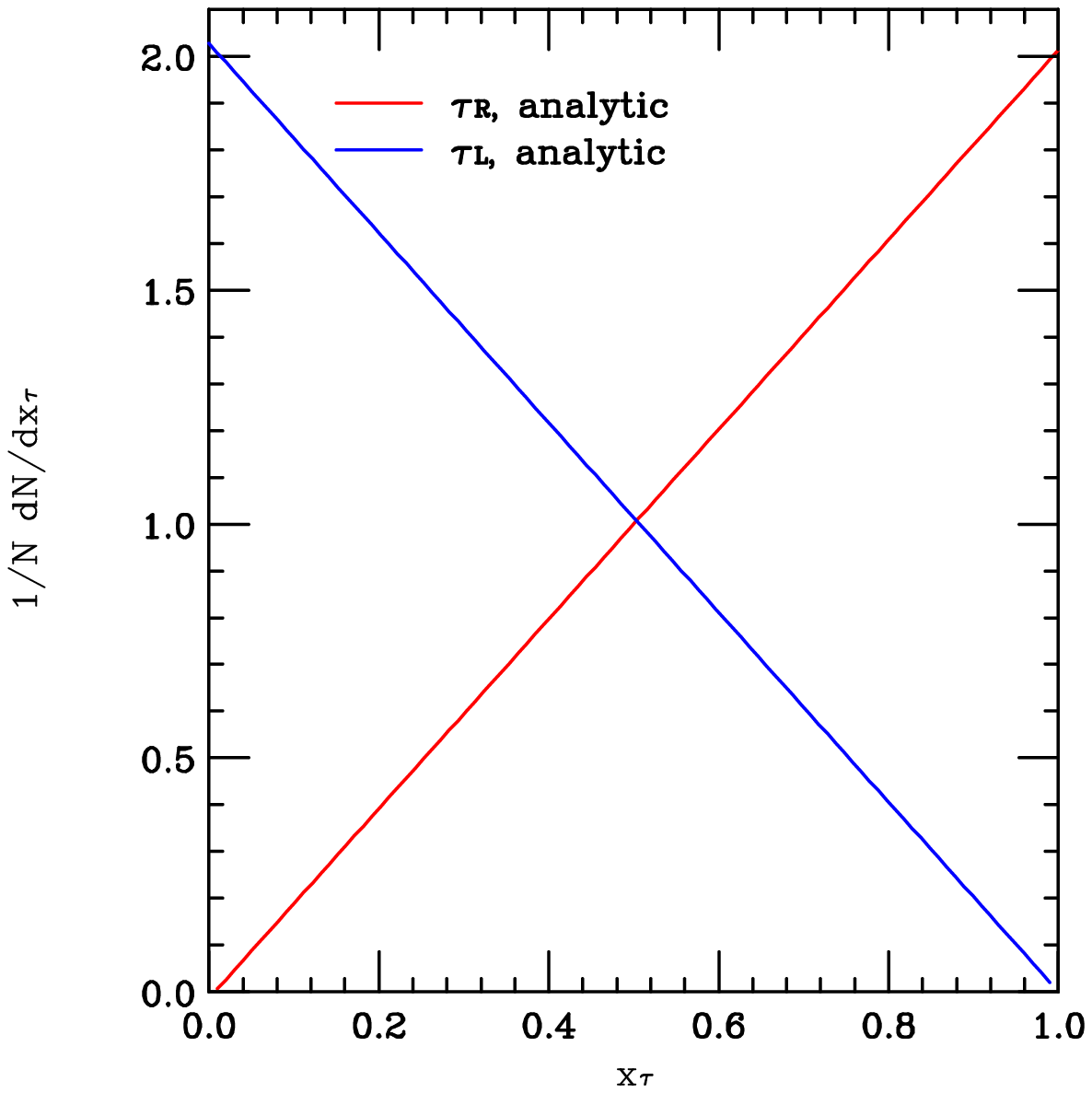}
  \hspace{3.0cm}
  \includegraphics[scale=0.6]{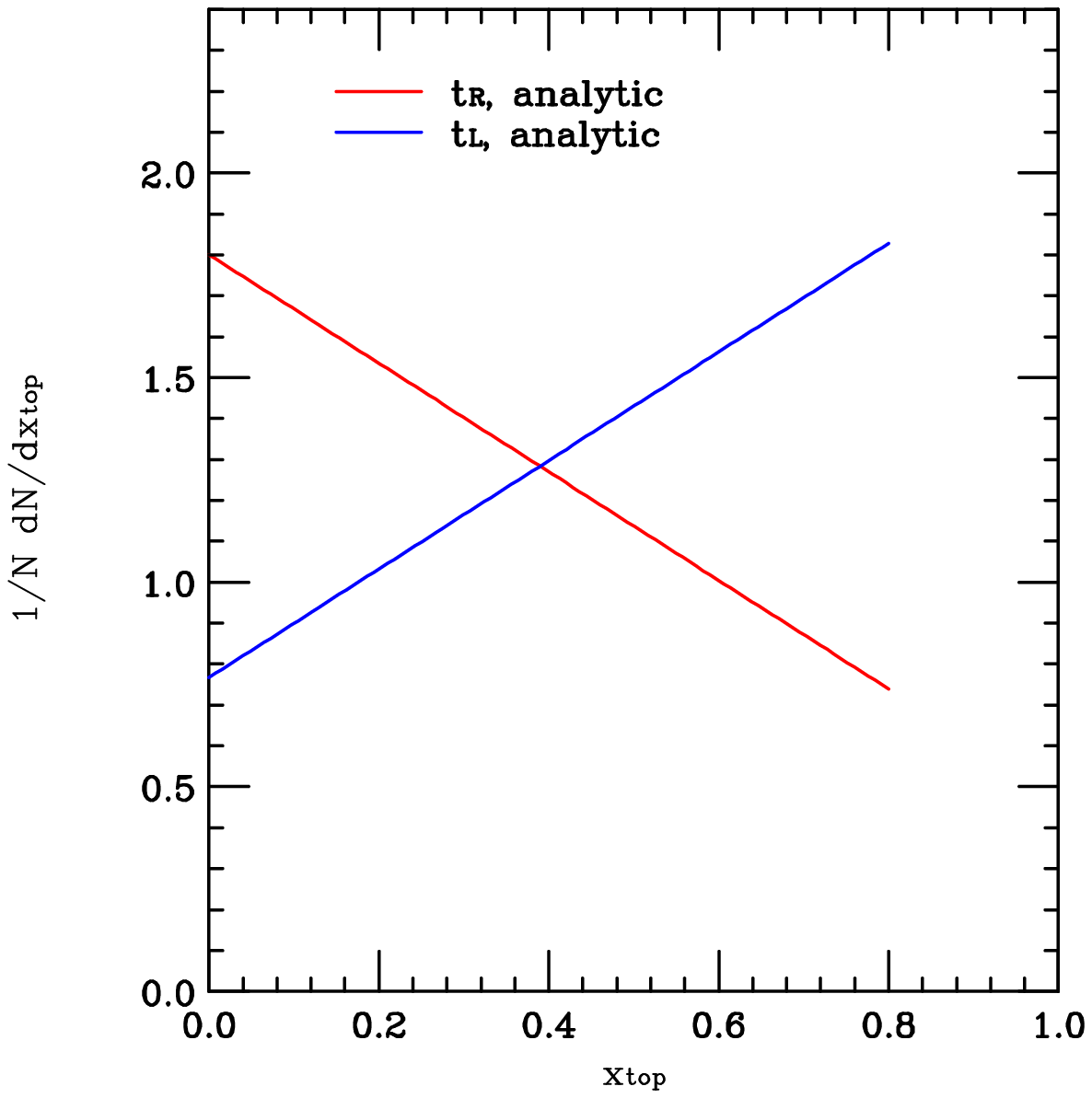}
  \caption{The predictions for the energy fractions $x_{\tau} = \mathcal{E}_{\mathrm{jet}} /
\mathcal{E}_{\tau}$ for the decay $\tau \rightarrow \pi \nu$ (left) and $x_{\mathrm {top}} = \mathcal{E}_{b} /
\mathcal{E}_{\mathrm{top}}$ (right) in the highly-boosted cases
($\beta_{\tau} = 1$ and $\beta_t = 1$).}
  \label{fig:pred}
\end{figure}

Analytic predictions for these variables can be derived using
Eq.~(\ref{eq:costhetabdist}), by transforming from the angular
variable $\cos \theta _i$ (defined in the centre-of-mass frame) to the
energy ratio (defined in the lab frame). In
the highly-boosted cases, where the boost factor, $\beta_k \equiv
|\vec{p}_k|/E_k$, is taken to be unity, their distributions are shown in Fig.~\ref{fig:pred} for the $\tau$ lepton decay mode
$\tau \rightarrow \pi \nu_\tau$ (left) and hadronic top (right). 
The distribution of $x_{\mathrm{top}}$ red has a cut-off at the maximum value of $x_{\mathrm{top,max}} = 1
- m_W^2 / m_{\mathrm{top}}^2 \sim 0.79$ and that of $x_{\tau}$ at a
minimum value of $x_{\tau,\mathrm{min}} = m_{\mathrm{\pi}}^2 /
m_{\tau}^2 $, as a result of the kinematic restrictions imposed by the
mass of the $W$ and the pion respectively. In the case of the $\tau$, we show
  only the $\tau \rightarrow \pi \nu_\tau$ mode, for which $m_\pi
  \simeq 0.14$ GeV, which results in a small
$x_{\tau,\mathrm{min}}$.
 
The exact analytic forms of the
distributions are given by~\cite{Shelton:2008nq}:\footnote{Note that in~\cite{Shelton:2008nq},
Eq.~(\ref{eq:xtop}) is missing a factor of $\frac{m^2_{\mathrm{top}} } { m^2_{\mathrm{top}} -
    m^2_W }$ in the second term. The
  corresponding distributions, however, appear to have been
  constructed with the correct formulae.} 
\begin{eqnarray}
\frac{1}{N} \frac{\mathrm{d} N} {\mathrm{d} x_{\tau}} &=&
\frac{1}{\beta_\tau}  \frac{ m^2_{\tau} } { m^2_{\tau} - m^2_{\mathrm{jet}}  } \left( 1 - \frac{1}{\beta_\tau}\frac{ P_\tau ( m^2_{\tau}  +  m^2_{\mathrm{jet}} ) } { m^2_{\tau}  - m^2_{\mathrm{jet}} } + \frac{1}{\beta_\tau}\frac{2 P_{\tau} m^2_{\tau} } { m^2_{\tau}  - m^2_{\mathrm{jet}} }x_\tau\right) \;\;, \label{eq:xtau}\\
\frac{1}{N} \frac{\mathrm{d} N} {\mathrm{d} x_{\mathrm{top}}} &=& \frac{1}{\beta_t} \frac{m^2_{\mathrm{top}} } { m^2_{\mathrm{top}} -
    m^2_W } \left( 1 - \frac{1}{\beta_t}\kappa_b P_t
   + \kappa_b P_t \frac{1}{\beta_t}\frac { 2 m^2_{\mathrm{top}} } { m^2_{\mathrm{top}} -
    m^2_W } x_{\mathrm{top}} \right) \;\;, \label{eq:xtop}
\end{eqnarray}
where $P_{i} = \pm 1$ (for $i = \tau, t$) represent right or left
helicities of the $\tau$ or top and:
\beq\label{eq:kappa_b}
\kappa_b = - \frac{ m_{\mathrm{top}}^2 - 2 m_W^2} { m_{\mathrm{top}}^2
  + 2 m_W^2 } \simeq -0.4\;,
\eeq
is~\cite{Shelton:2008nq} the resolving power of the $b$-quark at
leading order,\footnote{The authors of Ref.~\cite{Godbole:2011vw} present a study of polarisation observables
  at next-to-leading order with parton showers in $H^{-}t$ and $W^{-}t$
 production. They find that in those cases these observables are
  robust against higher-order corrections.} $m_{\mathrm{top}}$, $m_{W}$ and $m_\tau$ are the top quark,
$W$ boson and $\tau$ lepton masses respectively.  The approximation $\beta_k\rightarrow 1$ is almost always good for $\tau$
leptons, and with the current beyond-the-standard model (BSM) third generation limits rising as
the LHC experiments produce more exclusion regions (see,
e.g.~\cite{Aad:2011wc, Chatrchyan:2011ay}), it should be even more applicable for most BSM models that include heavy particles
that decay to tops. Note that in the opposite limit, of $\beta_t
\rightarrow 0$, the above distributions tend to
delta-functions, $\delta (x_k - x_{k,0})$, since the lab frame and parent top (or tau) decay
frames become identical. The values of $x_{k,0}$, determined by
the masses of the decay products in the two-body decay, are $x_{\tau,0} \simeq (m_{\tau}^2 - m_\pi^2)/(2m_{\tau})$ and
$x_{\mathrm{top}, 0} \simeq (m_{\rm top}^2 - m_W^2)/(2m_{\rm top})$
for the tau and top distributions respectively, neglecting jet and bottom quark masses.

\begin{figure}[!t]
  \centering 
  \vspace{5.5cm}
  \includegraphics[scale=0.60]{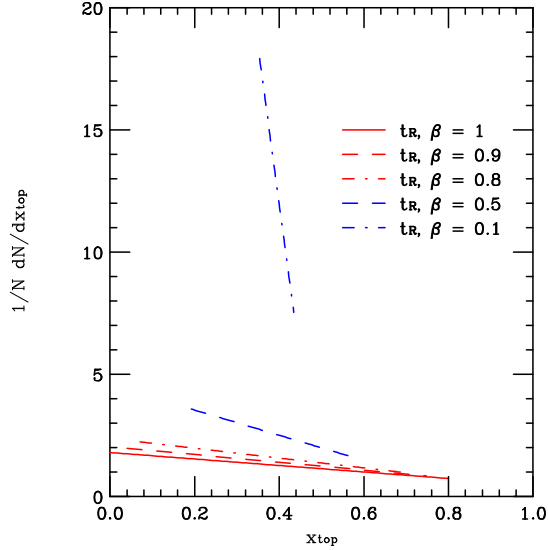}
  \caption{The predictions for the energy fraction $x_{\mathrm {top}} = \mathcal{E}_{b} /
\mathcal{E}_{\mathrm{top}}$ for the cases $\beta_t = 1.0, 0.9, 0.8, 0.5$ and
$0.1$. }
  \label{fig:pred2}
\end{figure}

It is obvious from Eqs.~(\ref{eq:xtau})
and~(\ref{eq:xtop}) that the effect of $\beta_t < 1$ is to alter the
predictions for the left- and right-handed distributions. The result 
of this effect is shown in Fig.~\ref{fig:pred2}, where we show the
$x_{\rm top}$ right-handed helicity distributions for fixed $\beta_t = 1.0, 0.9, 0.8, 0.5$ and
$0.1$. It is obvious that for $\beta = 0.9$ and $\beta = 0.8$, the
variation from the $\beta_t = 1$ line is small, whereas for $\beta_t = 0.5$ and
$\beta_t = 0.1$, the slope and position of the lines is dramatically altered.\footnote{Note that
  there is also a lower limit on allowed values of $x_{\mathrm{top}}$
  due to kinematics. The limits for arbitrary top quark boosts $\beta$
  are given by $(1-\beta_t) \leq \frac {m_{\rm top}^2}{m_{\rm top}^2  -
    m_W^2} x_{\mathrm{top}} \leq (1+\beta_t)$.} This is due to the
fact that the slope of Eq.~(\ref{eq:xtop}) is inversely proportional
to $\beta_t^2$. One may then safely consider the region $\beta_t \gtrsim 0.9$
as the `highly-boosted' region in the case of top quark decay. This
would correspond to top quarks with energies $E_t \gtrsim 2.5 m_{\rm
  top}$. In realistic scenarios in hadron colliders, $\beta_t$ is not
fixed, and one would need to integrate over the allowed values to
obtain the full result. 

\begin{figure}[!t]
  \centering 
  \vspace{1.5cm}
  \hspace{4cm}
  \includegraphics[scale=0.48, angle=90]{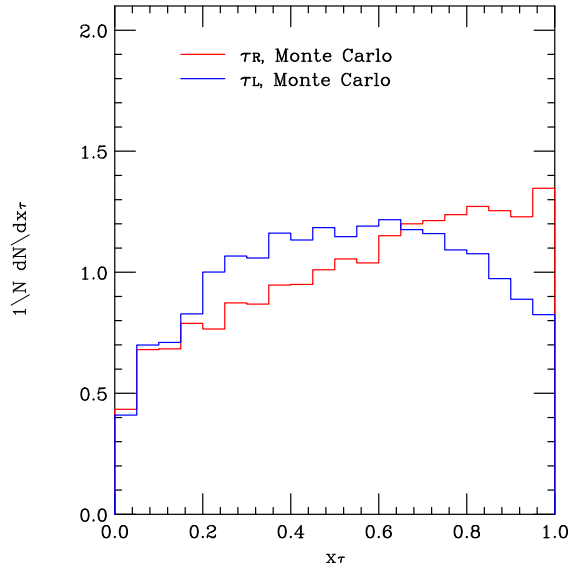}
  \vspace{0.5cm}
  \caption{The Monte Carlo predictions for the energy fractions $x_{\tau} = \mathcal{E}_{\mathrm{jet}} /
\mathcal{E}_{\tau}$ for all the decay modes when the $\tau$ is highly
boosted ($\beta_\tau = 1$).}
  \label{fig:xtau_full}
\end{figure}

The $\tau$ lepton can decay via multiple channels and each of these channels contributes to the total
distribution. We do not attempt here to reproduce its analytic form, as this would require calculating the
distributions of Eq.~(\ref{eq:xtau}) corresponding to each decay mode and integrating over
the distribution of the mass of the $\tau$ jet, $m_{\mathrm{jet}}$, with certain weight obtained by the matrix element squared.
Instead we present results constructed from the Monte Carlo-simulated decays of
the $\tau$ lepton using the \Herwigpp event generator. We show the
resulting distributions for the left- and right-handed
highly-boosted taus in Fig.~\ref{fig:xtau_full}. The contributions of
the multiple decay modes clearly modify the behaviour compared to that
of the single decay mode which appears in Fig.~\ref{fig:pred}.

\subsection{Semi-leptonic top variable, $u$}
\begin{figure}[!t]
  \begin{centering} 
  \hspace{3.0cm}
  \includegraphics[angle=90,scale=0.45]{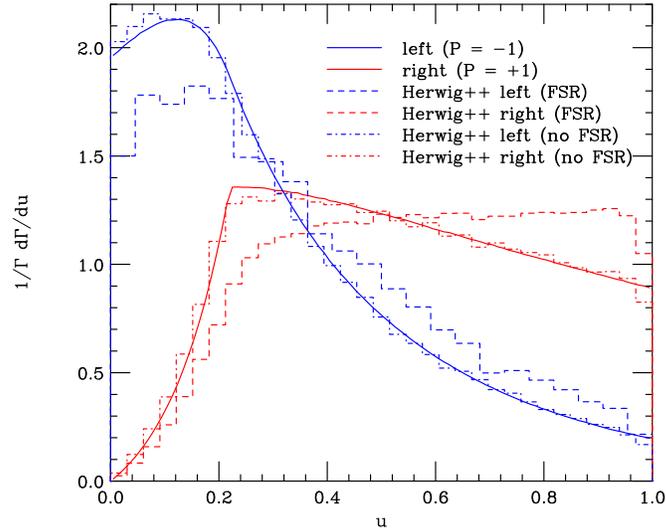}
  \vspace{0.5cm}
  \caption{The distribution $1/\Gamma \, \mathrm{d}\Gamma/\mathrm{d} u$ of the fraction of visible 
lab frame energy carried by the lepton in a highly-boosted
semi-leptonic top (i.e. $\beta_t = 1$),
$u = \mathcal{E}_\ell/(\mathcal{E}_\ell+\mathcal{E}_b) $ is shown.  The blue curve and red curves represent left- and
right-handed top quarks respectively.}
  \label{fig:udist}
  \end{centering}
\end{figure}

\begin{figure}[!t]
  \centering 
  \vspace{1.0cm}
  \hspace{3.5cm}
  \includegraphics[scale=0.60]{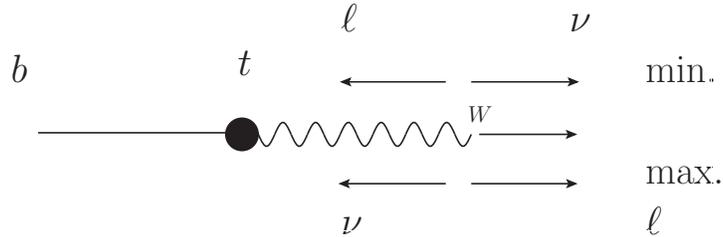}
  \vspace{0.5cm}
  \caption{A schematic diagram of the decay of the top in its rest
    frame into a $W$ and a $b$-quark, with subsequent decay of the $W$
  to a lepton and neutrino. The two configurations shown correspond to the minimum
  and maximum energy configurations of the lepton, corresponding to $E_{\ell,\rm min} =
m_W^2 / (2m_{\rm top})$ and $E_{\ell,\rm max} = m_{\rm top}/2$ respectively. The
minimum energy of the lepton causes the kink in the $u$ variable distribution.}
  \label{fig:minmaxellE}
\end{figure}

In the case of semi-leptonic top decays, $t\rightarrow b \ell
\nu_{\ell}$, where a $b$-jet is tagged and an electron or muon is
identified, we can calculate the fraction of the visible energy
carried away by the lepton, $u$:
\beq
u = \frac{\mathcal{E}_\ell}{ \mathcal{E}_\ell + \mathcal{E}_b }
\eeq
The resulting distributions of the variable $u$ for highly-boosted ($\beta_t\rightarrow 1$)~\cite{Shelton:2008nq} left-
and right-handed tops are shown in
Fig.~\ref{fig:udist}. Highly-boosted Monte Carlo-generated curves are
also shown for comparison, with and without final state radiation
(FSR).\footnote{Note that the curves of Fig.~2 in
Ref.~\cite{Shelton:2008nq} seem to fit the FSR \textit{on} case rather
than the one obtained by integrating the differential width directly,
which should fit the FSR \textit{off} case.} The kink at $u = m_W^2 /
m_{\rm top}^2 \sim
0.215$ is due to the fact that there exists a minimum possible value
of the lepton energy in the top rest frame, given by $E_{\ell,\rm min} =
m_W^2 / (2m_{\rm top})$, which arises when the lepton is anti-aligned with
the top boost direction. The maximum value of the energy is
$E_{\ell,\rm max} = m_{\rm top}/2$
and arises when the lepton is aligned with the boost direction. This
is clarified in Fig.~\ref{fig:minmaxellE}, where the schematic diagram demonstrates
the decay of a top in its rest frame. 

It is important to note that the variable $u$ has the advantage that there is no need to
explicitly reconstruct the top quarks in order to form it, even in the
case of $\beta_t \neq 1$. As a
result, it is expected to be less sensitive to the reconstruction systematics that
may enter other energy fraction variables.
\section{Applications}\label{sec:application}
We examine a model of a heavy vector boson ($Z'$) decaying to tops or
taus and a scenario of pair-production of third generation scalar
leptoquark states. In what follows we will neglect neutrino masses,
and use the following values for the top quark, $W$ boson and $\tau$
lepton masses respectively: $m_{\rm top} = 174.2 \gev$,  $m_W = 80.403 \gev$ and
$m_\tau = 1.777 \gev$. 


\subsection{Flavour-changing $Z'$}\label{sec:zprime}
We first examine the application of the variables $x_{\rm top}$ and $u$ that
have been defined in the
previous sections, on a model of a $Z'$ boson that possesses
flavour-changing quark couplings, described by the Lagrangian density:
\begin{eqnarray}\label{eq:zprimelag}
\mathcal{L} &=& g_{Z'}^{tuR} Z'^\mu \bar{u}_R \gamma_\mu t_R  + g_{Z'}^{tuL}
  Z'^\mu \bar{u}_L \gamma_\mu t_L \nonumber \\
&+&  g_{Z'}^{uuR} Z'^\mu \bar{u}_R \gamma_\mu u_R  + g_{Z'}^{uuL}
  Z'^\mu \bar{u}_L \gamma_\mu u_L + \mathrm{h.c.}\;\;,
\end{eqnarray}
where $g_{Z'}^{tuL}$ and $g_{Z'}^{tuR}$ are the flavour-changing left-
and right-handed parameters that we will
be varying.

It is evident that the $Z'$ bosons of this model would be produced at a hadron
collider via light $u\bar{u}$ initial states. They will decay to both
light $u\bar{u}$ and $t\bar{u}$ (and $u\bar{t}$). After discovery of
such a state, for example in dijet resonance searches, determining the
helicity structure of the $u\bar{u}Z$ vertex will be challenging, if not impossible. 
Hence one would concentrate on
determining the helicity structure of the vertex which involves the
top quark. 
\subsubsection{Parton-level results}
\begin{figure}[!t]
 \begin{center}
  \vspace{1.0cm}  
  \hspace{4.5cm}  
  \includegraphics[angle=90,scale=0.45]{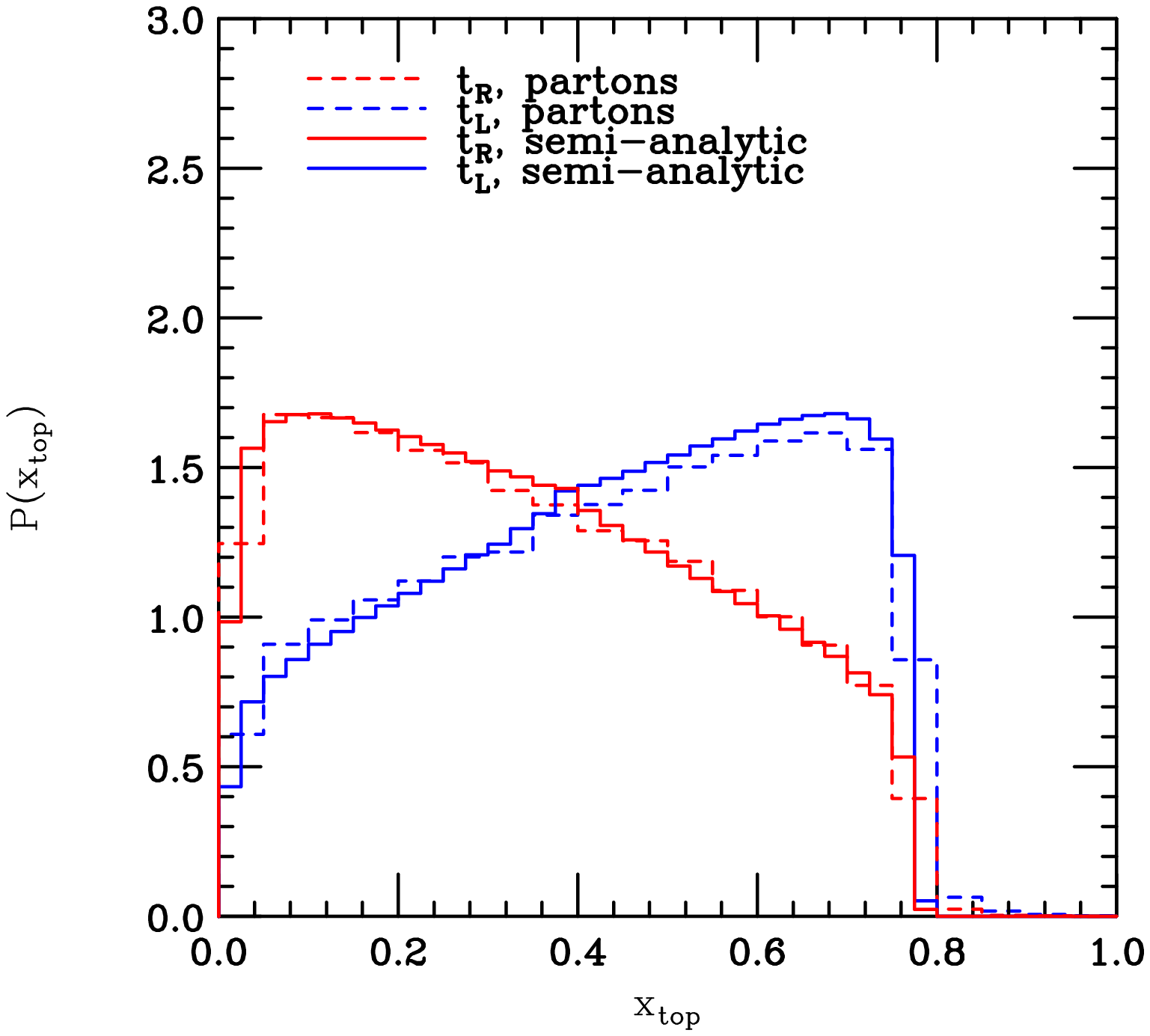}  
  \hspace{5.0cm}  
\includegraphics[angle=90,scale=0.45]{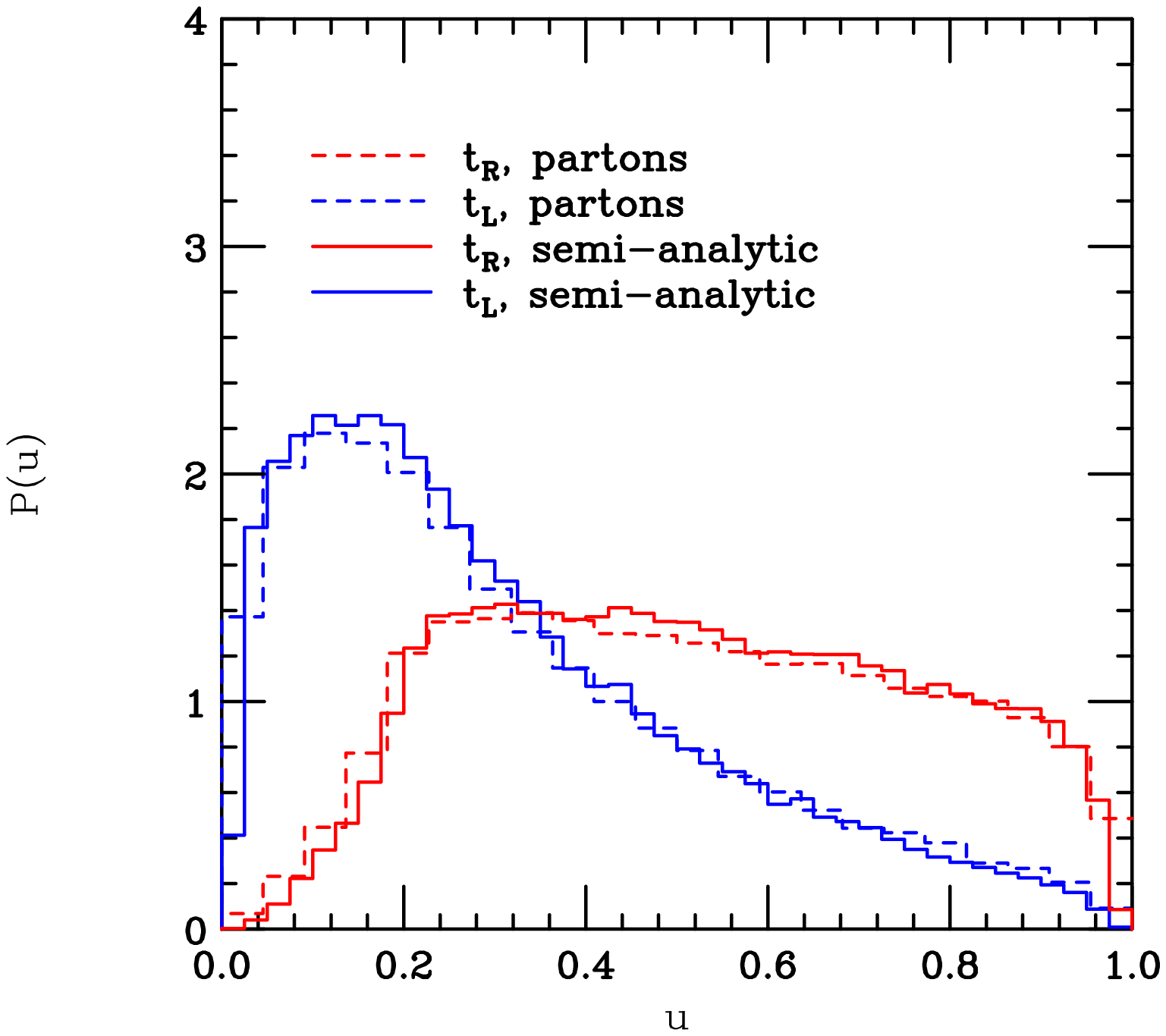}  
\caption{The $x_{\mathrm{top}}$ (left panel) and $u$ (right panel) variables for left- or right-handed
  $Z'$ bosons decaying to $u\bar{t}$ or $\bar{u}t$, at a 14 TeV LHC, obtained from
  parton-level events compared to a semi-analytic prediction as
  described in the main text.}
\label{fig:xudelpheszpparton}
  \end{center}
\end{figure}

We consider a $Z'$ described by the particular model given by the
Lagrangian of Eq.~(\ref{eq:zprimelag}), of mass $1.5$~TeV, choosing
either a purely left-handed third generation coupling:
$g_{Z'}^{tuR} = 0$, $g_{Z'}^{tuL} = 1$ or a purely right-handed one:
$g_{Z'}^{tuR} = 1$, $g_{Z'}^{tuL} = 0$, keeping in both cases
$g_{Z'}^{uuR} =  g_{Z'}^{uuL} = 1$.\footnote{This mass/coupling
  combination is currently being marginally excluded by results
  presented by the LHC experiments. The aim of the present study is to
  assess the viability of the reconstruction variables.}  We show
Monte Carlo results of the distributions of the variables $x_{\rm top}$ and $u$, for a 14
TeV LHC, in the dashed curves in
Fig.~\ref{fig:xudelpheszpparton}. These
results were obtained using parton-level events, ignoring
initial- and final-state radiation, hadronization, and applying no rapidity
or momentum cuts on the particles. The plots in
Fig.~\ref{fig:xudelpheszpparton} also contain semi-analytic predictions
that take into account the finite boost of the top quark in the lab
frame (solid curves). The semi-analytic $x_{\mathrm{top}}$ distribution
was produced by assuming that the cosine of the angle of the emitted $b$-quark in
the top quark rest frame, $\cos \theta_b$, was distributed
according to Eq.~(\ref{eq:costhetabdist}) with $\kappa_b \simeq -0.4$, and then boosted to the lab frame
according to the top quark boost ($\beta_t$) distribution. The
$\beta_t$ distribution was extracted from
the Monte Carlo event generator directly, but can be fitted using a
Gaussian distribution, yielding identical results (see appendix~\ref{app:beta}). Similarly,
the $u$ variable distribution was calculated first by distributing via a Monte
Carlo technique the energy and $z$-momentum of the lepton in the top centre-of-mass
frame ($E_\ell$ and $p^z_\ell$ respectively), according to the full matrix element (see appendix~\ref{app:polarizedtop}) and
then boosting to the lab frame using the top quark boost
distribution. The variable $u$ was then calculated by taking the
ratio:
\beq
u = \frac{E_\ell + \beta_t p^z_\ell}{E_\ell +  \beta_t p^z_\ell + E_b  + \beta_t p^z_b }\;.
\eeq
We ignore the mixing of helicities due to finite masses since the top quark is produced in association with a
light quark (for which $m_{Z'}  \gg m_u$).\footnote{These effects
  are small in the case of the $Z'$ model but have been
calculated in the leptoquark case in the following section, where the
reconstruction is explained in further detail.}

\subsubsection{Simulation and reconstruction}\label{sec:zprimereco}
We will assume that the new resonances have been
discovered, and that their mass has been measured to a satisfactory
accuracy (say, $\mathcal{O}$(a few \%)). We will also assume that the spin of the new resonance has
been determined by measuring the angular distributions of the
jets originating from the $u$ and $\bar{u}$ partons in the $u\bar{u}$
decay mode. It is necessary, however, to outline the details of the
method for reconstructing events of a particular topology, for which
we can form the variables we have been examining thus far at parton
level. 

\begin{figure}[!t]
  \centering 
    \includegraphics[scale=0.55]{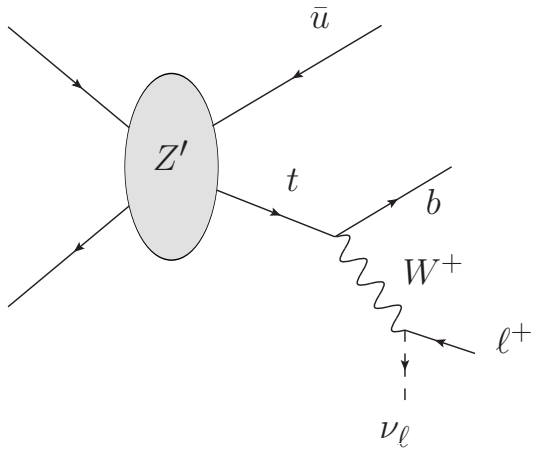}
\caption{Production of $\bar{u}t$ from the exchange of a $Z'$ and subsequent decay of the top into a leptonic
  $W$ and a $b$-jet. This mode can be fully reconstructed if one
  applies the $W$ mass shell condition, and chooses the solution which
  yields the `best' top mass.} 
\label{fig:zhstopology}
\end{figure}

We focus on LHC proton-proton collisions at 14~TeV, in which a $Z'$ is exchanged, producing a $u$ (or $\bar{u}$)-quark and
an anti-top (or top), with a subsequent semi-leptonic (restricted to $e$ or $\mu$) decay of the top. The topology is shown in
Fig.~\ref{fig:zhstopology}. The $u\bar{t}$ and $\bar{u}t$ decay modes
account for slightly less than $\sim 50\%$ of the
total decay widths, if only one helicity (left- or right-handed) is present: $\Gamma_{Z', M = 1.5~\mathrm{TeV}}
(u\bar{t}/\bar{u}t) = 236 \gev$. 
The leading-order cross sections for the specific
topology at a 14 TeV LHC are $\sigma(Z' \rightarrow t\bar{u}/\bar{t}u \rightarrow
b \ell^\pm \nu + \mathrm{jet}) = 6.3~\mathrm{pb}$, again, if only one
helicity is present, and including the branching ratio of the top
quark to electrons or muons. We checked that the obvious irreducible
background, SM QCD production of $t + q$, could be reduced to less
than $\sim 1~\mathrm{pb}$, with an appropriate invariant mass cut on
the visible decay products of the $tq$ pair, e.g. of around $500\gev$, with a smaller effect on the
$Z'$ signal, $\sim 60-70\%$. We expect that other, reducible, SM backgrounds should be easier
to reject in the helicity analysis.

If a $b$-jet is tagged and a high-$p_T$ lepton is found, along with
the high-$p_T$ jet originating from the $u$, the information left missing to fully reconstruct the final state are the three spatial
momentum components of the neutrino (assuming massless neutrino). We can obtain the transverse
components of the neutrino momentum to a reasonable accuracy
($\mathcal{O}$(~few \%)) by assuming that they are equal to
the components of the missing transverse momentum. Then, the only
remaining missing information is the $z$-component of the neutrino
momentum. In a hadron collider we do not possess any information on the initial $z$-momentum
of the system. However if we assume that the neutrino and the
lepton originated from the decay of an on-shell $W$ boson, we may apply the following mass-shell condition on their four-momenta:
\begin{equation}
(p_\ell + p_\nu)^2 = m_W^2\;\;,
\end{equation}
where $p_\ell$ and $p_\nu$ are the lepton and neutrino
four-momenta respectively, and $m_W$ is the on-shell $W$ boson
mass. This approximation is good, since the width of the $W$ boson is
small compared to its mass ($\Gamma_W \simeq 2.14~\mathrm{GeV}$ versus $m_W \simeq
80.40~\mathrm{GeV}$), and leaves us with a quadratic equation
for the $z$-component of the neutrino momentum. To pick one of the two
solutions, we choose the one that also yields a top mass closest to the
on-shell top mass, via:
\begin{equation}
m_{t, A/B}^2 = (p_\ell + p_{\nu,A/B} + p_b)^2\;\;,
\end{equation}
where $p_b$ is the $b$-jet four-momentum and $m_{t,A/B}$ is the top
quark mass obtained by using the solutions for the neutrino momentum
$p_{\nu,A/B}$. Once the `best' solution is chosen, we possess all
information required concerning the event, and we can thus calculate all the variables we have been examining at parton level.

The $Z'$ model has been implemented and has been simulated using the \Herwigpp
event generator with initial- and final-state radiation turned on, as
well as hadronization and the underlying event. We simulated 10
fb$^{-1}$ of data, a reasonable amount in a near- to mid-term LHC
run at 14~TeV. The events were then
processed through the Delphes detector simulation~\cite{Ovyn:2009tx}, where the
following minimum cuts are applied to the reconstructed objects:
\begin{itemize}
\item{$p_{T,\mathrm{min}}$ for jets of 20.0~GeV.}
\item{$p_{T,\mathrm{min}}$ for electrons and muons of 10.0~GeV.}
\end{itemize}
The default Delphes $b$-quark flat-$p_T$ tagging efficiency was replaced by a more
realistic function of jet transverse momentum, $p_T$, which has the form
\beq\label{eq:btagging}
P(p_{T,\mathrm{j}}) = 0.08 + 0.006\times p_{T,\mathrm{j}}\times \exp{(-3\times 10^{-5}p_{T,\mathrm{j}}^2)}\;.
\eeq
See for example Ref.~\cite{Bellan:2011vm} for further details. The dependence on jet
pseudo-rapidity remained flat. An additional cut requiring the total
missing transverse energy to be greater than 20~GeV was applied. The rest represent the default settings present in the Delphes ATLAS
detector card, where the triggering simulation has been turned
off. In this analysis, and the rest of this paper, we use the
anti-$k_T$ clustering algorithm with a radius parameter $R = 0.4$ to
construct jets.\footnote{It is advantageous to use a smaller radius parameter
  for the anti-$k_T$ clustering algorithm than the Delphes default one of $R=0.7$, since underlying
  event and pile-up contaminations are expected to be approximately proportional to
  $R^2$. See Ref.~\cite{Salam:2009jx} for further details on jet algorithms and
  the underlying event.}

\begin{figure}[!t]
\begin{tabular}{c}
\begin{minipage}{1.0\hsize}
  \begin{center}
  \vspace{4.5cm}
 \includegraphics[scale=0.52, angle=0]{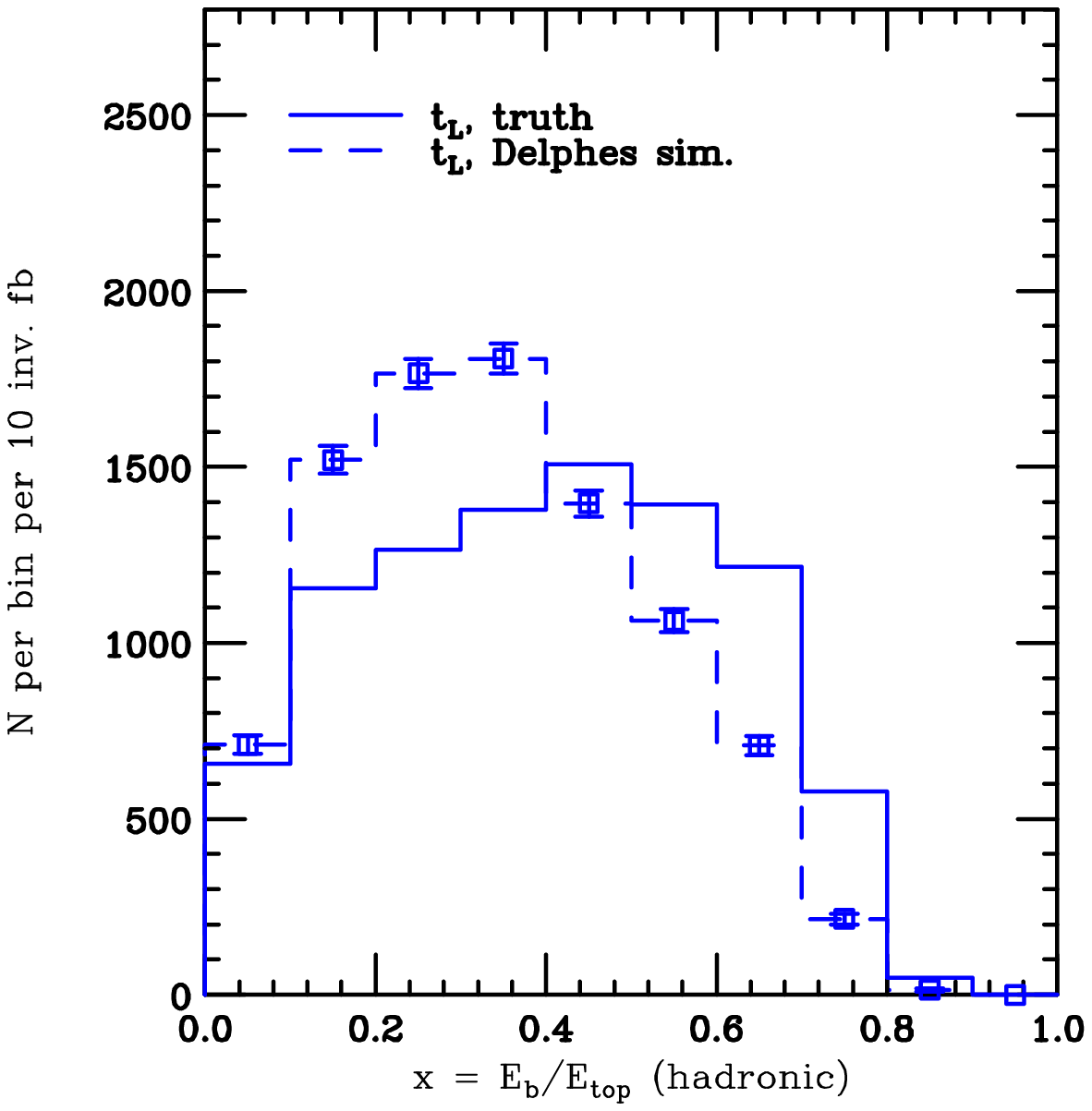}
  \hspace{2.5cm}
  \includegraphics[scale=0.52,angle=0]{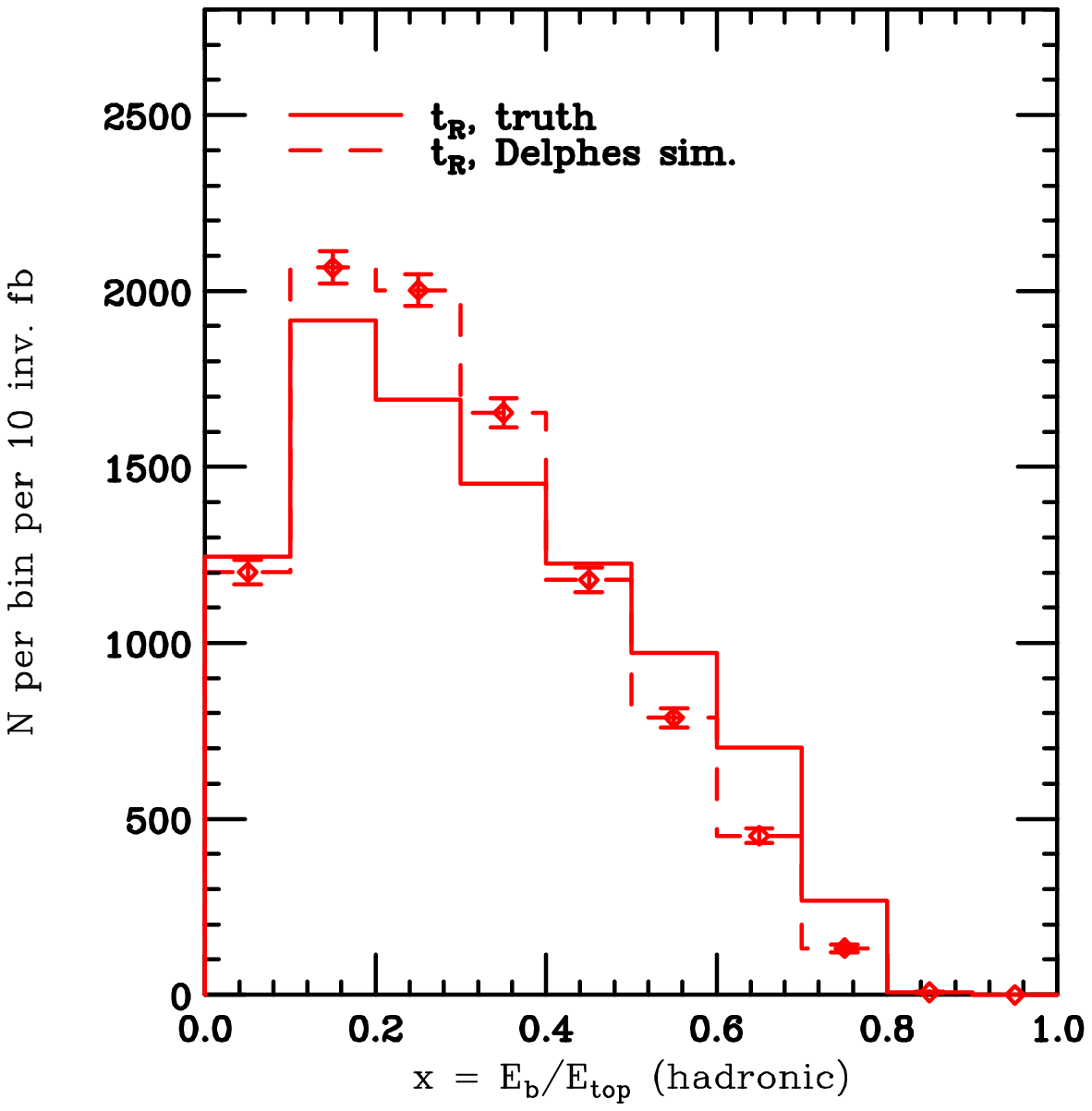}
  \caption{Shown in the figures is a comparison between the
results obtained for the $x_{\rm top}$ variable for the 1.5 TeV
flavour-changing $Z'$ model \textit{before} detector simulation (but applying all cuts and using equivalent
jet-finding) and \textit{after} the Delphes simulation for the left- and right-
handed fermions (blue and red respectively).}
\label{fig:xtoptruthzp}
  \end{center}
  \end{minipage}
\\
\begin{minipage}{1.0\hsize}
  \centering 
  \vspace{5.5cm}
 \includegraphics[scale=0.52, angle=0]{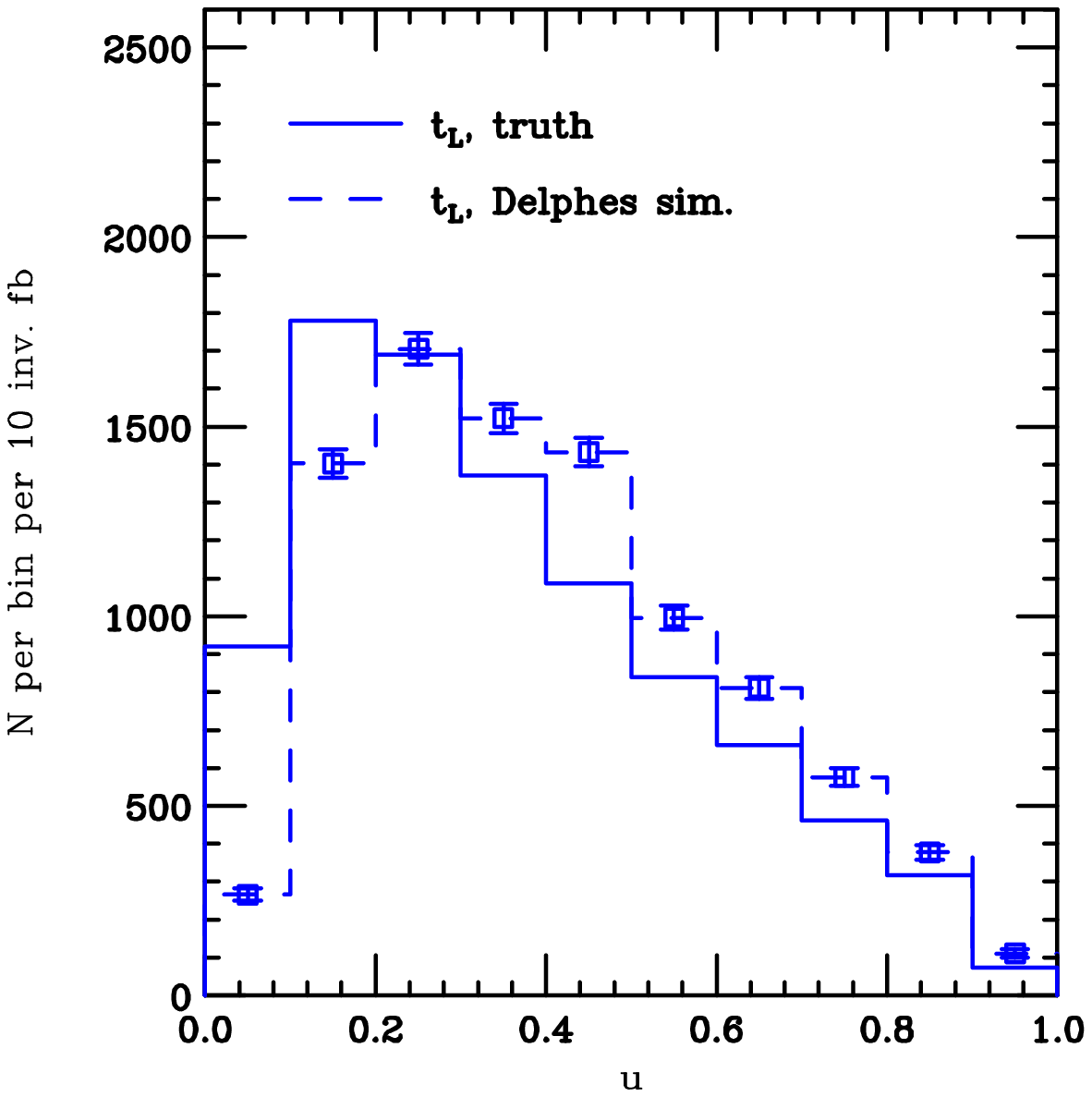}
  \hspace{2.5cm}
  \includegraphics[scale=0.52,angle=0]{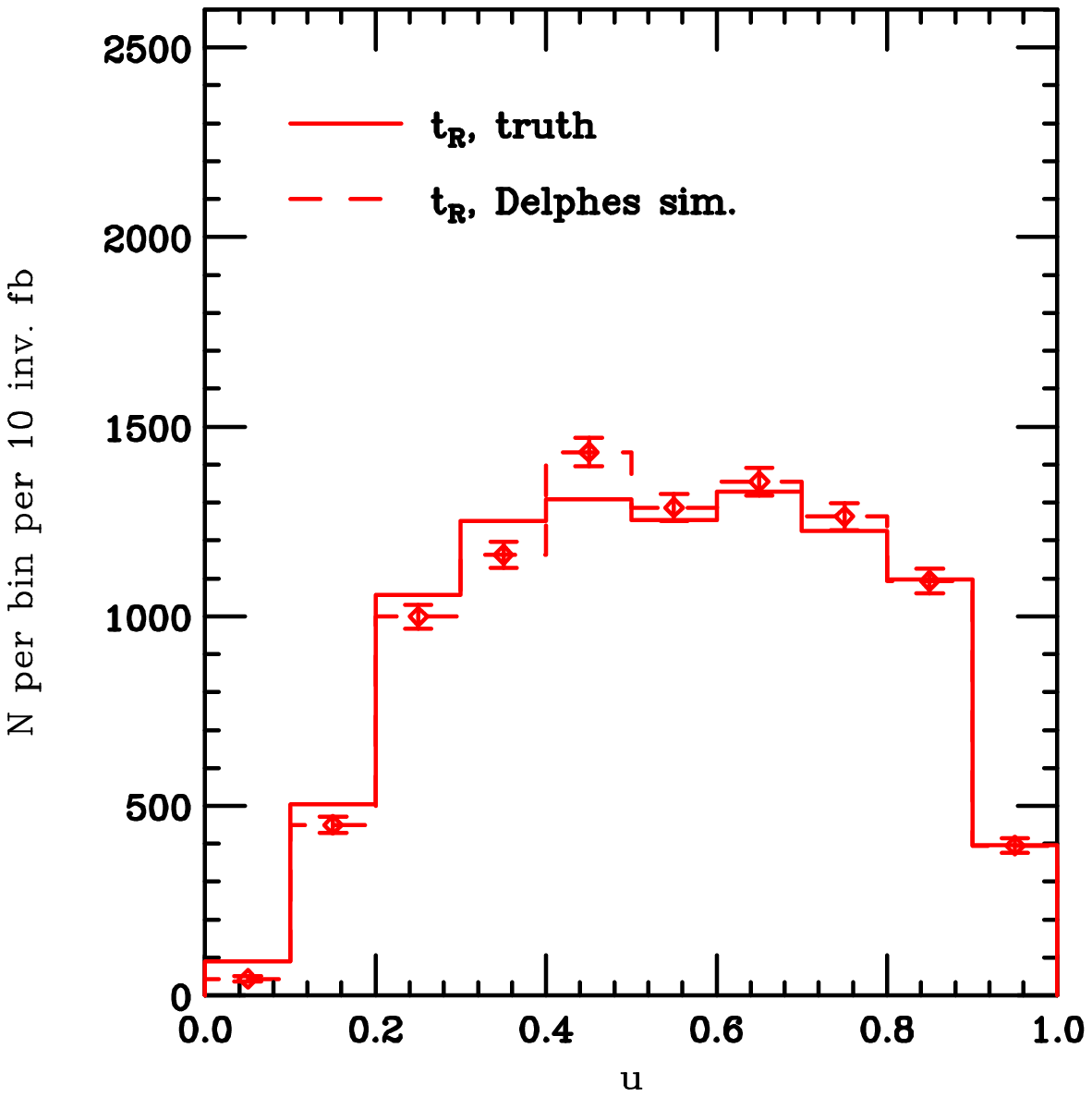}
  \caption{Shown in the figures is a comparison between the
results obtained for the $u$ variable for the 1.5 TeV
flavour-changing $Z'$ model  \textit{before} detector simulation (but applying all cuts and using equivalent
jet-finding) and \textit{after} the Delphes simulation for the left- and right-
handed fermions (blue and red respectively).}
\label{fig:utruthzp}
  \end{minipage}
  \end{tabular}
\end{figure}

\begin{figure}[!t]
\begin{tabular}{c}
\begin{minipage}{1.0\hsize}
  \centering 
    \vspace{5.5cm}  
   \includegraphics[angle=0,scale=0.52]{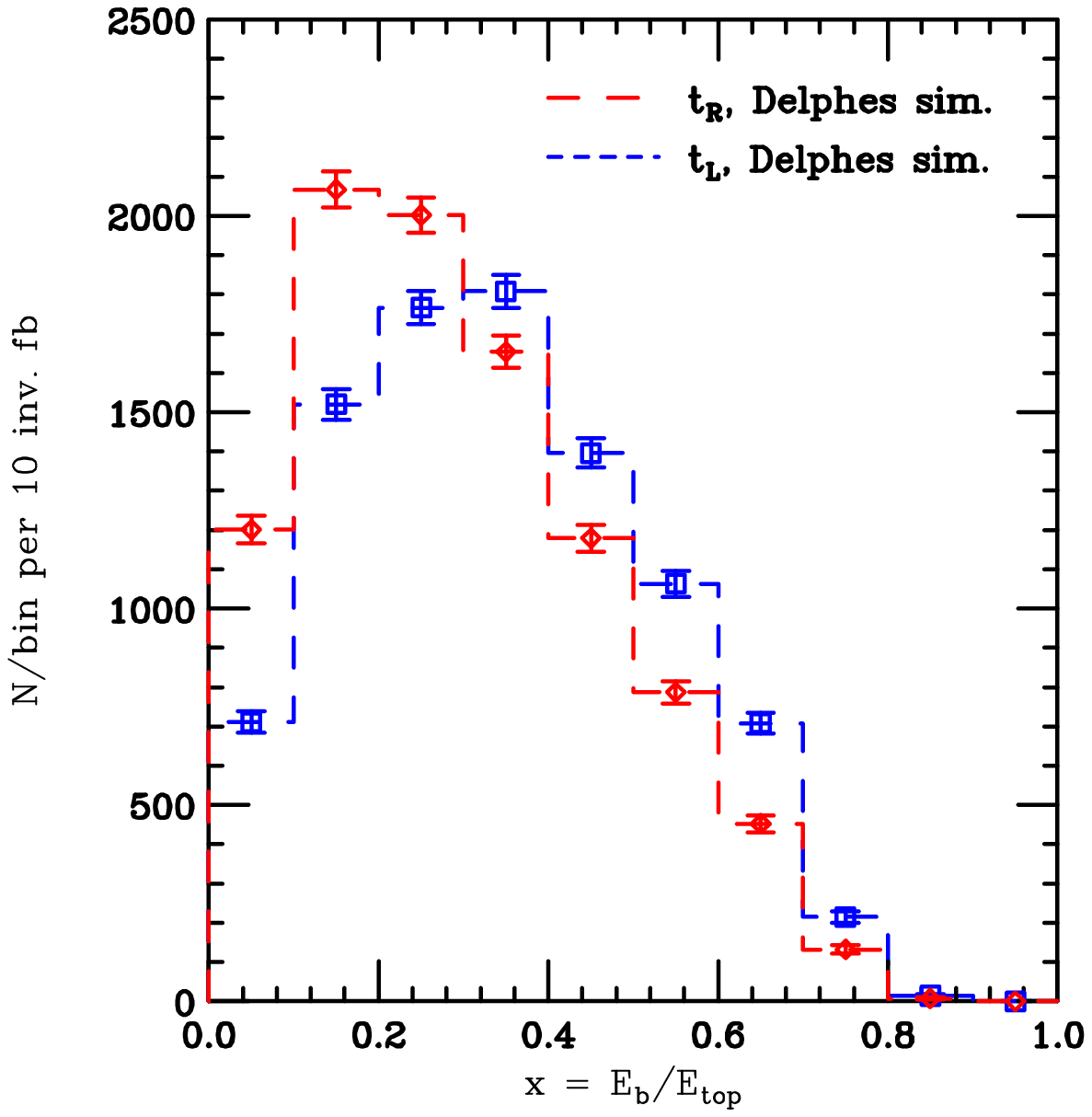}  
    \hspace{2.5cm}  
    \includegraphics[angle=0,scale=0.52]{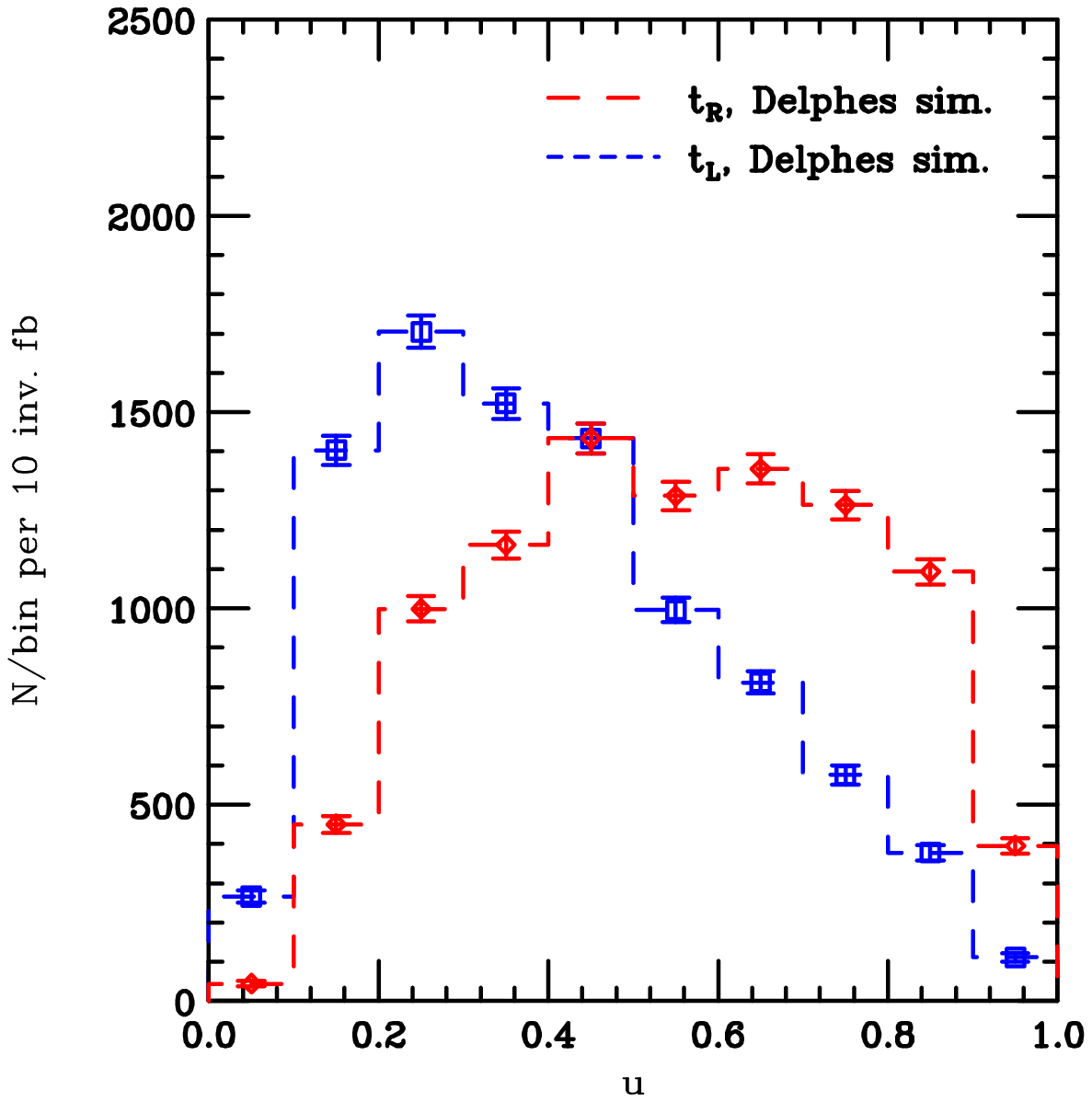}  
\caption{The $x_{\mathrm{top}}$ (left panel) and $u$ (right panel)  variables for left- or right-handed
  $Z'$ bosons decaying to $u\bar{t}$ or $\bar{u}t$, obtained from the reconstructed events for an LHC run at 14
  TeV, with 10 fb$^{-1}$.}
\label{fig:xudelpheszp}
\end{minipage}
\end{tabular}
\end{figure}

The `best' reconstructed top mass, after detector effects, was found to possess a peak at the correct
on-shell top mass, $\sim 174 \gev$, with an approximately Gaussian
distribution of width $\sim 20 \gev$. To illustrate the effect of the
Delphes detector simulation, we show a comparison of results before
and after Delphes processing in Figs.~\ref{fig:xtoptruthzp}
and~\ref{fig:utruthzp} for $x_{\rm top}$ and $u$ respectively. The comparisons between the left-
and right-handed variable distributions after detector effects are shown in
Fig.~\ref{fig:xudelpheszp}. It is clear
that, at least with the minimum cuts, the $u$ variable performs well
in discriminating between the pure left- or right-handed top
quarks. The $x_{\mathrm{top}}$ variable is less different
between the two top helicities, but the difference is still
statistically significant. The advantage possessed by the variable $u$ is
clear: since no explicit reconstruction is required, the approximations
that are associated with this process do not have a significant effect. As can be seen in the before/after plots in Figs.~\ref{fig:xtoptruthzp}
and~\ref{fig:utruthzp}, the $u$ variable is less sensitive to the
experimental effects, which seem to `squeeze' the distributions towards
middle values of the energy fraction more dramatically in the case of
$x_{\rm top}$. 

\begin{figure}[!t]
\begin{tabular}{c}
\begin{minipage}{1.0\hsize}
  \centering 
  \vspace{4.5cm}
 \includegraphics[scale=0.55, angle=0]{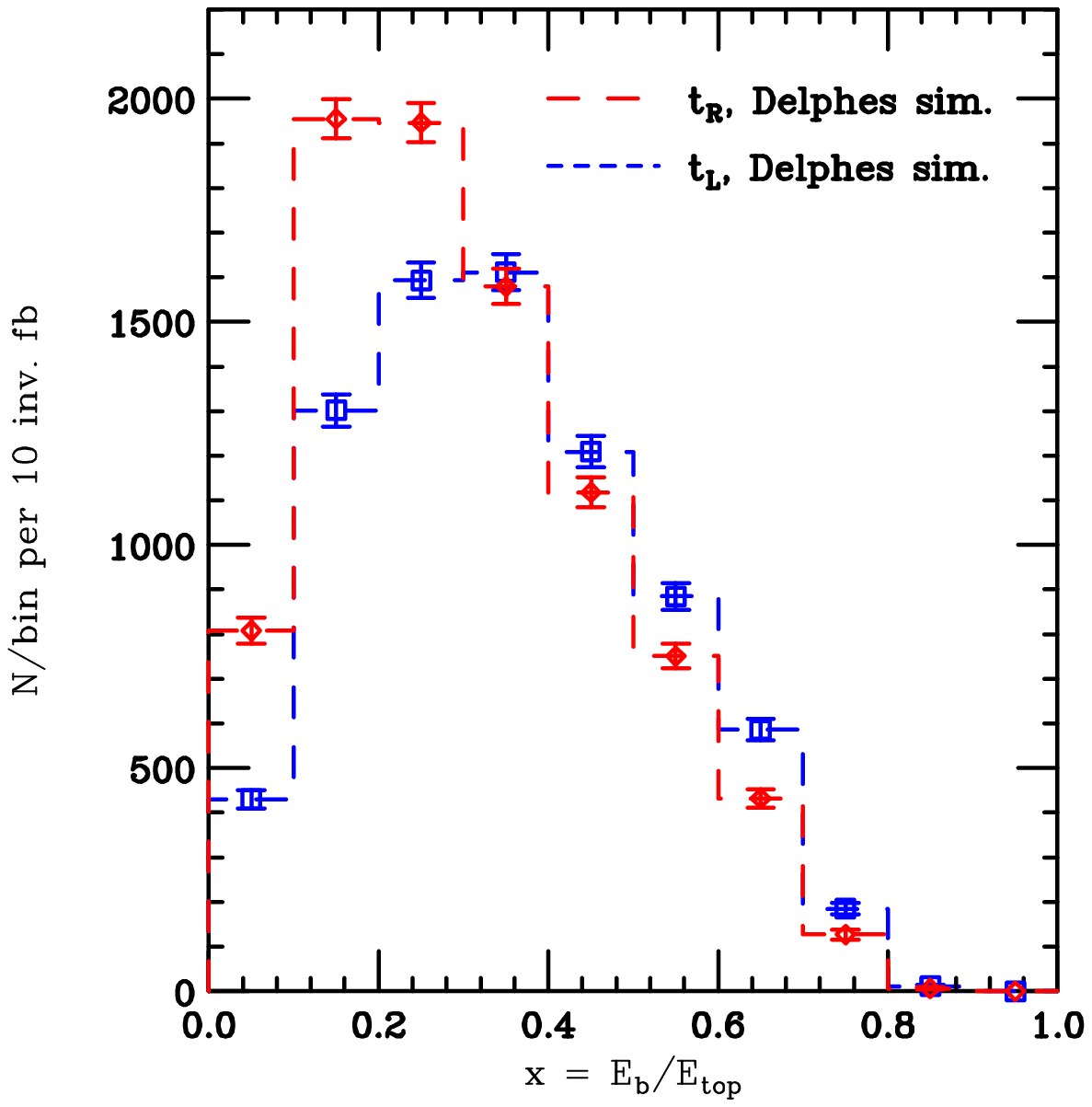}
  \hspace{1.5cm}
  \includegraphics[scale=0.55,angle=0]{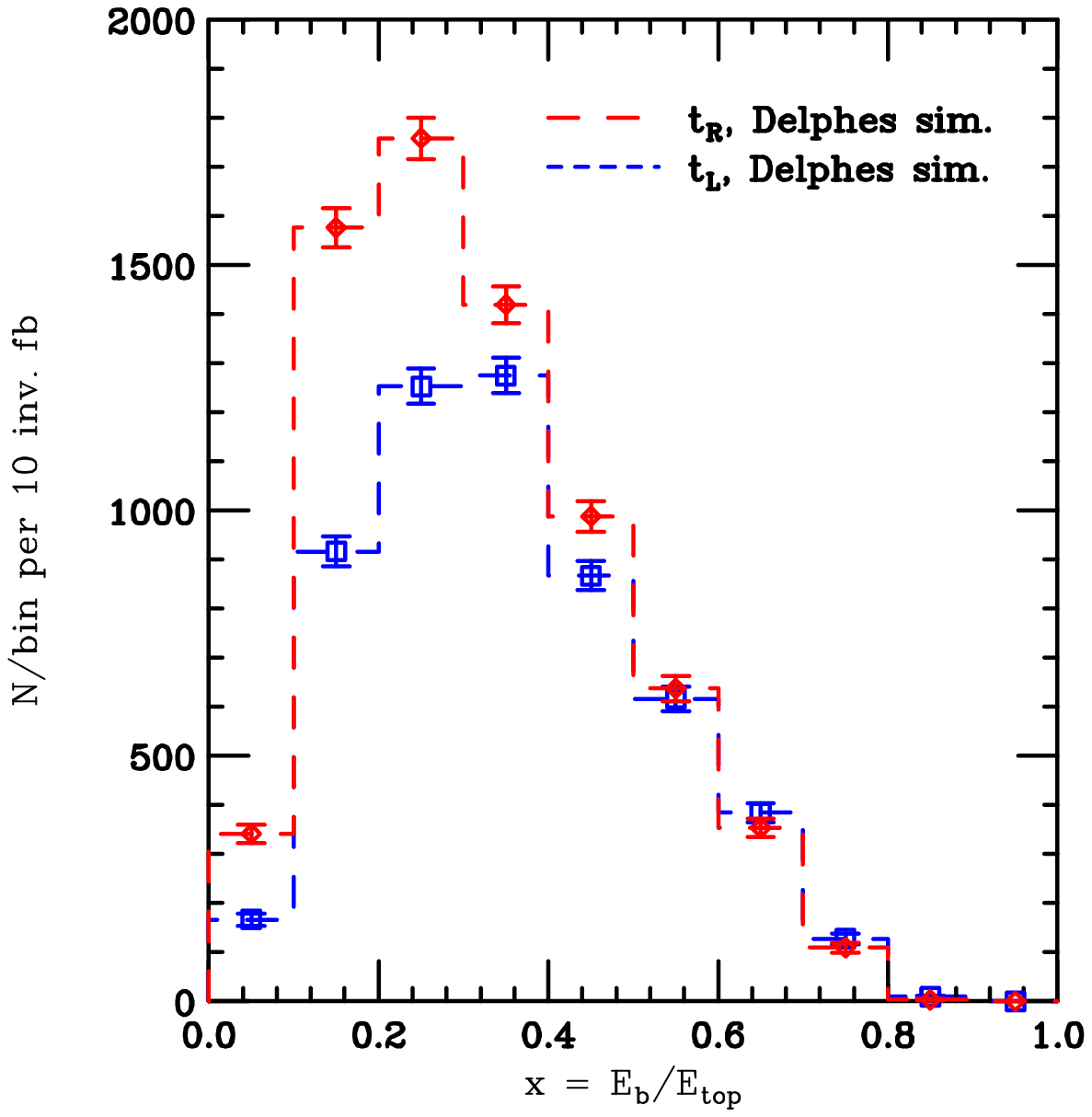}
  \caption{The $x_{\mathrm{top}}$ variable for a left- or right-handed
  $Z'$ bosons decaying to $u\bar{t}$ or $\bar{u}t$, obtained from the reconstructed events for an LHC run at 14
  TeV, with 10 fb$^{-1}$, with the set of cuts A (left) and B
  (right), as explained in the text.}
\label{fig:xtopzprimecuts}
\end{minipage}
\\
\begin{minipage}{1.0\hsize}
  \centering 
  \vspace{6.5cm}
 \includegraphics[scale=0.55, angle=0]{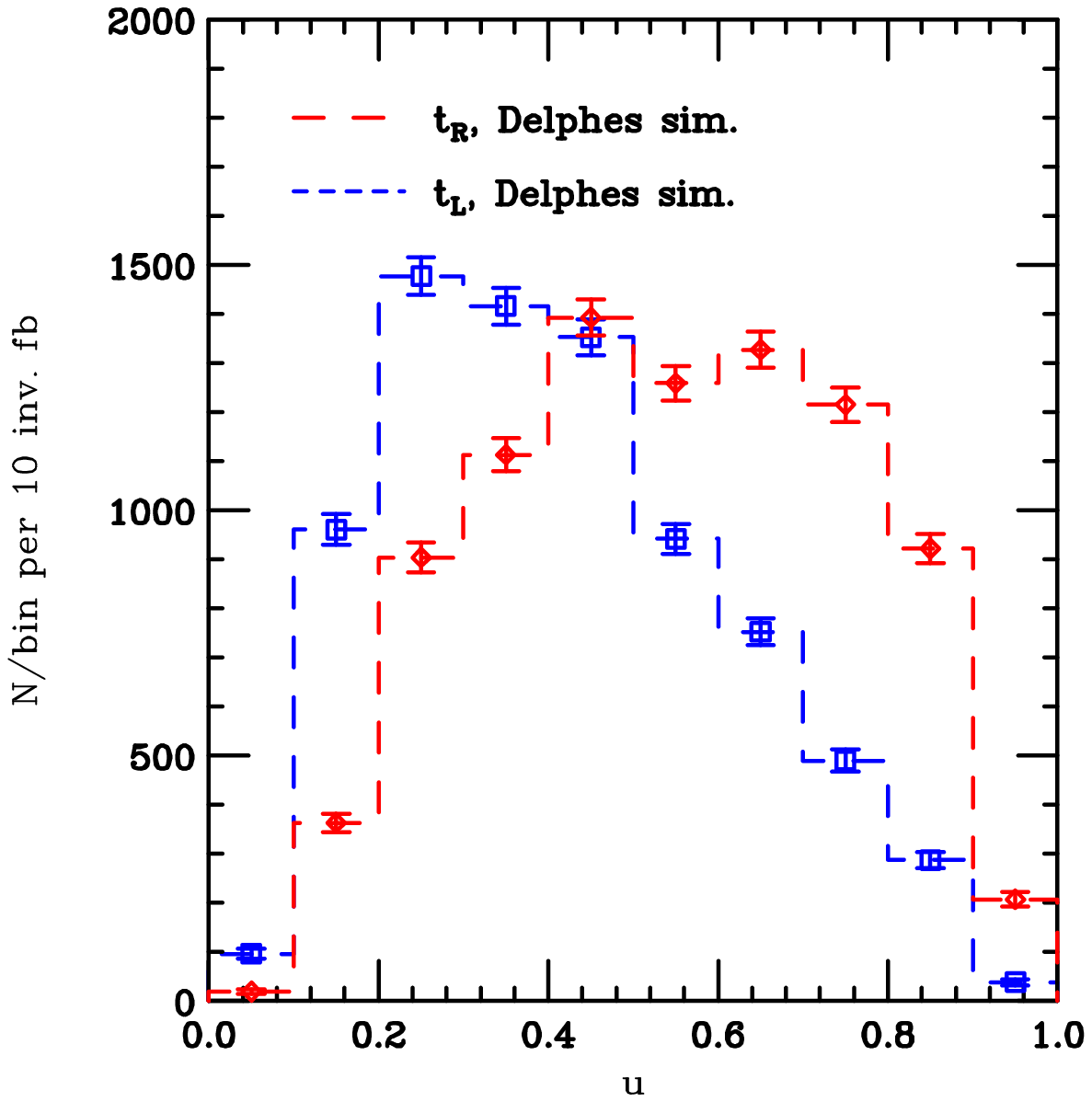}
  \hspace{1.5cm}
  \includegraphics[scale=0.55,angle=0]{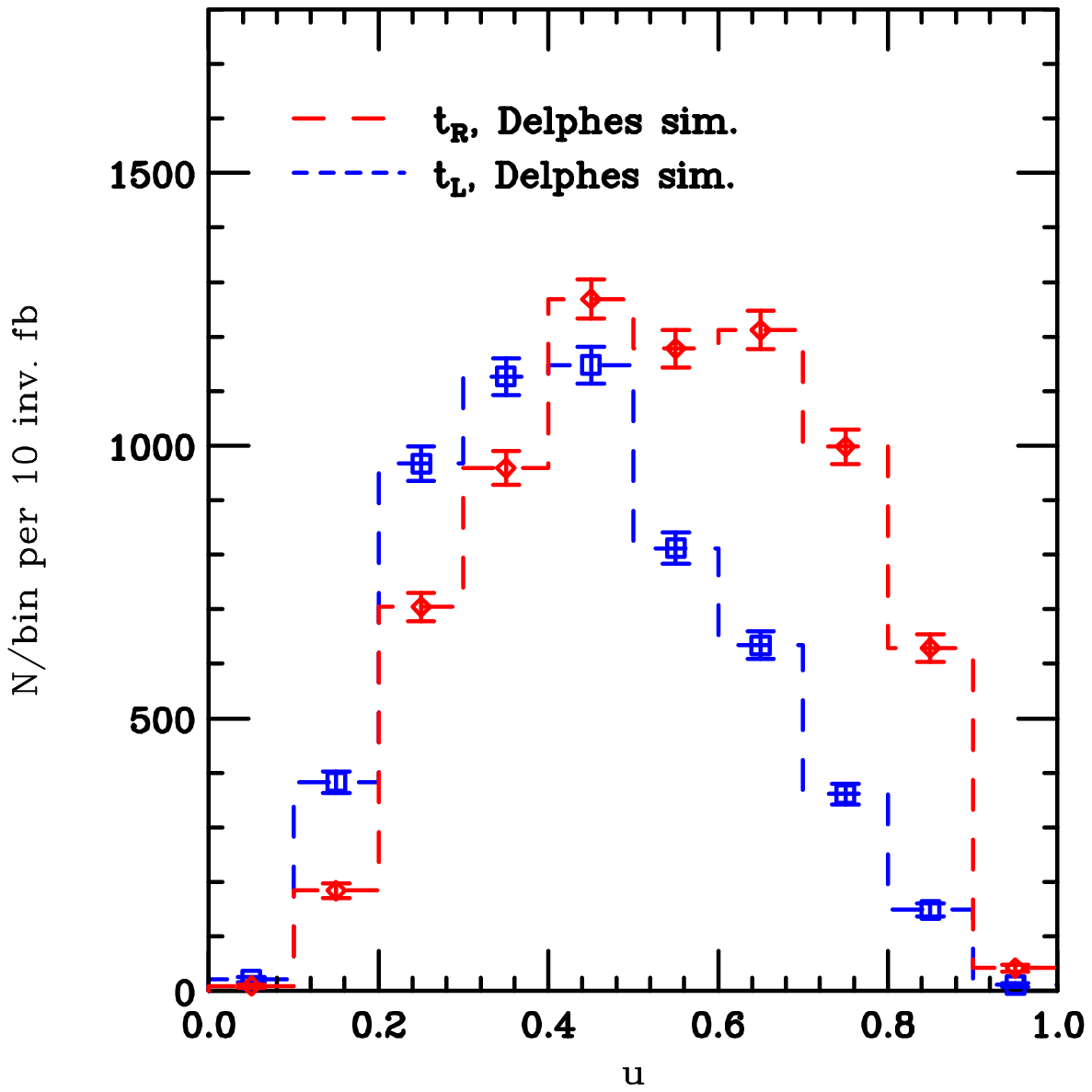}
  \caption{The $u$ variable for a left- or right-handed
  $Z'$ bosons decaying to $u\bar{t}$ or $\bar{u}t$, obtained from the reconstructed events for an LHC run at 14
  TeV, with 10 fb$^{-1}$, with the set of cuts A (left) and B
  (right), as explained in the text.}
\label{fig:uzprimecuts}
\end{minipage}
\end{tabular}
\end{figure}

To investigate the effect of higher transverse momentum cuts on the
objects used in calculating these variables, we constructed two further
sets of plots with higher cuts which we call `A' and `B' and are,
respectively, $p_T > 30\gev$ and $p_T > 50 \gev$ for both jets and
leptons. The resulting distributions are shown in Figs.~\ref{fig:xtopzprimecuts}
and~\ref{fig:uzprimecuts}, where the set of cuts A is shown on the left, and set B on
the right. Table~\ref{tb:chisqzprime} shows the value of $\chi^2 /
N_{\mathrm{d.o.f.}}$ between the left- and right-handed distributions
for the case of minimal cuts, as well as cuts `A' and
`B'.\footnote{For the comparison of two binned data sets, $\chi^2$ can be defined
as~\cite{numericalrec}:
\beq
\chi ^2 = \sum_i \frac{(R_i - S_i)^2}{ R_i + S_i } \nonumber\;, 
\eeq
where $R_i$ and $S_i$ are the number of events in $i^{\mathrm{th}}$
bin of the first and second data sets respectively.} The value
of $\chi^2 / N_{\mathrm{d.o.f.}}$ indicates how distinguishable the two distributions are statistically. It
  is evident that even in the case of higher cuts, discrimination in
  the specific scenario between the left- and right-handed modes is
  still possible. The higher cuts would also be beneficial for the rejection
  of proton-proton pile-up,\footnote{Pile-up is contamination originating from
    multiple secondary proton-proton collisions in the same bunch-crossing.} which would be detrimental at high instantaneous
  luminosity.

\begin{table}[!t]
\begin{center}
\begin{tabular}{|c|c|c|c|c|} \hline
Cut set & $x_\mathrm{top}$ & $u$  \\\hline \hline
Min. & 40.7  & 145.8 \\\hline 
A &  36.7 & 112.2 \\\hline 
B &   37.6 & 75.8 \\\hline 
\end{tabular}
\end{center}
\caption{The value of $\chi^2 / N_{\mathrm{d.o.f.}}$ between the left-
and right-handed distributions in the $Z'$ model with decays to
$t\bar{u}$ and $\bar{t} u$, for the three
different sets of cuts. It is evident that the distributions are
distinguishable even for the higher cuts, with the $u$ variable
distribution performing best in all cases.}
\label{tb:chisqzprime}
\end{table}

\subsection{$Z' \rightarrow \tau^+ \tau^-$}\label{sec:ztautau}

Another interesting example, which we can use to examine the effectiveness of the energy ratio
$x_\tau$, is a model of a heavy $Z'$ possessing a $\tau^+ \tau^-$ decay
mode. For this model, we define a Lagrangian density similar to the one given in the
previous section:
\begin{eqnarray}\label{eq:zprimetautau}
\mathcal{L} &=& g_{Z'}^{\tau\tau R} Z'^\mu \bar{\tau}_R \gamma_\mu
\tau_R  + g_{Z'}^{\tau\tau L}
  Z'^\mu \bar{\tau}_L \gamma_\mu \tau_L \nonumber \\
&+&  g_{Z'}^{uuR} Z'^\mu \bar{u}_R \gamma_\mu u_R  + g_{Z'}^{uuL}
  Z'^\mu \bar{u}_L \gamma_\mu u_L + \mathrm{h.c.}\;\;,
\end{eqnarray}
We examine discrimination between two cases, $Z' \to \tau^+_R
\tau^-_R$ ($g_{Z'}^{\tau\tau R} = 1$, $g_{Z'}^{\tau\tau L} = 0$) and
$Z' \to \tau^+_L \tau^-_L$ ($g_{Z'}^{\tau\tau R} = 0$,
$g_{Z'}^{\tau\tau L} = 1$), using the $x_\tau$ variable defined in
Eq.\,(\ref{eq:x}).

We checked that the irreducible backgrounds to the $\tau^+ \tau^-$
decay mode, e.g. originating from the $Z^0/\gamma \rightarrow \tau^+ \tau^-$, can be rejected
using an invariant mass cut on the $\tau$-jet pair. For example, an invariant mass
cut of $\sim 500 \gev$ offers a rejection factor of $\mathcal{O}(10^{-4})$
for these backgrounds, bringing them to cross sections
$\mathcal{O}(0.1~\mathrm{pb})$, while reducing the signal only by a
factor of $\sim 0.7$ for the case of a 1.5 TeV $Z'$, maintaining a cross section of a few picobarns. Possible backgrounds to the helicity determination may also arise from fake
$\tau$-tagged QCD events, and these would need to be assessed by the
individual experimental collaborations.

To compute the $x_\tau$ variable, one needs to reconstruct the neutrino energies
from the $\tau$ decays.  
For this purpose, the authors of Ref.\cite{Anderson:1992jz} have used 
the collinear approximation to reconstruct the neutrino momenta.
In the collinear approximation, the neutrinos are assumed to be collimated to 
the associated $\tau$-jets.
This assumption is almost always good in cases of a heavy resonance
decaying to $\tau$ leptons. 
Once the neutrino momentum directions are determined, the neutrino energies can be calculated 
using the missing transverse momentum constraint:
\begin{equation}
\begin{pmatrix}
p_{\rm miss}^x \\ p_{\rm miss}^y
\end{pmatrix} 
=
\begin{pmatrix}
\sin\theta_{\rm jet 1} \cos\phi_{\rm jet 1} &&  \sin\theta_{\rm jet 2}
\cos\phi_{\rm jet 2} \\
\sin\theta_{\rm jet 1} \sin\phi_{\rm jet 1} &&  \sin\theta_{\rm jet 2}
\sin\phi_{\rm jet 2} 
\end{pmatrix} 
\begin{pmatrix}
E_{\nu1} \\ E_{\nu_2}
\end{pmatrix} \,,
\label{eq:collmat}
\end{equation}
where $\theta_{\rm{jet} i}$, $\phi_{\rm{jet} i}$ are the polar and
azimuthal angles respectively, related to jet $i$, $p^{x,y}_{\rm
  miss}$ are the missing transverse momentum components and $E_{\nu
  i}$ is the energy of neutrino $i$.  

However, when the two jets are back-to-back, i.e. $\phi_{\rm{jet} 1} =
\phi_{\rm{jet 2}} + \pi$,  the inverse of the matrix in Eq.\,(\ref{eq:collmat}) becomes 
singular and any small mismeasurement on the missing transverse energy 
or jet momentum directions would cause a very large error on the
reconstructed neutrino energy~\cite{Elagin:2010aw}. The back-to-back configuration is strongly preferable if a heavy resonance, such as 
the standard model Higgs boson or a $Z'$, is considered. 

\begin{figure}[!t]
\centering 
\vspace{2.5cm}
\hspace{0.0cm}
\includegraphics[scale=0.45, angle=0]{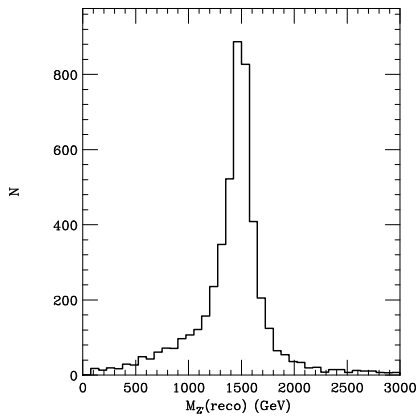}
 \caption{The real part of the complex solutions that arise after
   applying the two linear missing-$p_T$ constraints and the two
   quadratic $\tau$ mass-shell constraints. A clear peak structure is
   seen at the input mass of 1.5 TeV.}
\label{fig:from_trueR}
\end{figure}

One can avoid the use of the collinearity assumption by instead using information on the $\tau$ decay vertices
\cite{Gripaios:2011jm}. 
The most useful and best-measured attribute of these is their impact parameter. 
The impact parameter {\bf b} is the displacement of a decay vertex 
in a direction perpendicular to that of the visible decay momentum, 
in this case the $\tau$ jet momentum ${\bf p}_j$. 
Then the invisible momentum ${\bf p}_\nu$ must lie in the (${\bf b}, {\bf p}_j$) plane, 
so we can write ${\bf p}_\nu = x{\bf b} + y {\bf p}_j$. 
For hadronic $\tau$ decays, the invisible momenta are carried by single neutrinos 
and so their four-momenta are fixed by $x$ and $y$ for each decay. In
this section we focus on hadronic $\tau$ decays, by including a
lepton veto in our event selection criteria. 
These four quantities are subject to two linear missing-$p_T$ constraints 
and two quadratic $\tau$ mass-shell constraints, 
giving four (complex) solutions for the neutrino momenta which allow us to
compute the invariant mass of the $\tau$ pair. In
Fig.~\ref{fig:from_trueR}, we plot the real part of the invariant
mass, where we use the true jets and missing transverse momenta for $10^4$ events.
A distinct peak structure is seen at the input $Z'$ mass of 1.5\,TeV.

\begin{figure}[!t]
  \centering 
  \vspace{3.5cm}
 \includegraphics[scale=0.38, angle=0]{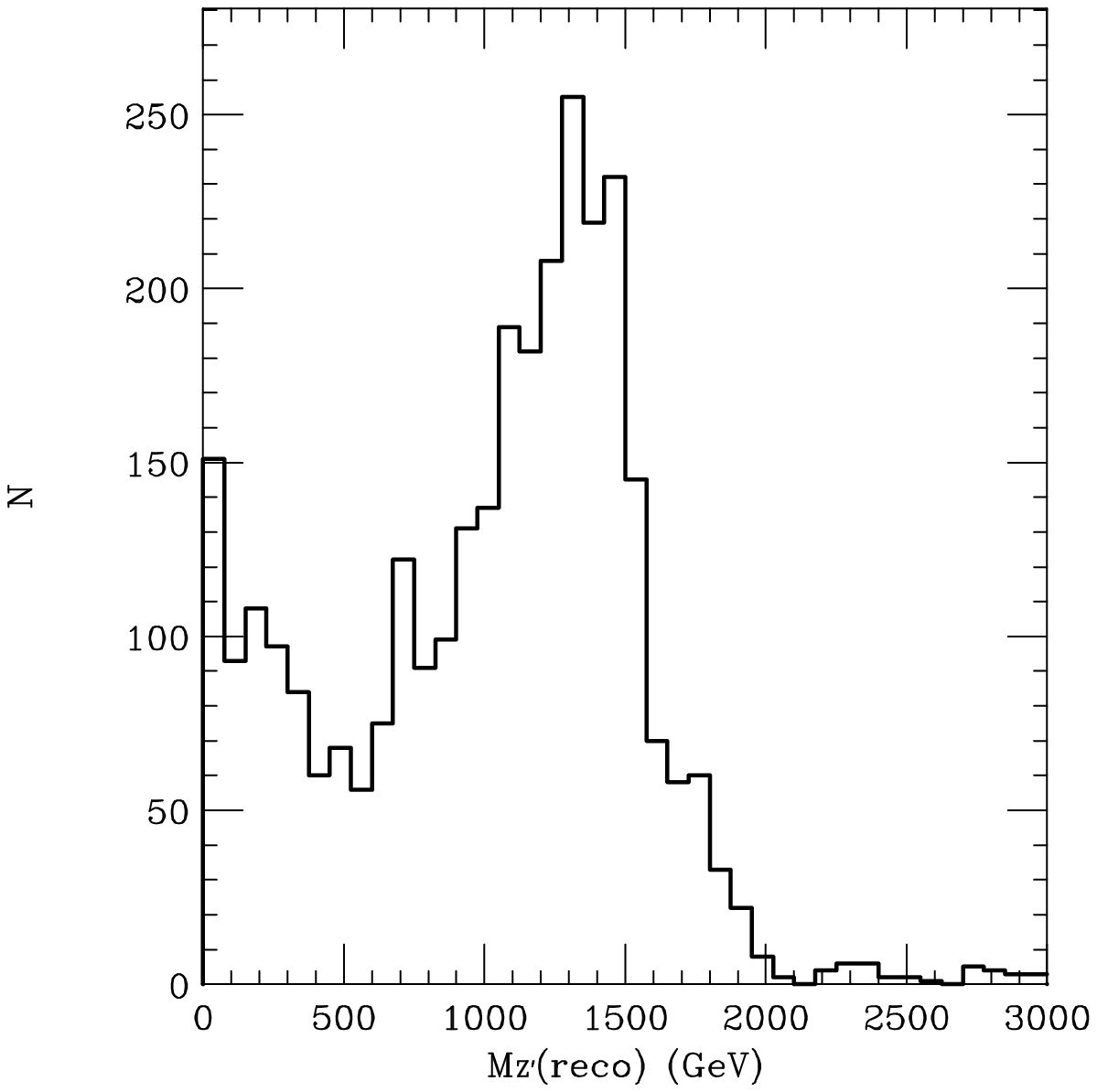}
  \hspace{1.0cm}
 \includegraphics[scale=0.38,angle=0]{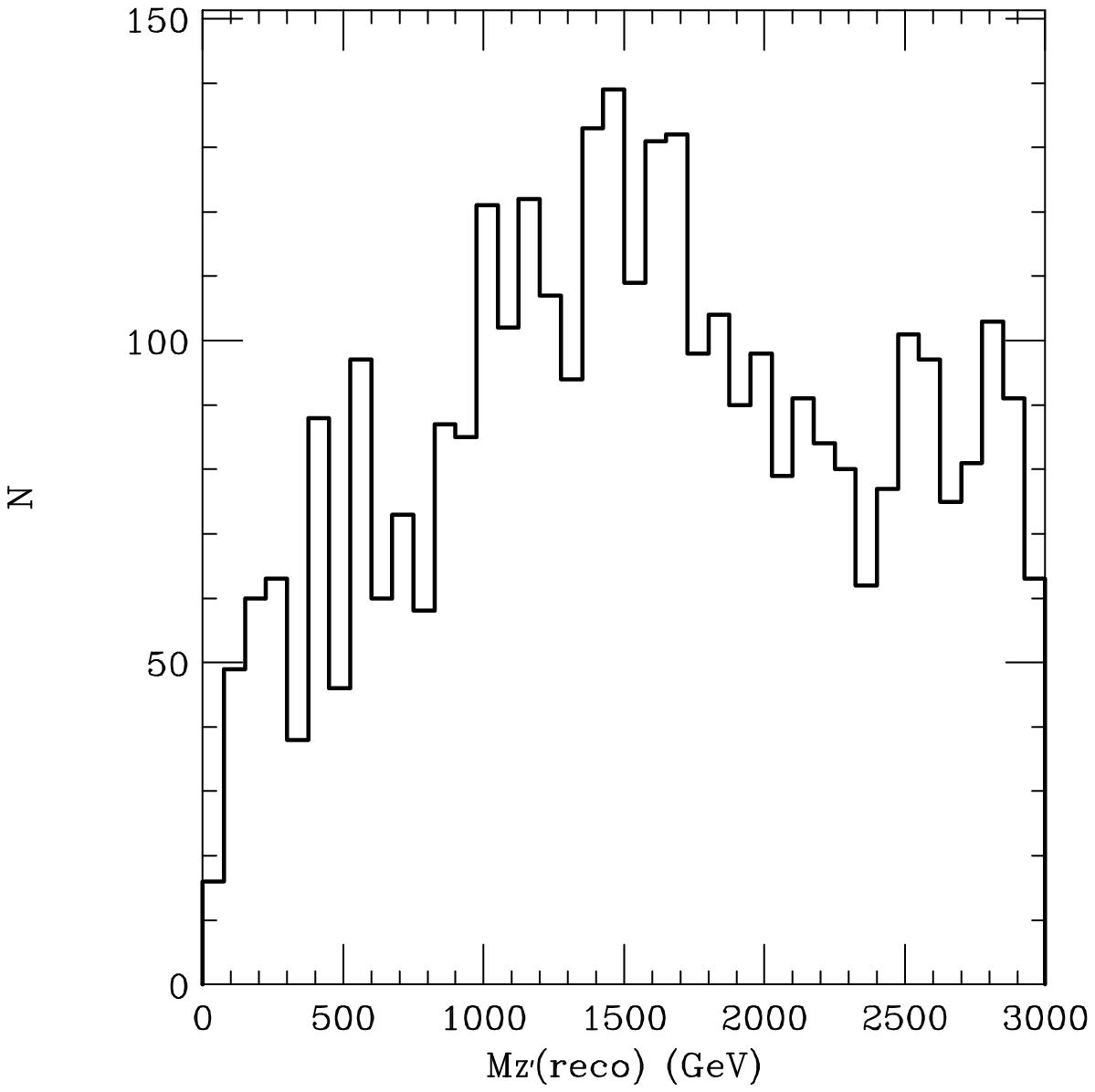}
 \hspace{1.0cm}
\includegraphics[scale=0.38,angle=0]{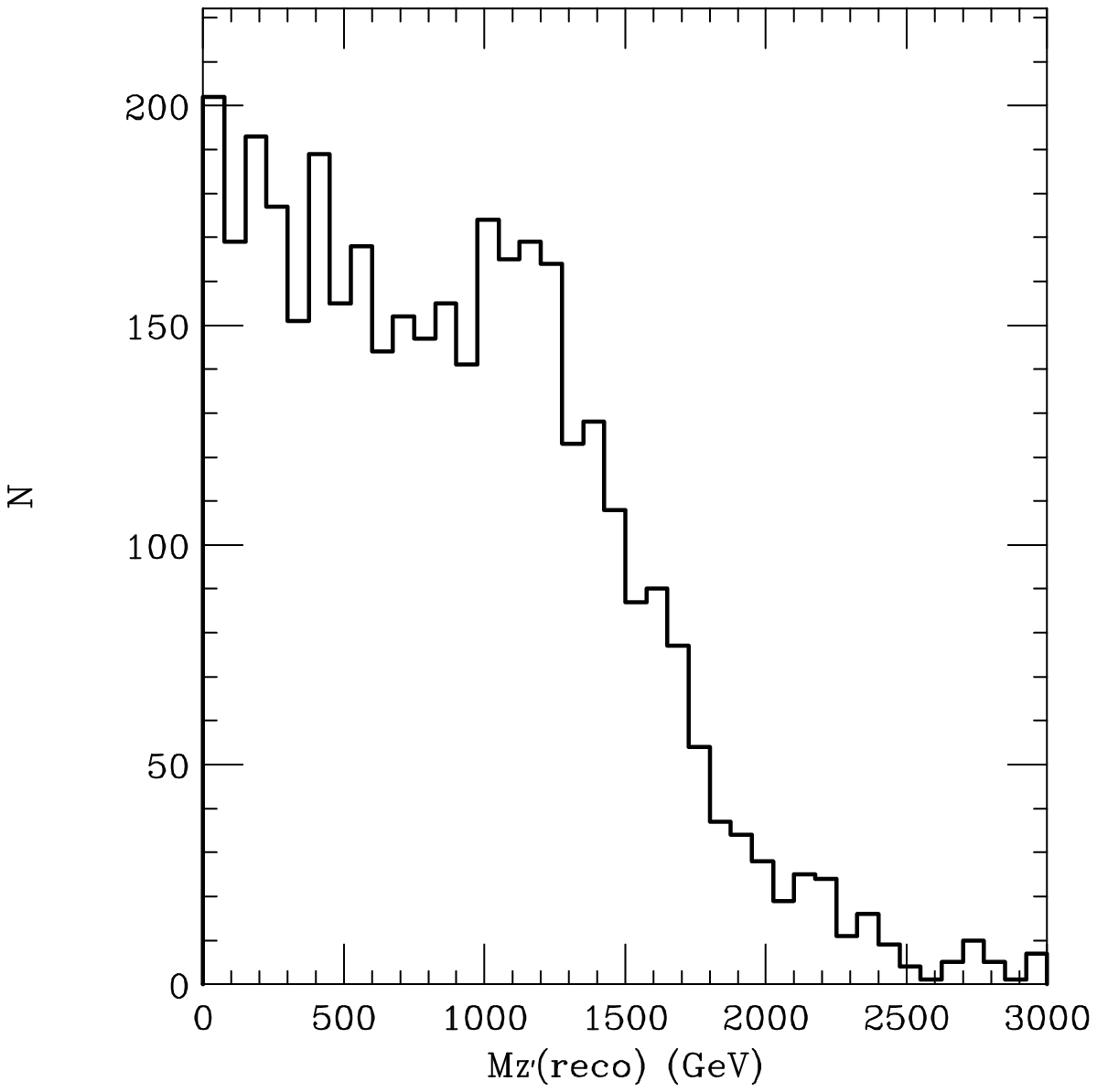}
  \caption{The distributions of the real part of the four
    complex solutions resulting from the reconstruction of the events
    after using detector-level jets instead of parton-level jets
    (left plot), detector-level missing transverse momentum instead of
    the true one
    (middle plot) and using completely detector-level objects (i.e. both
    for jets and MET, right plot). In all plots non-smeared $\tau$ vertices
    have been used. }
\label{fig:Mz_sim}
\end{figure}

Unfortunately, our reconstruction method is still sensitive to the momentum mismeasurement.
In Fig.~\ref{fig:Mz_sim}, we show the hadronization and detector effects on the
$Z'$ mass reconstruction.
In the left plot, the parton-level jets are replaced with the
detector-level jets obtained from the Delphes simulation, in the
middle plot only the
parton-level missing transverse momentum is replaced with the
detector-level one and in the right plot, we use all the detector-level
objects. In all of the plots in Fig.~\ref{fig:Mz_sim}, the true impact
parameter was used.  As can be seen, the peak structure is completely lost if one uses the
detector objects. The event selection cuts here, and in all the results
that follow in this section, are the default Delphes cuts for the ATLAS
detector ($p_{T,\ell} > 10 \gev$, $p_{T,\mathrm{jet}} > 20 \gev$)
along with the addition of the lepton veto (in this case rejecting all
events with an identified lepton). 

However it is important to realise that, in the model we are
considering, the $Z'$ mass can be measured independently from the decay modes 
containing the light quarks. In this case, we can use the mass shell constraint, $(p_{\tau_1}+p_{\tau_2})^2 = m^2_{Z'}$,
as an extra condition to correct for the detector resolution.
To accomplish this we construct the following likelihood function:
\begin{equation}
L_{\rm tot} = L_{\rm detector} \times L_{\rm phys} ,
\end{equation}
where we have defined
\begin{eqnarray}
L_{\rm detector} &\equiv& \Pi_{i=1}^2 (P_{E_i} P_{\theta_i} P_{\phi_i}
) \times P_E^{\rm miss} \times P_{\phi}^{miss}, \nonumber \\
L_{\rm phy} &\equiv& P_{m_{Z'}} \times \Theta({\rm Re}[E_{\nu_1}]) \times  \Theta({\rm Re}[E_{\nu_2}]).
\end{eqnarray}
The $P_{E_i}$, $P_{\theta_i}$,  $P_E^{\rm miss}$, $P_{\phi}^{miss}$ are Gaussian 
probability functions centred at the origin, with arguments $(E_{j_i} - E^{\rm obs}_{j_i})/E^{\rm obs}_{j_i}$,
$(\theta_{j_i} - \theta^{\rm obs}_{j_i})$, 
$(p_{\rm Tmiss} - p_{\rm Tmiss}^{\rm obs})/p_{\rm Tmiss}^{\rm obs}$, 
$(\phi_{p_{\rm Tmiss}} - \phi_{p_{\rm Tmiss}}^{\rm obs})$, respectively.
We also use the flat probability distribution, $P_{\phi_i}$, 
for the $(\phi_i - \phi_i^{\rm obs})$ with a range [-0.9:0.9]
so that the probability function matches the actual probability 
of mismeasurement, which is simulated by Delphes.
The likelihood $L_{\rm phys}$ allows us to correct the mismeasured observable 
by requiring physical conditions with some probability. $P_{m_{Z'}}$ is a Gaussian probability function with an argument of 
($m_{\tau\tau} - m_{Z'}$), where $m_{Z'}$ is the true $Z'$ mass assumed to be
measured through some other decay mode.
$\Theta(x)$ is 1 if $x>0$, 0 otherwise.    
The probability functions we use and the ones that appear in Delphes
are shown in appendix~\ref{app:delphesVparton}.

\begin{figure}[!t]
\centering 
\vspace{2.5cm}
\includegraphics[scale=0.45, angle=0]{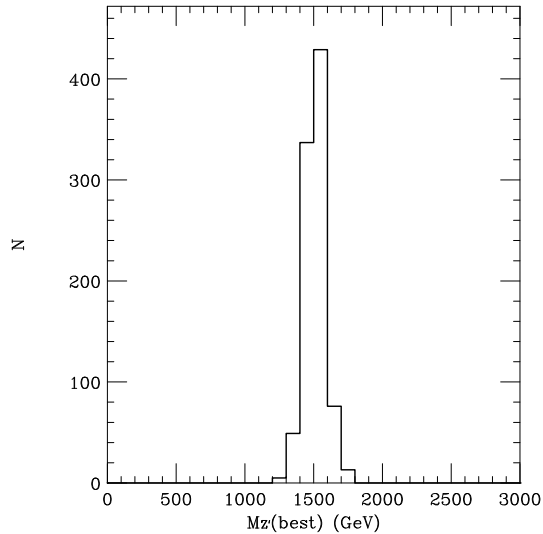}
 \caption{The $\tau\tau$ invariant mass of the maximum likelihood
   sample as described in the text. The distribution is sharply peaked
 at the input $Z'$ mass of 1.5 TeV.}
\label{fig:mZ_best}
\end{figure}

For each event, we generate 1000 pseudo-events, in which the observed momenta are
slightly shifted in a random `direction' according to the same probability function.
We only keep the pseudo-event, $i_{\rm max}$ (corresponding to $L_{\rm tot}^{\rm max}$), that
provides the maximum likelihood. We show the $\tau\tau$ invariant mass
distribution obtained from $i_{\rm max}$ sample
in Fig.~\ref{fig:mZ_best}.  

\begin{figure}[!t]
  \centering 
  \vspace{2.5cm}
  \hspace{3.5cm}
 \includegraphics[scale=0.45, angle=90]{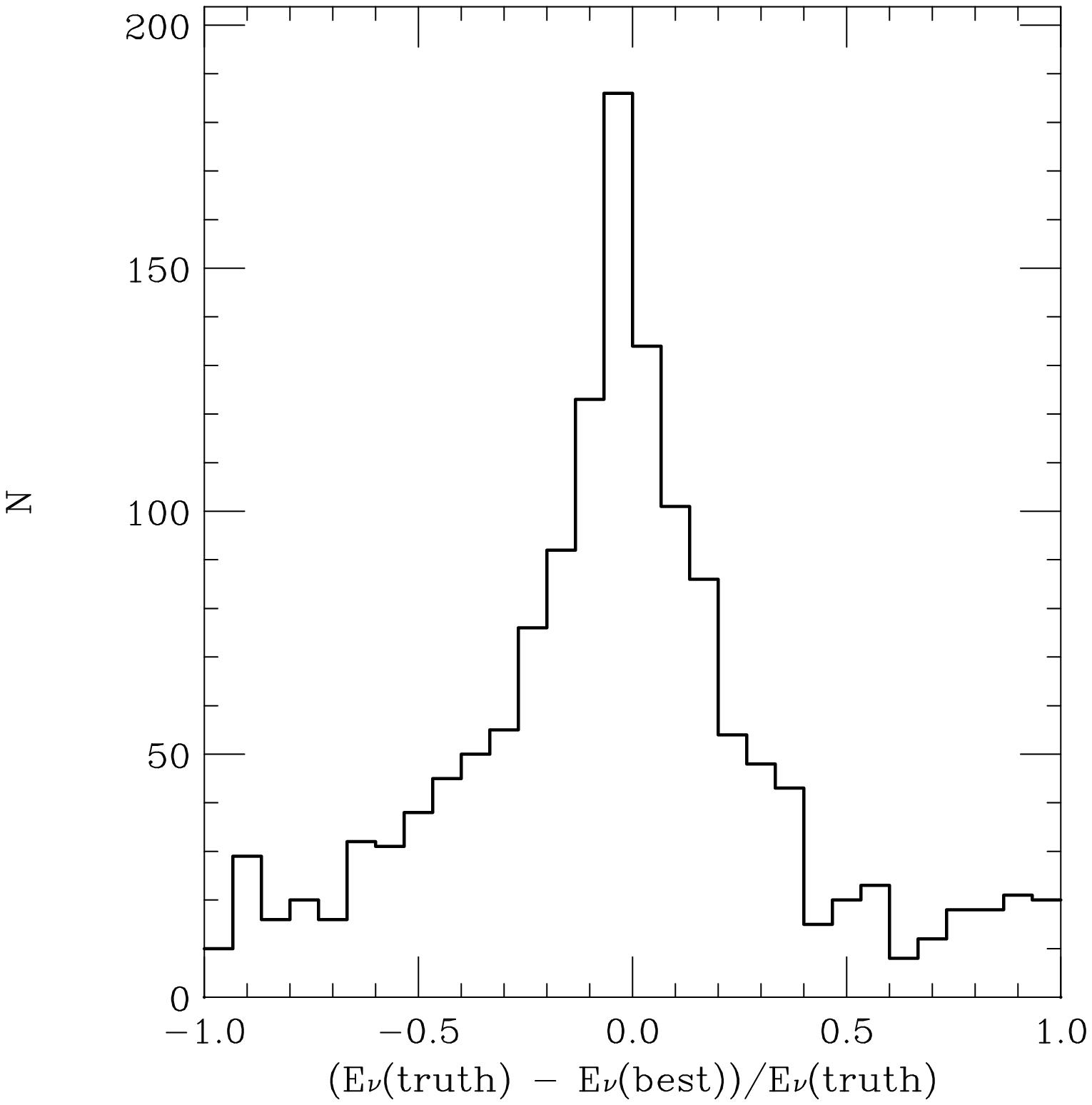}
  \hspace{1.5cm}
  \includegraphics[scale=0.59,angle=0]{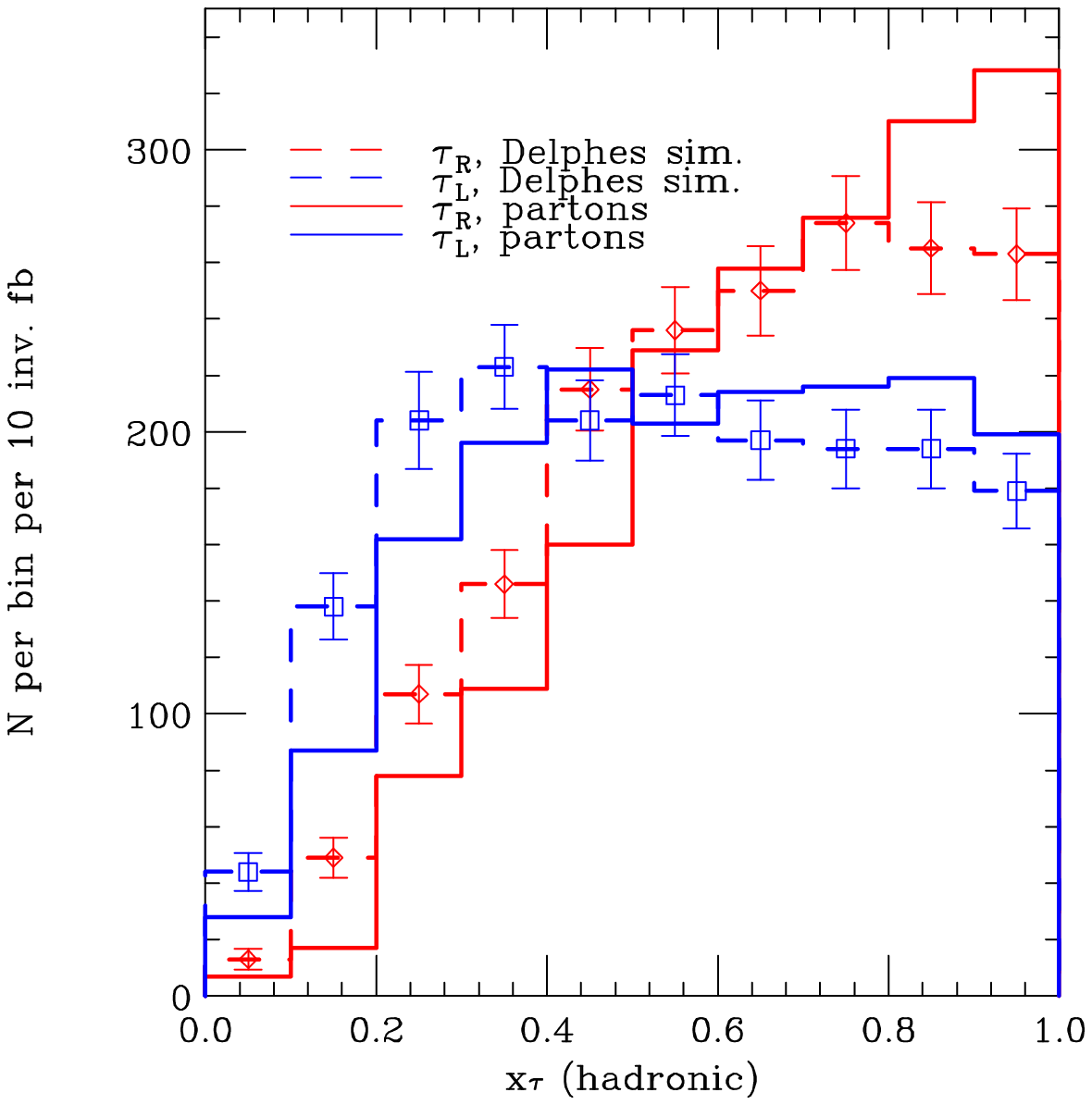}
  \caption{The relative difference between the true neutrino energy
    and the reconstructed neutrino energy using the likelihood method
    described in the text (left panel) and the resulting energy fraction,
    $x_\tau$ distributions for the purely left- and purely
    right-handed cases (right panel) for 1~fb$^{-1}$ of integrated
    luminosity at 14~TeV.}
\label{fig:Xrec_best}
\end{figure}

In the left panel in Fig.~\ref{fig:Xrec_best},
we show the relative difference between the true neutrino energy and
the reconstructed neutrino energy by the likelihood method.
As can be seen, the true neutrino energy is well reconstructed on an event-by-event basis
with about 50\% error.
The right panel in Fig.~\ref{fig:Xrec_best} shows the reconstructed
$x_\tau$ variable in this method, using 1~fb$^{-1}$ of integrated
luminosity at a 14~TeV LHC (corresponding to $\sim 7000$ events before
cuts). The red (blue) solid histogram is obtained from
$Z' \to \tau_R^+ \tau_R^- ~(Z' \to \tau_L^+  \tau_L^-)$ sample. The
lepton veto in this realistic case was applied by requiring no leptons
with $p_T > 10 \gev$ in $|\eta| < 2.4$. The dashed histogram is the corresponding parton-level distribution of $x_\tau$.
It is obvious that the reconstructed $x_\tau$ has a very similar distribution to the parton-level one,
and the difference between the left and right-handed $x_\tau$ distributions  is visible even
after the effects of detector resolution. The value of $\chi^2 /
N_{\mathrm{d.o.f}}$ was found to be $\sim 13.8$, indicating the high
difference between the left- and right-handed $x_\tau$ histograms. 

\subsection{Third-generation leptoquark pair-production}\label{sec:leptoquark}
Methods for reconstructing third-generation scalar leptoquark states
in events where they were pair-produced have been studied in
Ref.~\cite{Gripaios:2010hv}. There, different mass variables were
constructed and all the possible combinations of decay modes were
studied, including QCD and detector effects. The main background,
$t\bar{t}$ production, was shown in Ref.~\cite{Gripaios:2010hv} to produce negligible contribution in
the suggested reconstruction variables in the vicinity of the leptoquark
signal. We will focus here on the leptoquark types that can decay to a top quark and a tau
lepton. 

For a list of states that can decay to the $t\tau$ modes see
Tables 1 and 2 in Ref.~\cite{Gripaios:2010hv}. Instead of focusing on
a specific leptoquark type, we consider a general scalar leptoquark
which possesses a branching fraction of 1 to a top quark and a $\tau$
lepton. We call the leptoquarks of this type, with electromagnetic charge $\pm5/3$, $S_{LL}$ and $S_{RR}$, where the index
indicates  that the leptoquark will decay either to $t_L\tau_L$ or
$t_R\tau_R$ respectively. In appendix~\ref{app:leptoalternative} we consider leptoquarks of
electromagnetic charge $\pm1/3$ decaying
to the mixed combinations $\bar{t}_R\tau_L$ or $\bar{t}_L \tau_R$
($S_{RL}$ and $S_{LR}$ respectively). To obtain results for other
scenarios of leptoquarks that decay into this mode, one has to simply
rescale the results to account for the appropriate cross sections. 

\subsubsection{Parton-level results}
We produce Monte Carlo distributions of the variables outlined in
section~\ref{sec:vardef} for leptoquark
states to compare to the predicted distributions at parton level. We
do not present parton-level distributions for the $\tau$, as the corresponding
parton-level results have been already extracted from the Monte Carlo event
generator itself and appear in Fig.~\ref{fig:xtau_full}.

\begin{figure}[htb]
  \centering 
  \vspace{1.5cm}
  \hspace{5.0cm}
 \includegraphics[scale=0.44,angle=90]{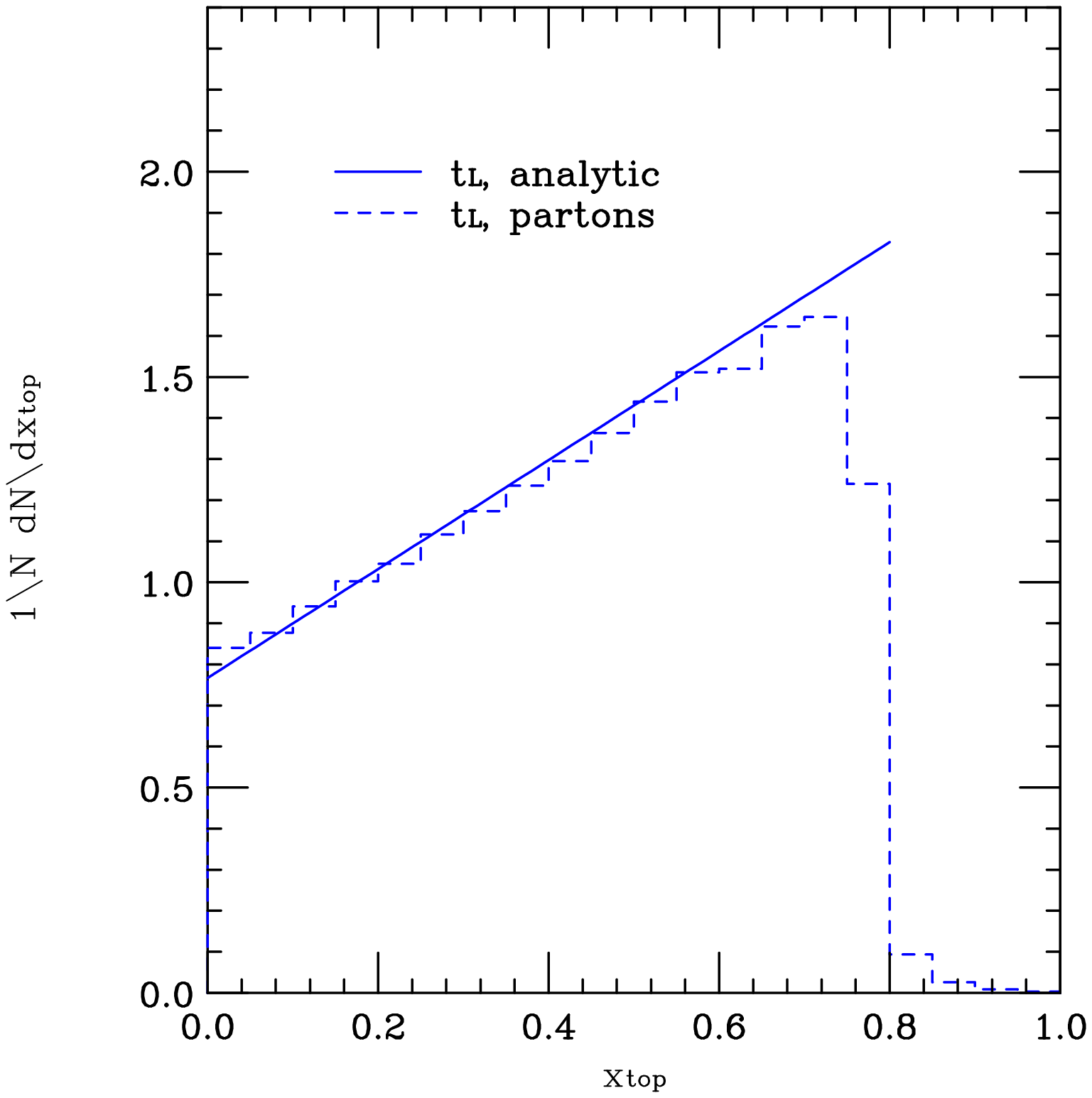}
  \hspace{4.5cm}
  \includegraphics[scale=0.44,angle=90]{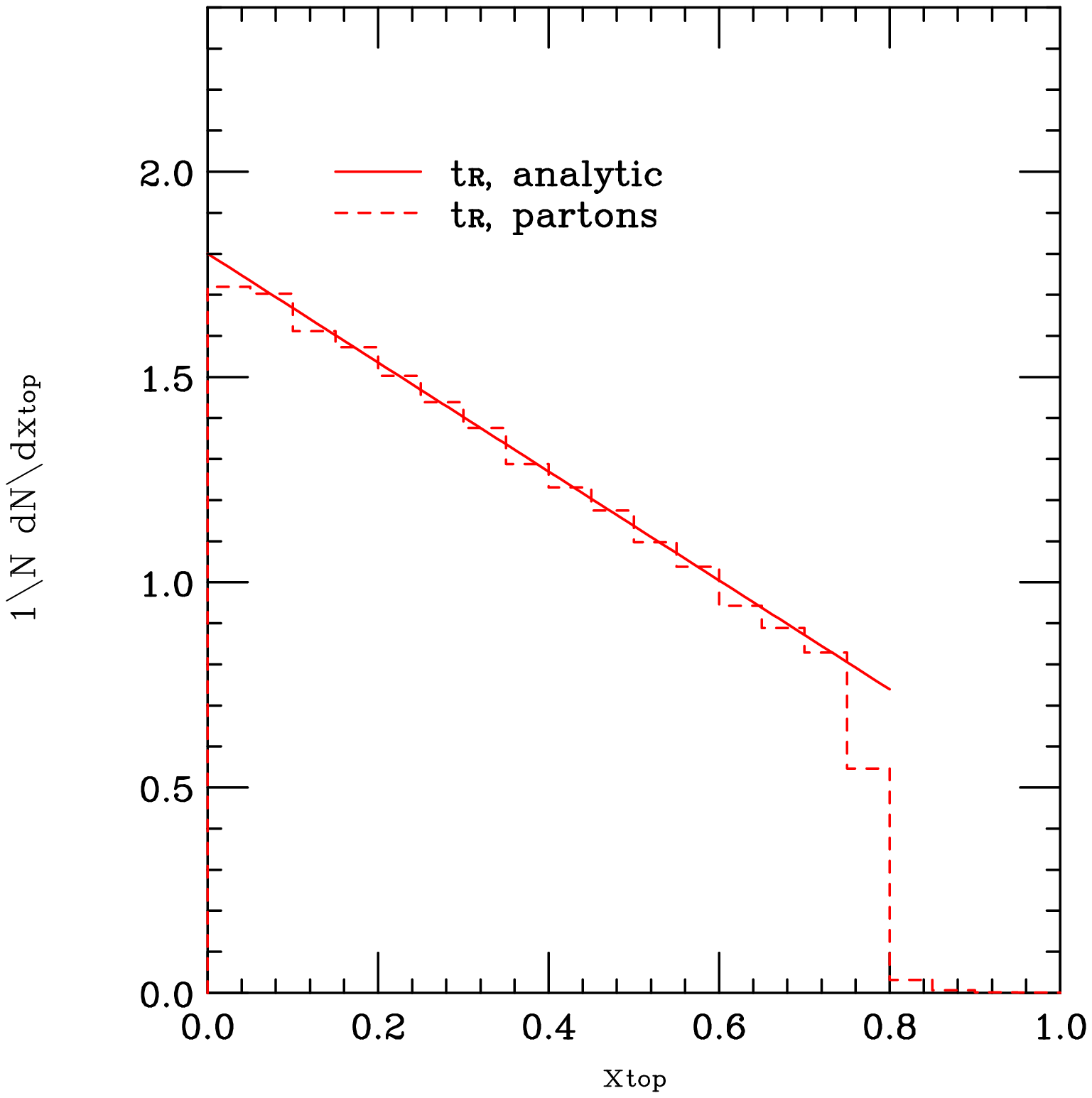}
  \vspace{0.5cm}
  \caption{The Monte Carlo results for the energy fractions $x_{\mathrm {top}} = \mathcal{E}_{b} /
\mathcal{E}_{\mathrm{top}}$ for $\tau_L t_L$
(left) and $\tau_R t_R$ (right) in the highly-boosted case. These are compared to analytical predictions as described in the text.}
\label{fig:ds0ts0}
\end{figure}

The $S_{XX}$ ($X \in \{R,L\}$) leptoquark can decay to $t\tau$ modes, described by the
Lagrangian terms:
\beq
g_{RR} \bar{t}^c_R \tau_R  S_{RR} +
g_{LL} \bar{t}^c_L \tau_L  S_{LL} + \mathrm{h.c.}\;.
\eeq
In the present study we set either $g_{RR} = 0$, $g_{LL}=1$ (which we
call purely left-handed) or vice
versa: $g_{RR} = 1$, $g_{LL}=0$ (which we call purely right-handed). 
The result for the highly-boosted $x_\mathrm{top}$ distributions obtained for
purely left- and right-handed events in \Herwigpp is shown
in Fig.~\ref{fig:ds0ts0} with the appropriate analytic prediction. 
The proton-proton centre-of-mass energy was set to $140$
TeV and the leptoquark mass was set to $20$ TeV,
so that the top quarks are well within the highly-boosted region.

\begin{figure}[!htb]
  \centering 
  \vspace{4.5cm}
 \includegraphics[scale=0.55, angle=0]{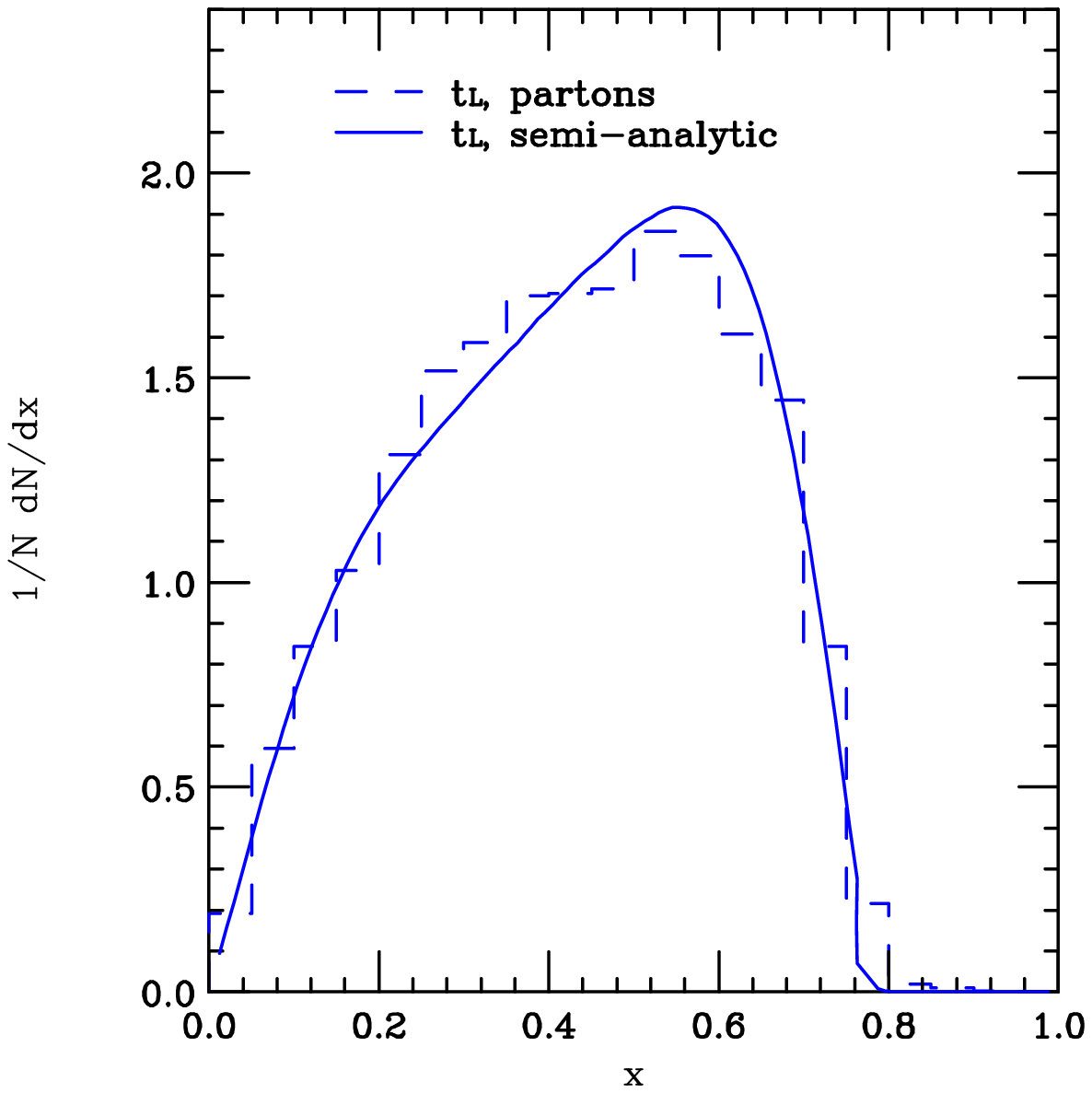}
  \hspace{2.5cm}
  \includegraphics[scale=0.55,angle=0]{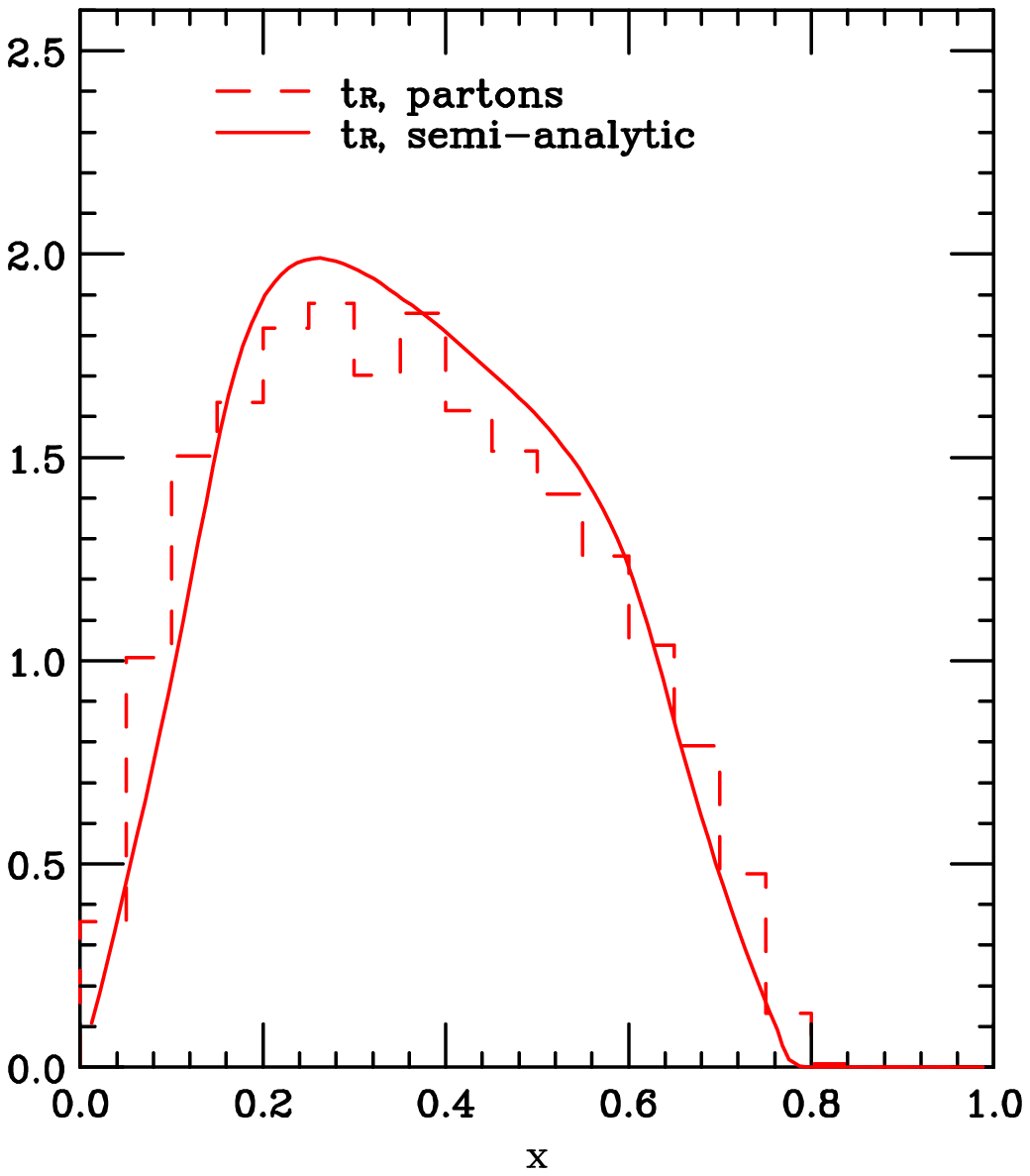}
  \caption{The results for the energy fractions $x_{\mathrm {top}} = \mathcal{E}_{b} /
\mathcal{E}_{\mathrm{top}}$ for $\tau_L t_L$
(left) and $\tau_R t_R$ (right) for a $400$~GeV mass at $14$ TeV $pp$ centre-of-mass energy. These are compared to
semi-analytical predictions as described in the text.}
\label{fig:xtop14}
\end{figure}
\begin{figure}[!htb]
  \centering 
  \vspace{4.5cm}
 \includegraphics[scale=0.52,angle=0]{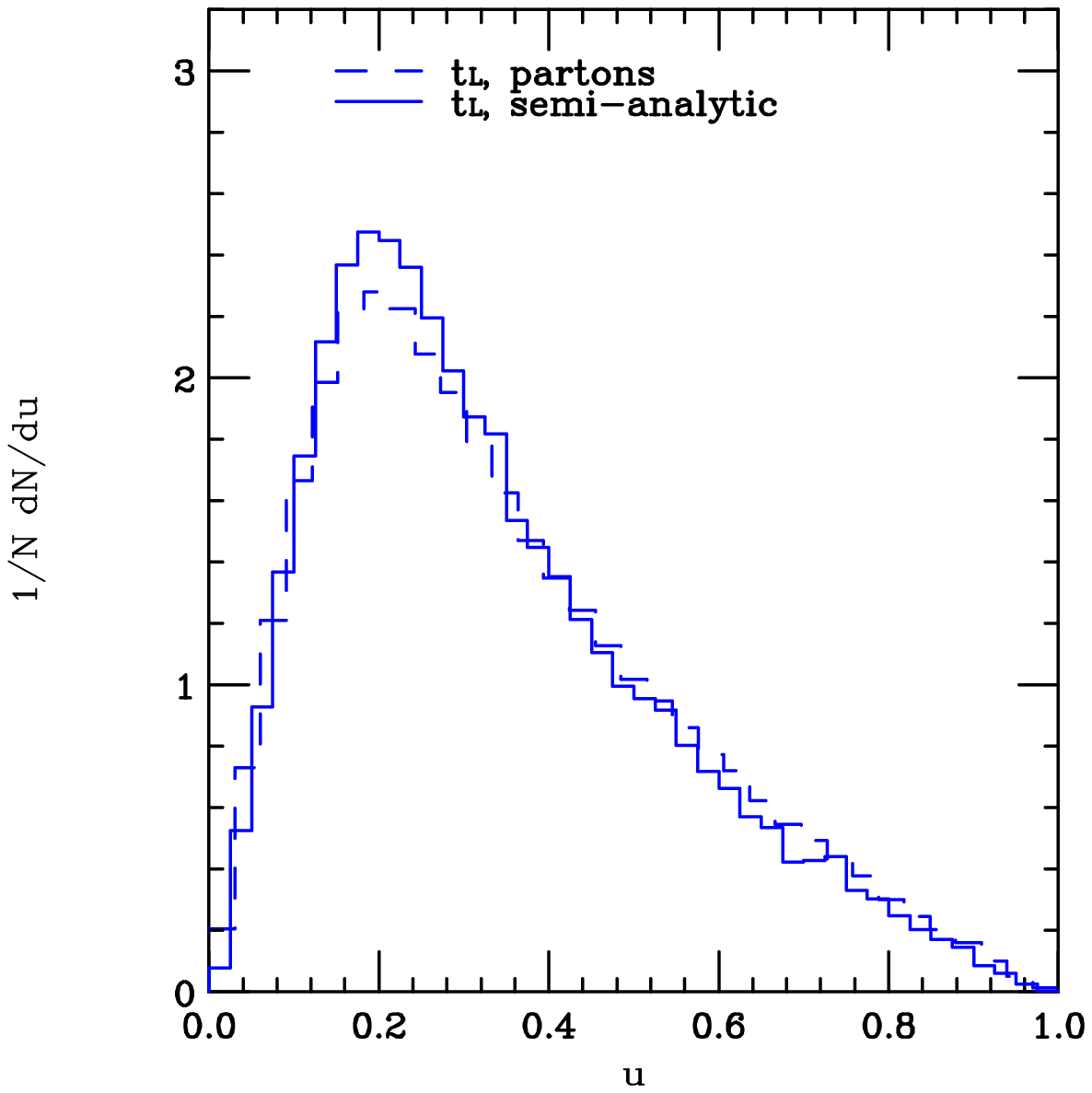}
 \hspace{2.5cm}
  \includegraphics[scale=0.52, angle=0]{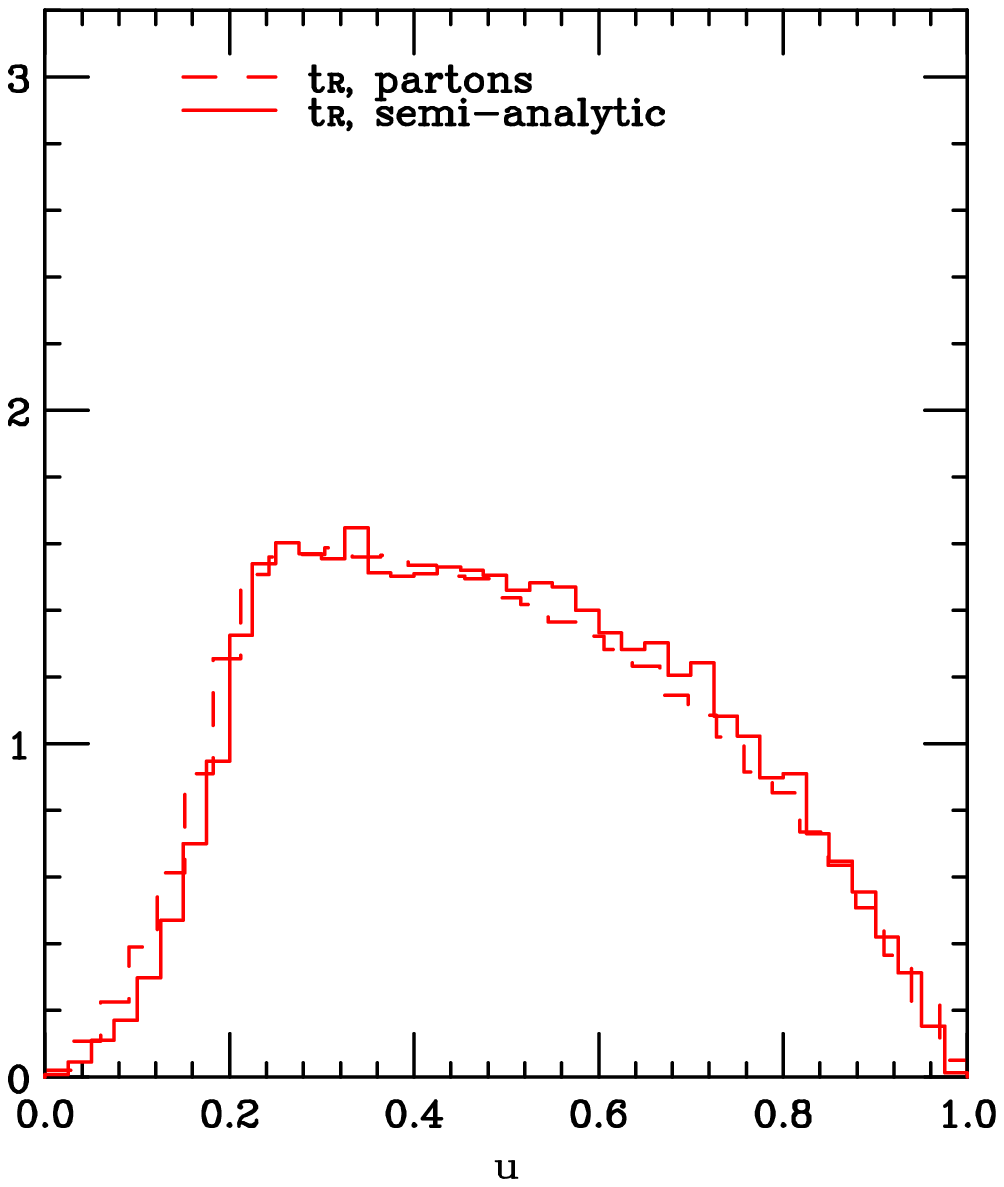}
 \caption{The Monte Carlo results for the energy fractions $u = \frac{\mathcal{E}_\ell}{ \mathcal{E}_\ell + \mathcal{E}_b }
$ for $\tau_L t_L$
(left) and $\tau_R t_R$ (right) for a $400$~GeV mass at $14$ TeV $pp$ COM energy. These are compared to
semi-analytical predictions as described in the text.}
\label{fig:utop14}
\end{figure}

We also present the top quark results originating from the decay of
scalar leptoquarks of mass $400 \gev$ at the LHC with $pp$ centre-of-mass
energy of 14 TeV. Figures~\ref{fig:xtop14} and~\ref{fig:utop14} show
the left- and right-handed distributions of the variables
$x_{\mathrm{top}}$ and $u$ respectively, produced using the \Herwigpp Monte Carlo event
generator and including semi-analytical predictions. These include the effects of the finite
top and tau masses, which introduce a mixture of helicities even
though the Lagrangian terms are purely chiral, and the effect of the
variation of the top boost in the lab frame, $\beta_t$.

The semi-analytical predictions have been produced using Monte Carlo
techniques as described in section~\ref{sec:zprime}: the $x_\mathrm{top}$
distribution was produced by distributing $\cos \theta_b$ according to 
$P(\cos \theta_b) = ( 1 + P_t k_b \cos \theta_b )$, and the $\beta_t$ distribution for the top quark boost was extracted from the
Monte Carlo event generator (see appendix~\ref{app:beta}). Using either a fit or the extracted
distribution yields indistinguishable results.

Since the boost of the parent leptoquark in the lab frame and the
daughter top quark in the lab frame are correlated, we extracted the
two-dimensional distribution $P(\beta_p,\beta_t)$ from the \Herwigpp
event generator (appendix~\ref{app:beta}). In principle, the $\beta_p$ distribution can be
calculated for any process using the hadron parton density functions and an assumption for the
hard process, in this case scalar $SU(3)_c$ triplet pair-production. The
variable $\beta_p$ in the case of pair-produced leptoquarks of mass
$M_{LQ}$ is then related to the centre-of-mass energy $Q$ by $\beta_p
= \sqrt{  1 - 4 M_{LQ}^2 /Q^2}$. The method described in
Ref.~\cite{Shelton:2008nq} was then used to calculate the detected polarisation of the top for each event. The
effect reduces the polarisation on average by less than 10\%. The
mixing of helicities due to the mass of the top quark is low (less
than 0.1\%) due to the fact that it is produced along with a very
light fermion (the $\tau$ lepton) in the case we are considering.  

The $u$ distribution was produced in a similar way, using the full
polarised top matrix element (see
appendix~\ref{app:polarizedtop}). The $W$ decay to a lepton and a
neutrino was set up in the $W$
rest frame using a polar angle $\bar{\theta}$ and an azimuthal angle
$\bar{\phi}$ for the lepton and neutrino momenta. These were then
boosted to the top frame, where the $b$ quark
and $W$ boson momenta were distributed in the top frame using a single
polar angle $\tilde{\theta}$.\footnote{Initially there are $9$ degrees
  of freedom coming from the momenta of the $b$, $\ell$ and
  $\nu$. Four-momentum conservation offers four constraints and the
  mass-shell conditions for the top and $W$ offer a further two. This
  leaves us with the three degrees of freedom: $\bar{\theta}$,
  $\bar{\phi}$ and $\tilde{\theta}$.} The $W$ mass was distributed
according to a Breit-Wigner, centred about $m_W$. The effect of the $W$ width was found to
be small. The distribution was then calculated by taking the ratio: 
\beq
u = \frac{E_\ell + \beta_t p^z_\ell}{E_\ell +  \beta_t p^z_\ell + E_b +  \beta_t p^z_b }\;,
\eeq
where $E_\ell$ and $E_b$ are the lepton and $b$-quark energies in the
top rest frame and $\beta_t$ is again the boost of the top, sampled from
either from the fit (Eq.~(\ref{eq:boostdist})) or the Monte Carlo distribution directly. The calculations of the effect of the finite top and $\tau$
masses in the decay of a scalar and the relation of the top axis of
polarisation and direction of motion follow those which appear in Ref.~\cite{Shelton:2008nq} and are described briefly
in appendix~\ref{app:pd}. 

\subsubsection{Simulation and reconstruction}
\begin{figure}[!t]
  \centering 
    \includegraphics[scale=0.60]{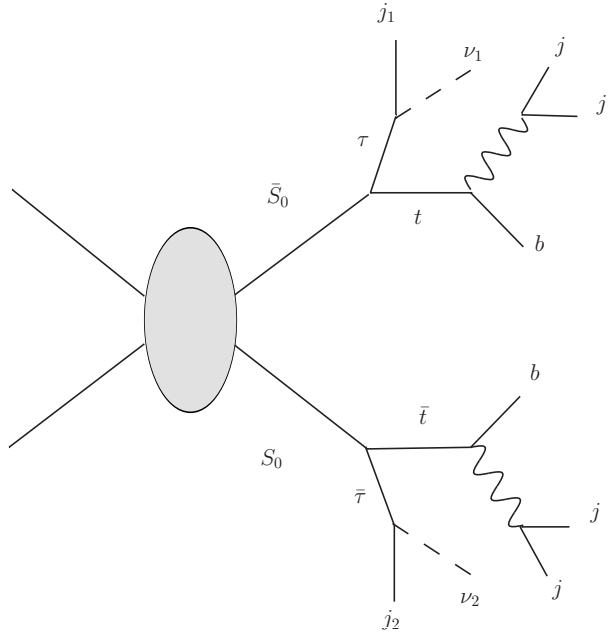}
\caption{Pair-production of a leptoquark pair with decay to
  $(t\tau)(t\tau)$, followed by two fully hadronic top decays.} 
\label{fig:s0topology}
\end{figure}

A mass reconstruction strategy for the $(t\tau)(t\tau)$ decay mode is
described in Ref.~\cite{Gripaios:2010hv}. The reconstruction there focuses on the modes $\bar{S} (S) \rightarrow bj(j)j_1 \nu_1$,
$S (\bar{S}) \rightarrow b \ell \nu_3 j_2 \nu_2$.  We call this the hadronic/semi-leptonic
mode, as opposed to the fully hadronic mode which we will examine
below. An important assumption, that we have already discussed in section~\ref{sec:ztautau}, which allowed for
the full reconstruction of this decay mode, is the collinearity of the
decay products of the tau leptons, owing to the fact that they are
highly-boosted in the lab frame. This has been tested for different
leptoquark masses in Ref.~\cite{Gripaios:2010hv}. The assumption can be
applied by the relation $p_{\tau_i} = z_i p_{j_i}$, where $i = 1,2$ and the
energy ratios imply that $z_i \geq1$.\footnote{Since the events we are considering are not a $\tau \tau$ resonance, 
there is no issue with
back-to-back $\tau$ leptons as the one which previously appeared in
section~\ref{sec:ztautau}.  
} 
In Ref.~\cite{Gripaios:2010hv} a quartic equation was obtained for the energy ratio
$z_2$ and each solution is an unique reconstruction of the whole
event. This method provides a clean way to discriminate the leptoquark
signal from the background. However, our preliminary attempts to reconstruct the
helicity of this mode have indicated that both the number of events, and
the quality of the individual four-momenta reconstruction are
insufficient for detailed determination of the top or $\tau$
helicities. Hence, we focus instead on the topology that contains two fully hadronic tops, shown in
Fig.~\ref{fig:s0topology}.

The signal was generated using the \Herwigpp event generator,
including initial- and final-state radiation (ISR and FSR),
hadronization effects and the underlying event (multiple parton interactions) and the detector response was simulated using the Delphes package with the default
ATLAS settings, modified by the $b$-tagging function of
Eq.~(\ref{eq:btagging}), without the trigger simulation. Here, we also used
the anti-$k_T$ algorithm with radius parameter $R=0.4$. The following cuts were applied on data corresponding
to an integrated luminosity of 100 fb$^{-1}$ at 14 TeV:
\begin{itemize}
\item{A minimum of 6 jets (since the jets originating from the $W$
    could be identified as one jet in the cases where the $W$ is
    highly boosted).}
\item{The missing transverse momentum in the event, $\slashed{E}_T > 20
    \gev$.}
\item{Two $\tau$-tagged jets \textit{and} two $b$-tagged jets, all with the
    extra requirement that they have $p_{T} > 20\gev$.}
\end{itemize}
Note that these are the minimal cuts that one could impose in principle experimentally, and are
lower than those imposed in the reconstruction techniques proposed in Ref.~\cite{Gripaios:2010hv}.

Full reconstruction of the decay topology shown in
Fig.~\ref{fig:s0topology} is possible, since we would only be missing the
neutrino momenta originating from the decays of the highly-boosted
$\tau$ leptons, once the hadronic tops have been reconstructed. Using the approximation for the tau collinearity, one
is left with only two unknowns, the $z_1$ and $z_2$ energy
fractions. By assuming that the $x$ and $y$ components of the missing
momentum are equal to those of the sum of the two neutrino components, we
obtain two linear equations:
\begin{eqnarray}
p_{\rm miss}^x &=& p^x_{j1} ( z_1 - 1 ) + p^x_{j2} (z_2 - 1)\;,\nonumber \\
p_{\rm miss}^y &=& p^y_{j1} ( z_1 - 1 ) + p^y_{j2} (z_2 - 1)\;,
\end{eqnarray}
which can be solved to give:
\begin{eqnarray}\label{eq:z1z2btaubtau}
z_1 &=& 1+ \frac{ p^y_{j2} p^x_{\rm miss} - p^x_{j2} p^y_{\rm miss} } { p^x_{j1} p^y_{j2} - p^y_{j2} p^x_{j2} }\;, \nonumber \\
z_2 &=& 1 -\frac{ p^y_{j1} p^x_{\rm miss} - p^x_{j1} p^y_{\rm miss} }
{ p^x_{j1} p^y_{j2} - p^y_{j2} p^x_{j2} }\;.
\end{eqnarray}
The invariant mass of each of the two leptoquarks may be written as $m_S^2 = (p_t + p_\tau)^2$, resulting in the following expression:
\begin{equation}
m_S^2 = 2 z_i p_{ti} \cdot p_{ji} + m_{\rm top}^2\;,
\end{equation}
where we have neglected the $\tau$ mass term. Using
Eqs.~(\ref{eq:z1z2btaubtau}), we obtain two values of $m_S$ per
event. Since this analysis would be performed \textit{after} potential
discovery, we would already have a measurement of the mass of the leptoquark. This would
allow for elimination of backgrounds that may contribute and alter the energy
fraction distributions. 

\begin{figure}[!t]
  \centering 
  \vspace{5.0cm}  
  \hspace{0.0cm}     
    \includegraphics[scale=0.60, angle = 0]{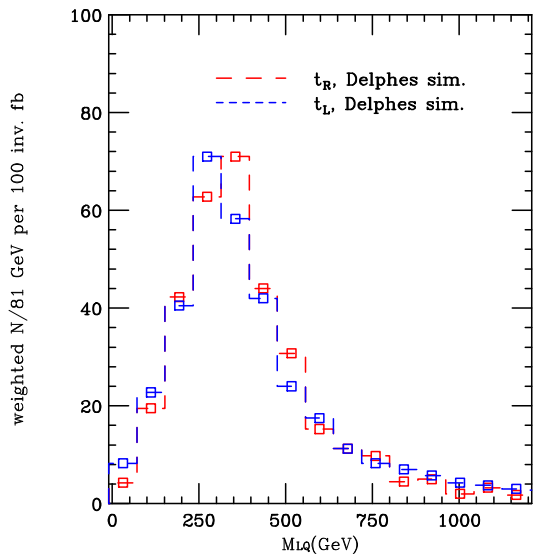}
    \vspace{0.5cm}  
\caption{Mass reconstruction in the fully hadronic mode for the
  purely left- and right-handed cases for an integrated luminosity of
  100 fb$^{-1}$. There are 4 entries per event corresponding.} 
\label{fig:ms0ttauttauhad}
\end{figure}

\begin{figure}[!t]
  \centering 
  \vspace{4.5cm}
 \includegraphics[scale=0.52, angle=0]{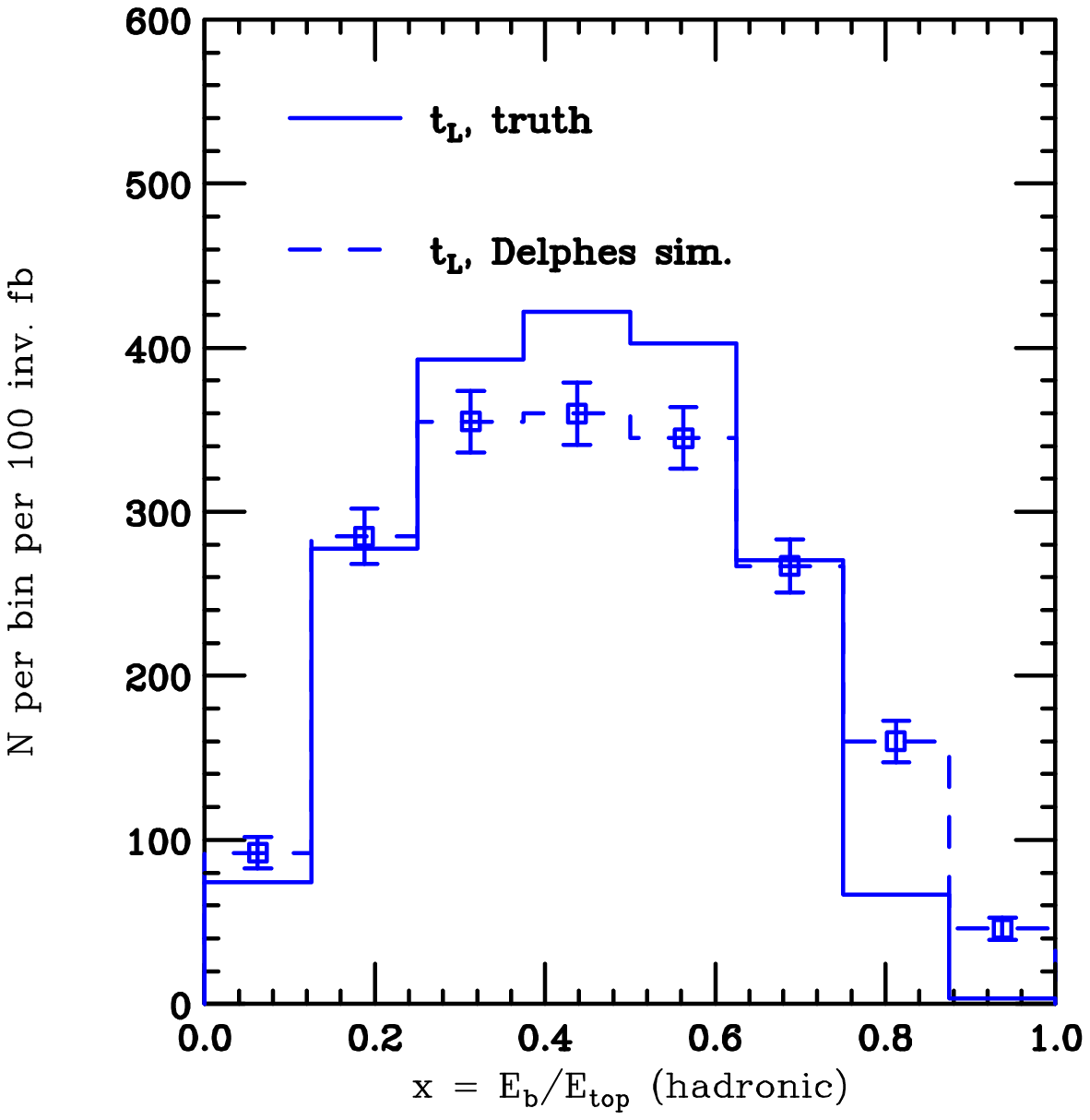}
  \hspace{2.5cm}
  \includegraphics[scale=0.52,angle=0]{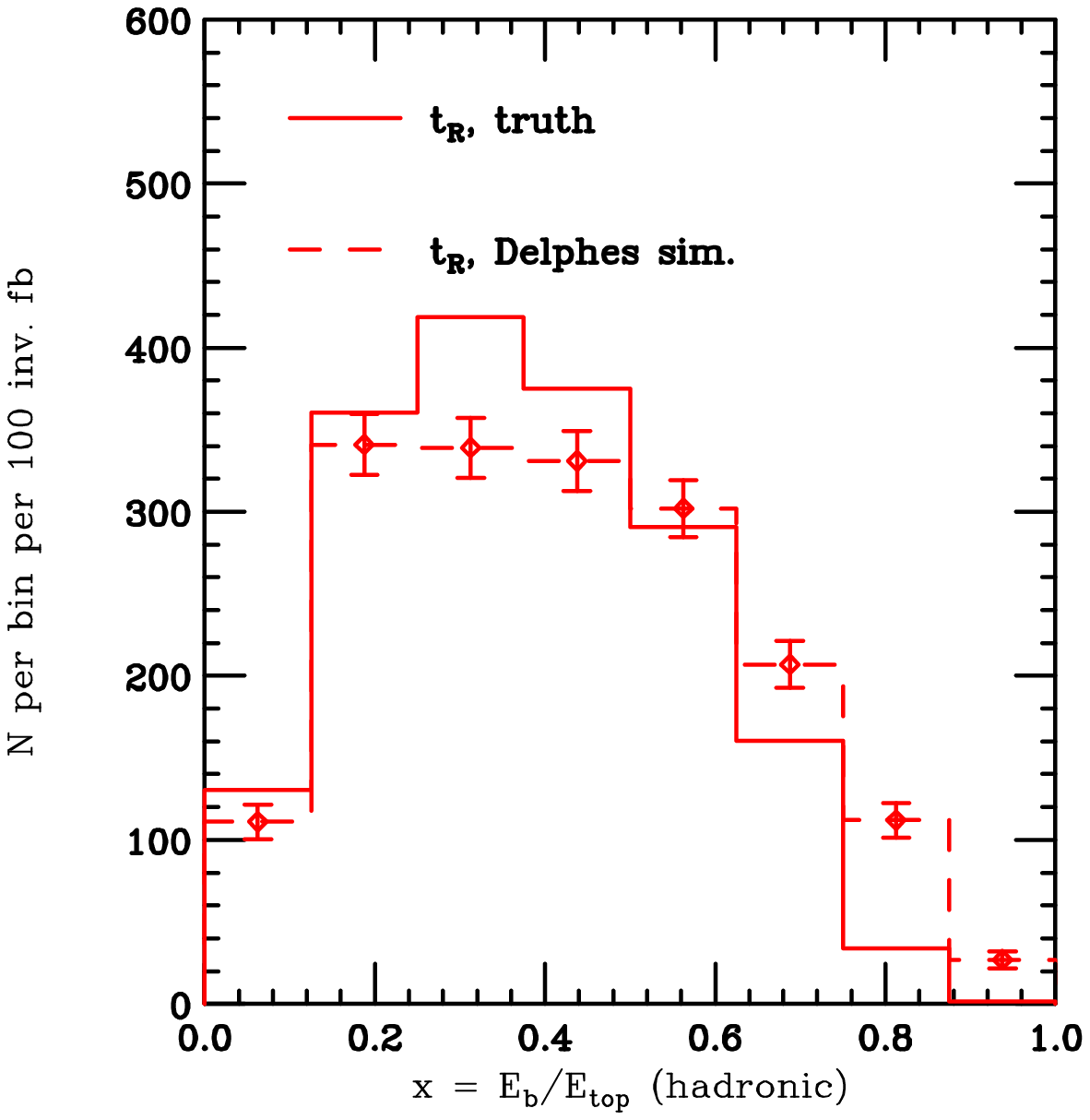}
  \caption{Shown in the figures is a comparison between the
results obtained for the $x_{\rm top}$ variable for 400 GeV leptoquarks \textit{before}
detector simulation (but applying all cuts and using equivalent
jet-finding) and after the Delphes simulation for the left- and right-
handed fermions (blue and red respectively).}
\label{fig:xtoptruth}
\end{figure}

\begin{figure}[!t]
  \centering 
  \vspace{4.5cm}
 \includegraphics[scale=0.52, angle=0]{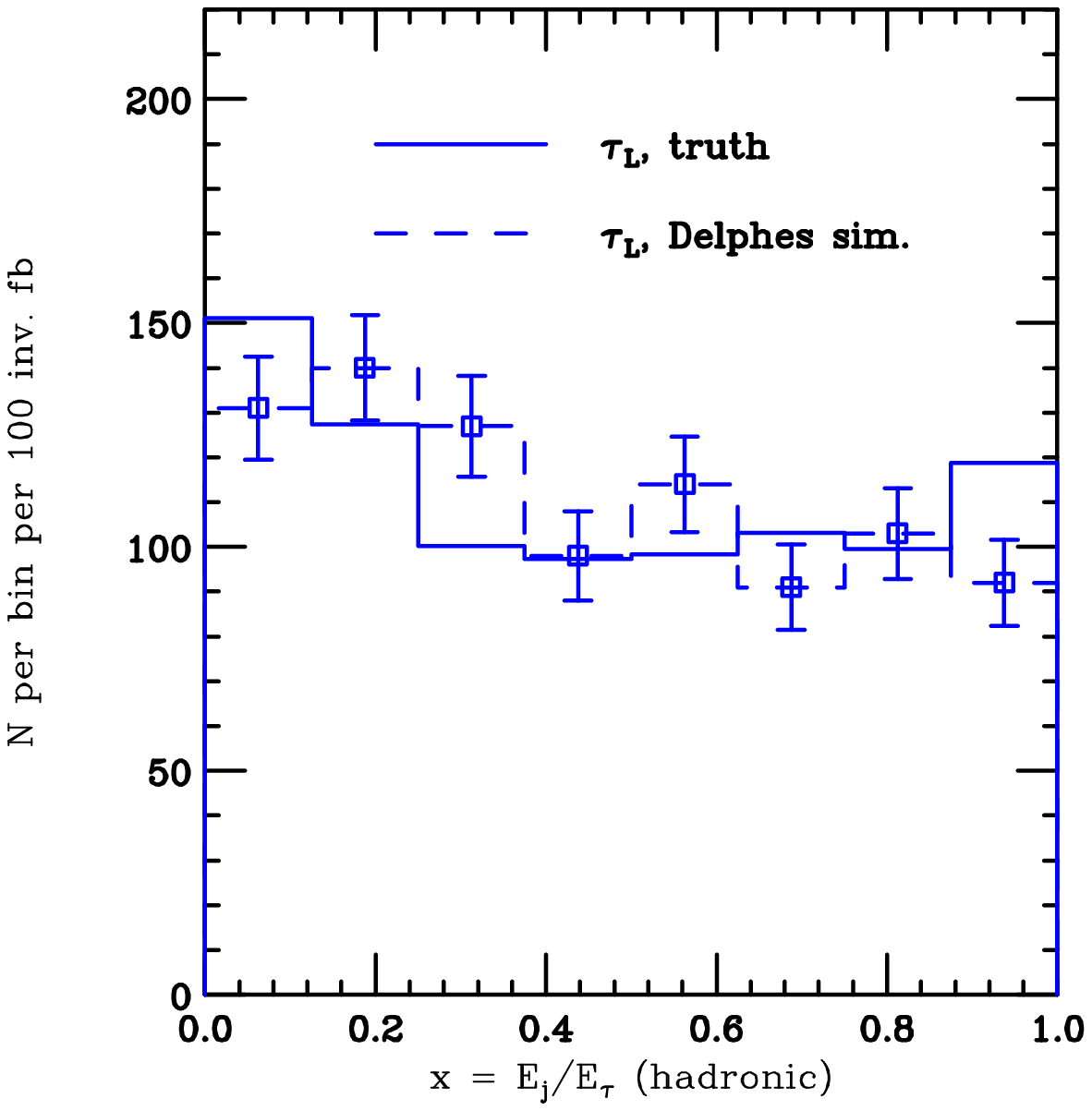}
  \hspace{2.5cm}
  \includegraphics[scale=0.52,angle=0]{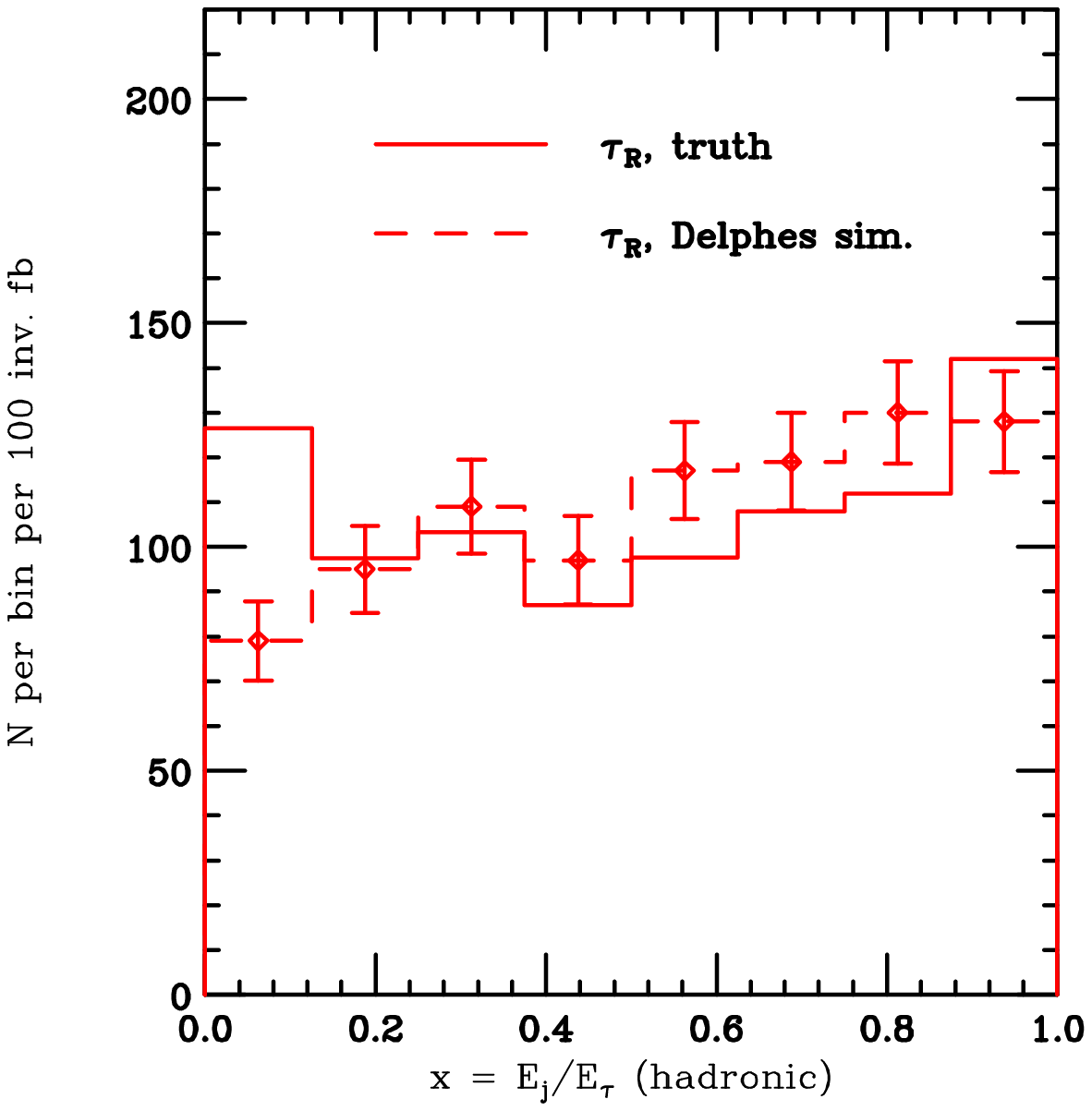}
  \caption{Shown in the figures is a comparison between the
results obtained for the $x_{\tau}$ variable for 400 GeV leptoquarks \textit{before} detector
simulation (but applying all cuts and using equivalent
jet-finding) and after the Delphes simulation for the left- and right-
handed fermions (blue and red respectively).}
\label{fig:xtautruth}
\end{figure}

\begin{figure}[!t]
  \centering 
  \vspace{4.5cm}  
   \includegraphics[angle=0,scale=0.52]{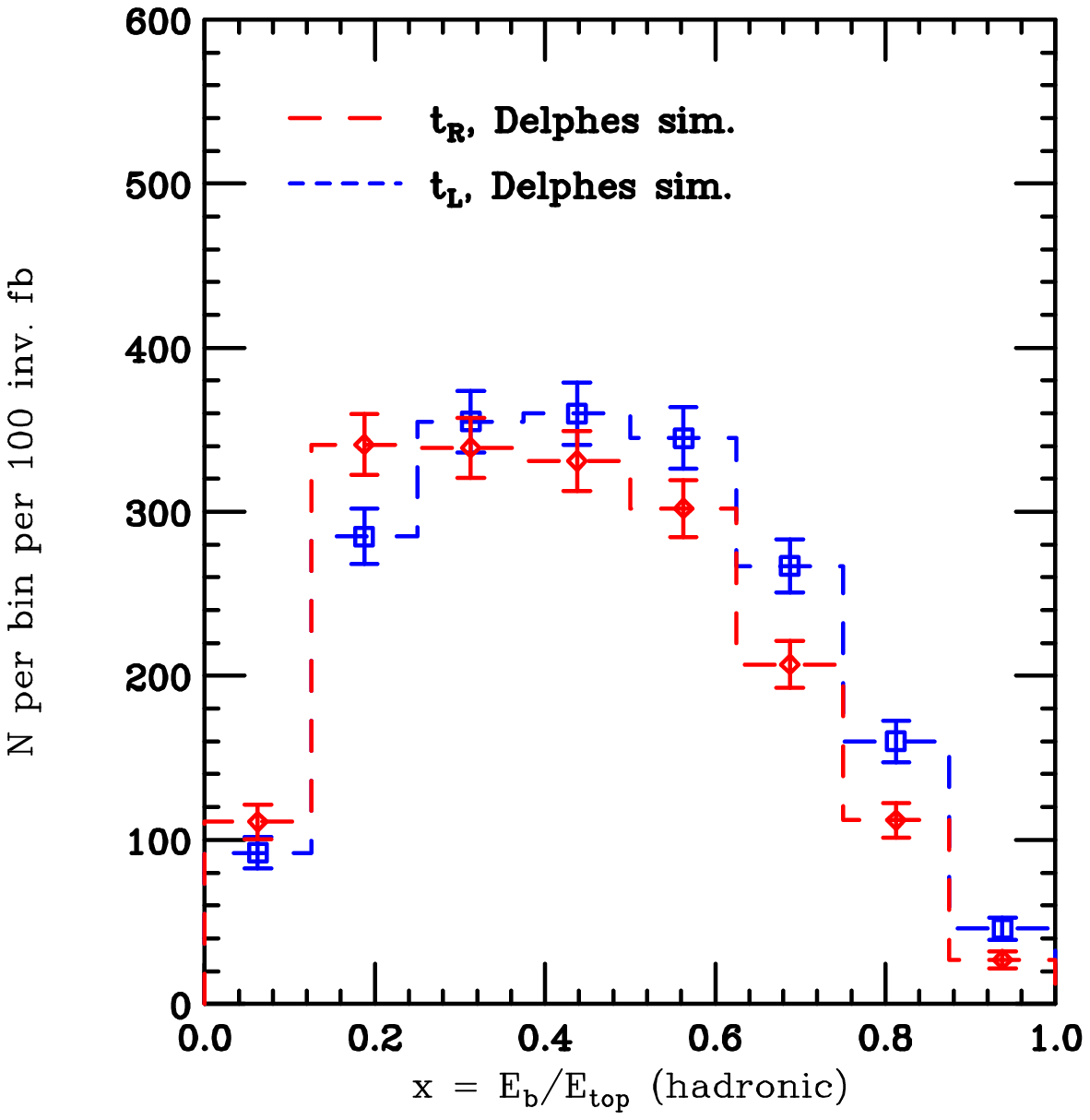}  
    \hspace{2.5cm}  
    \includegraphics[angle=0,scale=0.52]{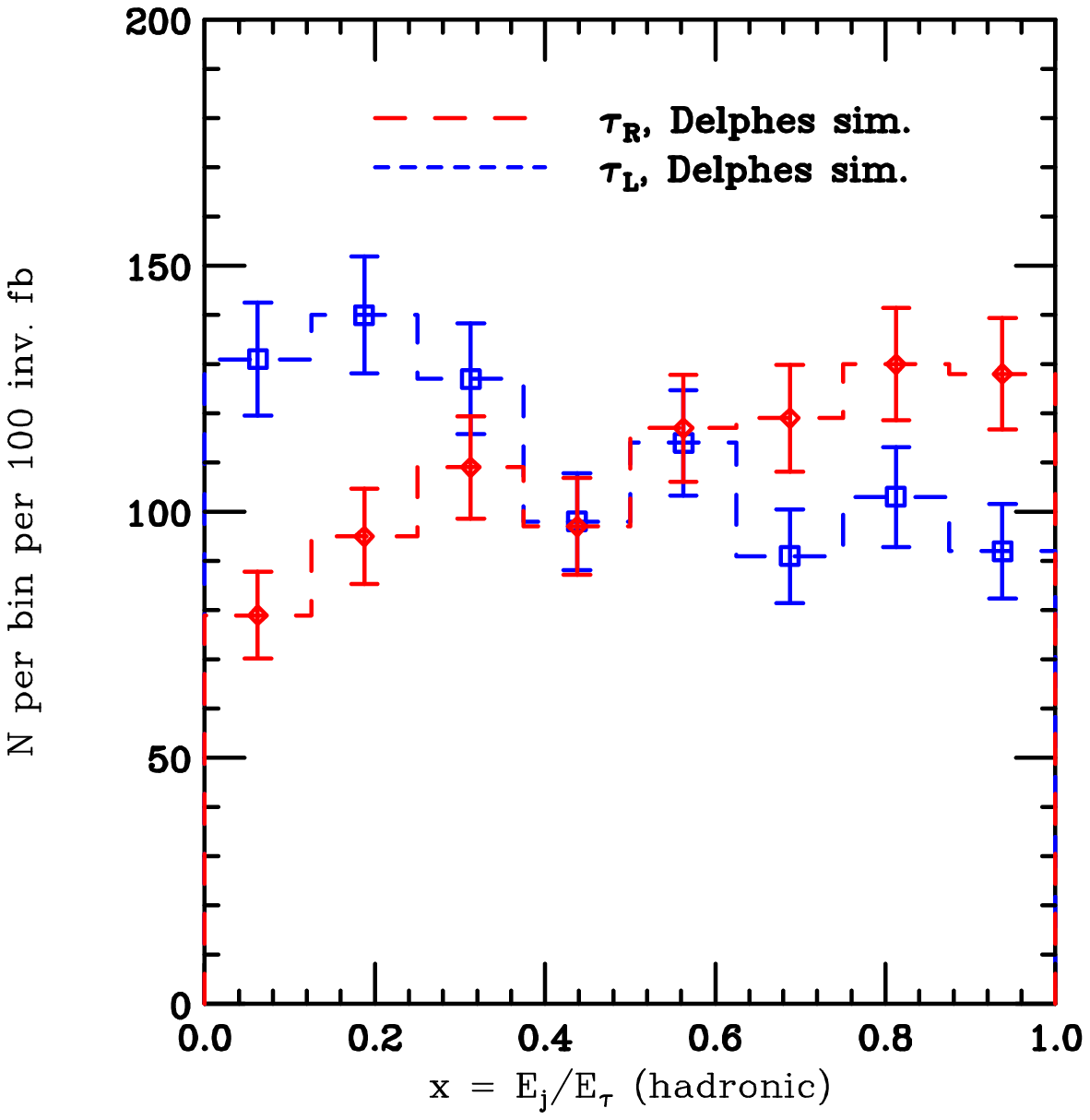}  
\caption{The $x_{\mathrm{top}}$ (left panel) and the $x_\tau$  (right
  panel) variables for left- or
  right-handed $t\tau$ for an integrated luminosity of
  100 fb$^{-1}$ modes originating from the reconstructed events in the fully
  hadronic case,
  using the method described in the text.} 
\label{fig:xtaudelphesttau_had}
\end{figure}

To assess the possibility of measuring the helicity of the top quarks
and tau leptons, we generated 100~fb$^{-1}$ of a fully hadronic sample for purely left-handed or right-handed couplings and passed them through the
Delphes simulation. We
then analysed events which contained 2 $\tau$-tagged jets and 2
$b$-tagged jets. We looked for 1 or 2 jets which reconstructed the top mass in conjunction with the tops, within an $80 \gev$
window.\footnote{To further improve the `top-tagging' capabilities of
  the analysis, one can employ a more advanced tagging algorithm such
  as the one presented in Ref.~\cite{Plehn:2011sj}. For our purposes,
  the simpler reconstruction method of requiring combinations of jets
  to satisfy the top mass is sufficient to provide good results.} For completeness, we show in Fig.~\ref{fig:ms0ttauttauhad} the
resulting reconstructed masses using the above method, without any
attempt to optimise the resulting values. In the figure, we plot all
the 4 entries per event, which are a consequence of the two leptoquark
masses ($m_S$, $m_{\bar S}$) and the two-fold combinatoric ambiguity,
arising from the possible pairings of the top quarks
and the $\tau$ leptons.
To obtain the best value of the variables $x_\mathrm{top}$ and
$x_\tau$, we choose the combination which yields the best leptoquark mass,
that is, the one closest to the true mass. We then obtain two values
of the $x_\tau$ variable: $x_{\tau,1/2} = 1/z_{1/2}$ and two values of
the $x_{\mathrm{top}}$ variable by using the energies of the two
$b$-jets and the reconstructed top energy. A comparison between the
results obtained \textit{before} detector simulation, but applying all cuts and using equivalent
jet-finding, and after the Delphes simulation, is shown in
Figs.~\ref{fig:xtoptruth} and \ref{fig:xtautruth} on the left- and right- handed fermions, for
the variables $x_{\rm top}$ and $x_\tau$. The results without detector simulation have
been normalised to the number of events resulting after Delphes
simulation. The differences that arise at low and high energy fractions and can be
attributed primarily to the efficiency of the tagging algorithms and the
overall differences to the smearing of the four-momenta due to the
simulation of the response of the detector. Comparisons of the Delphes
results are shown in Fig.~\ref{fig:xtaudelphesttau_had} for the purely left- and
right-handed cases.

To assess the discrimination capabilities of the distributions, we
calculated the $\chi^2 / N_{\mathrm{d.o.f.}}$ between the
left- and right-handed distributions. To investigate the effect of
higher transverse momentum cuts, we re-ran the analysis with two higher
$p_T$ cuts on the jets and the missing transverse momentum: set `A-prime' with $p_T$ and $\slashed{E}_T > 25 \gev$ and set `A'
with $p_T$ and $\slashed{E}_T > 30 \gev$. The results are shown in
Table~\ref{tb:s0chisq}.
\begin{table}[!htb]
\begin{center}
\begin{tabular}{|c|c|c|} \hline
Cut set & $x_\mathrm{top}$  & $x_\tau$ \\\hline \hline
Min. &  4.0 & 4.5 \\\hline 
A-prime  & 2.9  & 3.8 \\\hline 
A &  1.9 & 2.8  \\\hline 
\end{tabular}
\end{center}
\caption{The value of $\chi^2 / N_{\mathrm{d.o.f.}}$ between the left-
and right-handed distributions in the leptoquark $t\tau$ decay
scenario, for the three
different sets of cuts. It is evident that the distributions become
more difficult to distinguish for the higher cuts at the given integrated
luminosity of 100~fb$^{-1}$.}
\label{tb:s0chisq}
\end{table}
\section{Conclusions}\label{sec:conclusions}
We have investigated variables that have been defined for the purpose of
determining the helicity of top quarks and tau leptons, in a more
realistic setting than what has been done thus far in the literature. We first
examined these analytically in the highly-boosted case, where the
calculation is simplified, and no explicit event reconstruction is
required for the $u$ variable. Subsequently we focused on two specific models:
one containing a new heavy vector boson, $Z'$, with decays to
either a light jet and a top quark or two taus, and a specific scenario in scalar
leptoquark pair-production, in which the decay of both leptoquarks is
into a top quark and a $\tau$ lepton. We examined the flavour-changing $Z'$ model at
parton level, producing the relevant distributions semi-analytically and comparing these directly to the Monte
Carlo-generated distributions. We considered experimental and
reconstruction effects for the case of a LHC at proton-proton centre-of-mass
energy of 14~TeV and 10~fb$^{-1}$ of integrated luminosity, and we investigated the applicability of the
helicity discrimination variables. We found that in the case of a 1.5
TeV $Z'$ that can decay into an up quark
and a top quark, the top quark helicity can be determined, even
for higher momentum cuts.  For the $Z'$ model
with decays to $\tau^+\tau^-$ we used the $\tau$ decay vertex information along
with a likelihood method to correct for detector resolution effects,
resulting in good discrimination between the left- and right-handed
modes for 1~fb$^{-1}$ of integrated luminosity. For the leptoquark pair-production model, for 400~GeV
leptoquarks decaying into a top quark and a $\tau$ lepton each, we examined reconstruction of the case when both the decaying fully
hadronically. Discrimination in this scenario is more challenging, but
values of $\chi^2 / N_{\rm d.o.f.} \sim 2-3$ can be obtained even with
higher than minimal cuts for an integrated luminosity of 100~fb$^{-1}$.

To summarise, we have assessed the magnitude of the effects of QCD, cuts on the
transverse momentum, detector effects and finally the reconstruction
issues that arise in this phenomenological
study of helicity variables. This work is indicative of the difficulties that arise in `measuring'
the helicities to determine the form of the interactions of new
particles to quarks and leptons of the third generation. To fully
determine the potential performance of these variables, the next step
would be for the experimental collaborations to perform similar
analyses, with realistic detector response and particle identification
techniques. 

\section{Acknowledgements}
We would like to thank the Cavendish Laboratory's High Energy Physics
group for allowing continuous use of their computing facilities, as well as the
Cambridge Supersymmetry Working Group for useful suggestions and
discussion related to this paper. We would also like to thank Jos\'e
Zurita for his useful comments and suggestions. This research is supported in part
by the Swiss National Science Foundation (SNF) under contract
200020-138206 and the Research Executive Agency (REA) of the
European Union under the Grant Agreement number PITN-GA-2010-264564
(LHCPhenoNet).

\appendix
\section{Angular variables}\label{app:angular}
\subsection{Definitions}
\begin{figure}[!t]
  \centering 
  \includegraphics[scale=0.60]{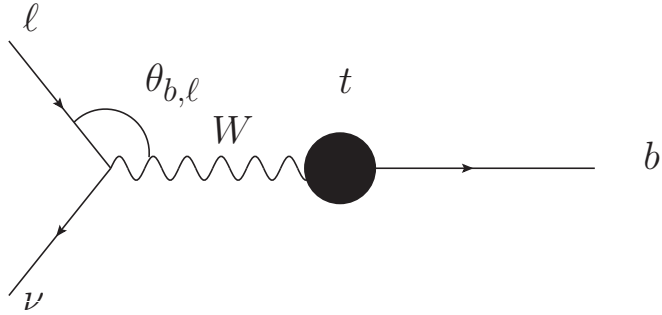}
  \vspace{0.5cm}
  \caption{The angle $\theta_{b,\ell}$, between the $b$ quark and
    lepton in top decay, shown in the centre-of-mass
    frame of the top for illustration purposes. The angle is
    calculated in the lab frame throughout this paper.}
  \label{fig:thetabl}
\end{figure}

We define here two `angular' variables, the first of which is the
angle $\theta_{b,\ell}$, defined between the
$b$-quark and the lepton, $\ell$, in semi-leptonic top decays as
shown in Fig.~\ref{fig:thetabl}. The angle is defined in the lab
frame, but is shown in the figure in the
centre-of-mass frame of the decaying top quark for illustration
purposes. We consider a function of this variable defined
by:\footnote{This form of the function is chosen to resemble the factor $0.25 \times (1 +
  \cos \theta_{b,\ell}^* )^2$, where $\theta_{b,\ell}^*$ is defined in
  the $W$ rest frame, which appears in the top quark decay
  differential cross section.}
\beq
f(\cos \theta_{b,\ell}) = 0.25 \times (1 + \cos \theta_{b,\ell} )^2\;.
\eeq
We will also be considering the distance between the lepton and neutrino for semi-leptonic top decays, given by $\Delta R(\ell,\nu) = \sqrt{ \delta \eta_{\ell,\nu}^2 +
  \delta \phi_{\ell,\nu}^2 }$, where $\delta \eta_{\ell,\nu}$ and
$\delta \phi_{\ell,\nu}$ are the distances in the pseudo-rapidity,
$|\eta_\ell - \eta_\nu|$, and
transverse plane angle, $|\phi_\ell - \phi_\nu|$.
\subsection{Angular variables in the $Z'$ flavour-changing model}
\begin{figure}[!htb]
  \centering 
  \vspace{5.0cm}  
    \includegraphics[angle=0,scale=0.55]{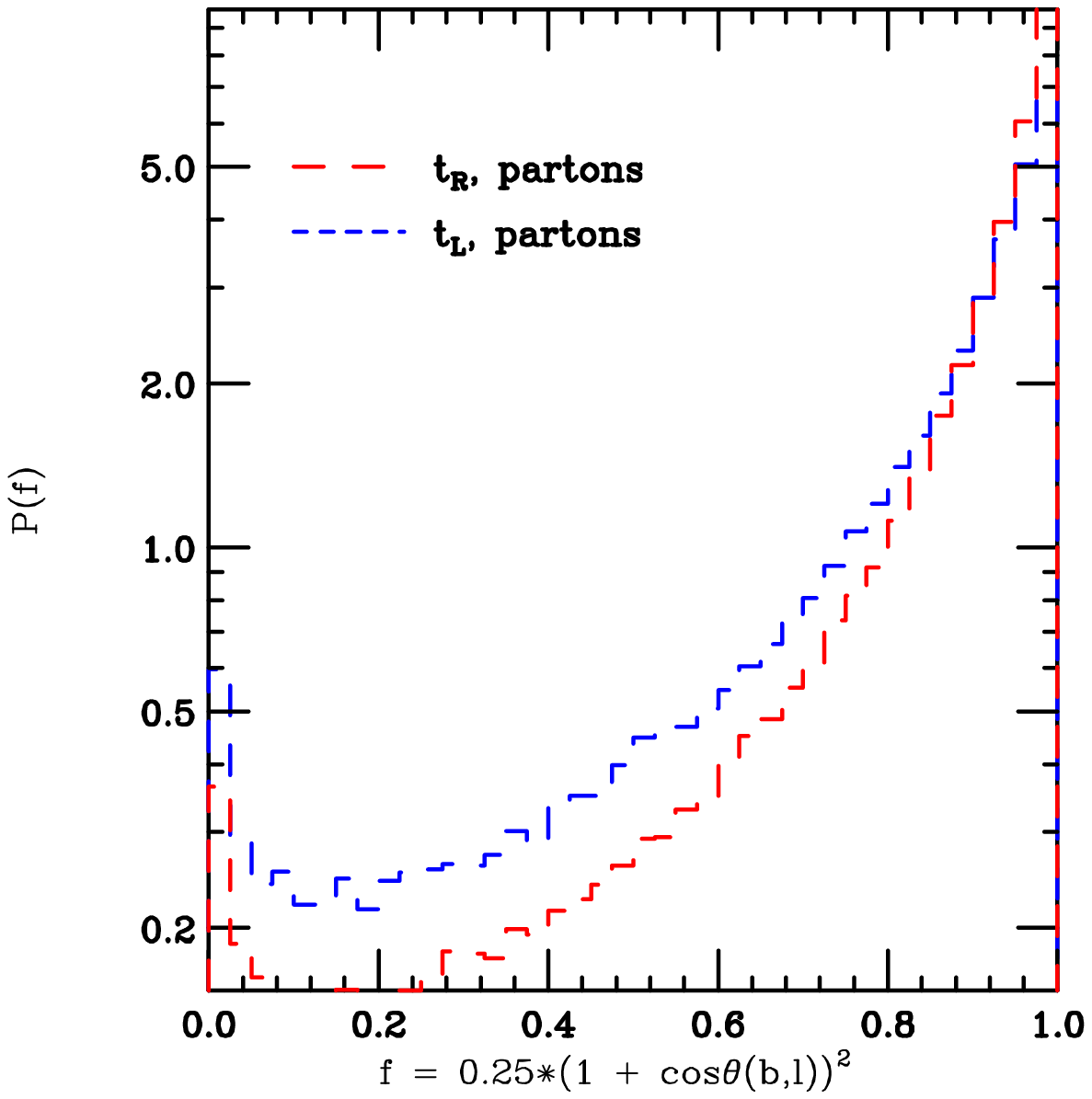}  
   \hspace{2.5cm}  
   \includegraphics[angle=0,scale=0.55]{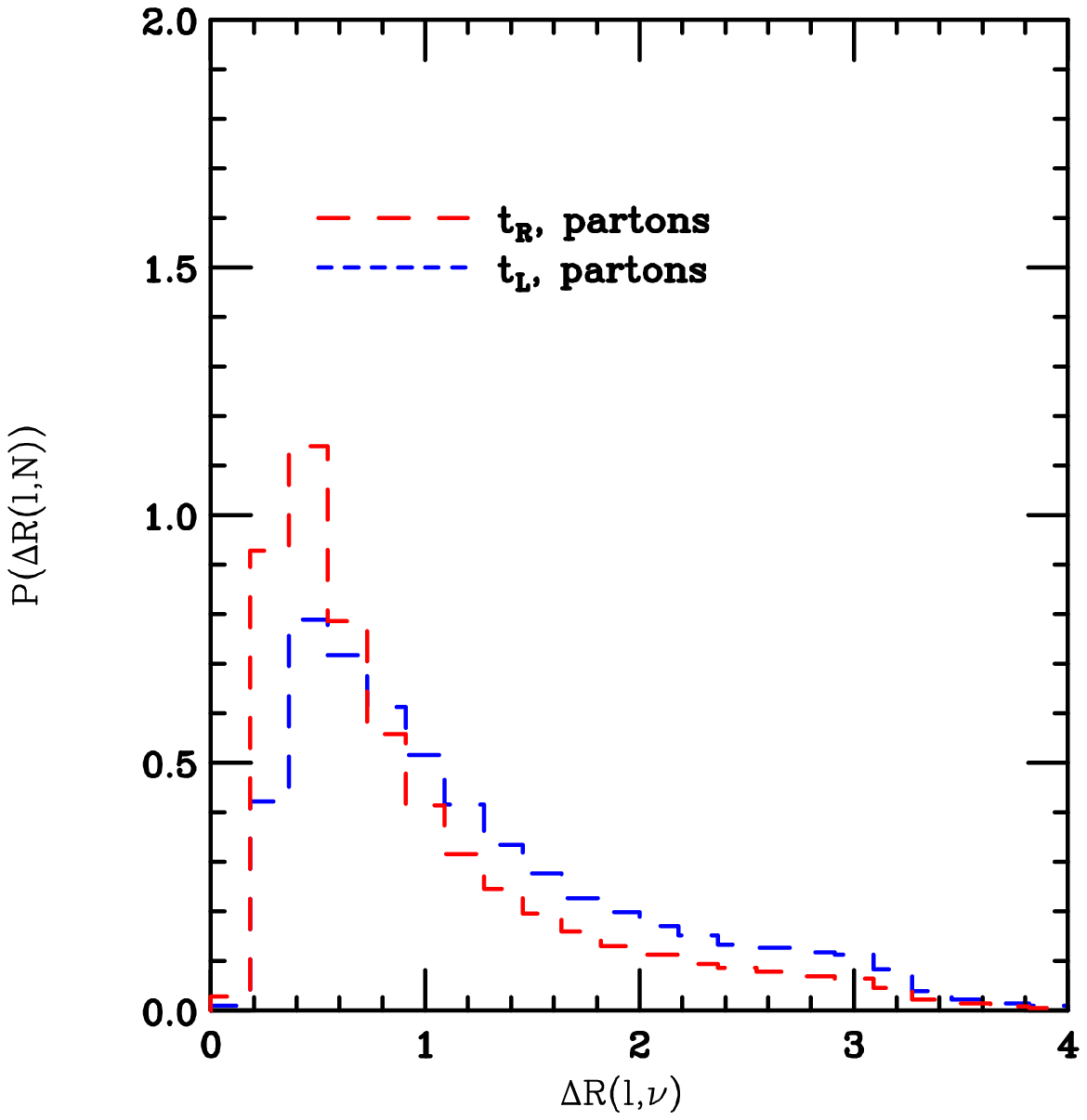}  
  \caption{The $f(\cos \theta_{b,\ell})$ (left panel) and  $\Delta R(\ell,\nu)$ (right panel) variables for left- or right-handed $Z'$ bosons decaying to $u\bar{t}$ or $\bar{u}t$, obtained from
  parton-level events} 
  \label{fig:angdelpheszpparton}
\end{figure}

\begin{figure}[!htb]
  \centering 
  \vspace{5.0cm}  
  \includegraphics[angle=0,scale=0.55]{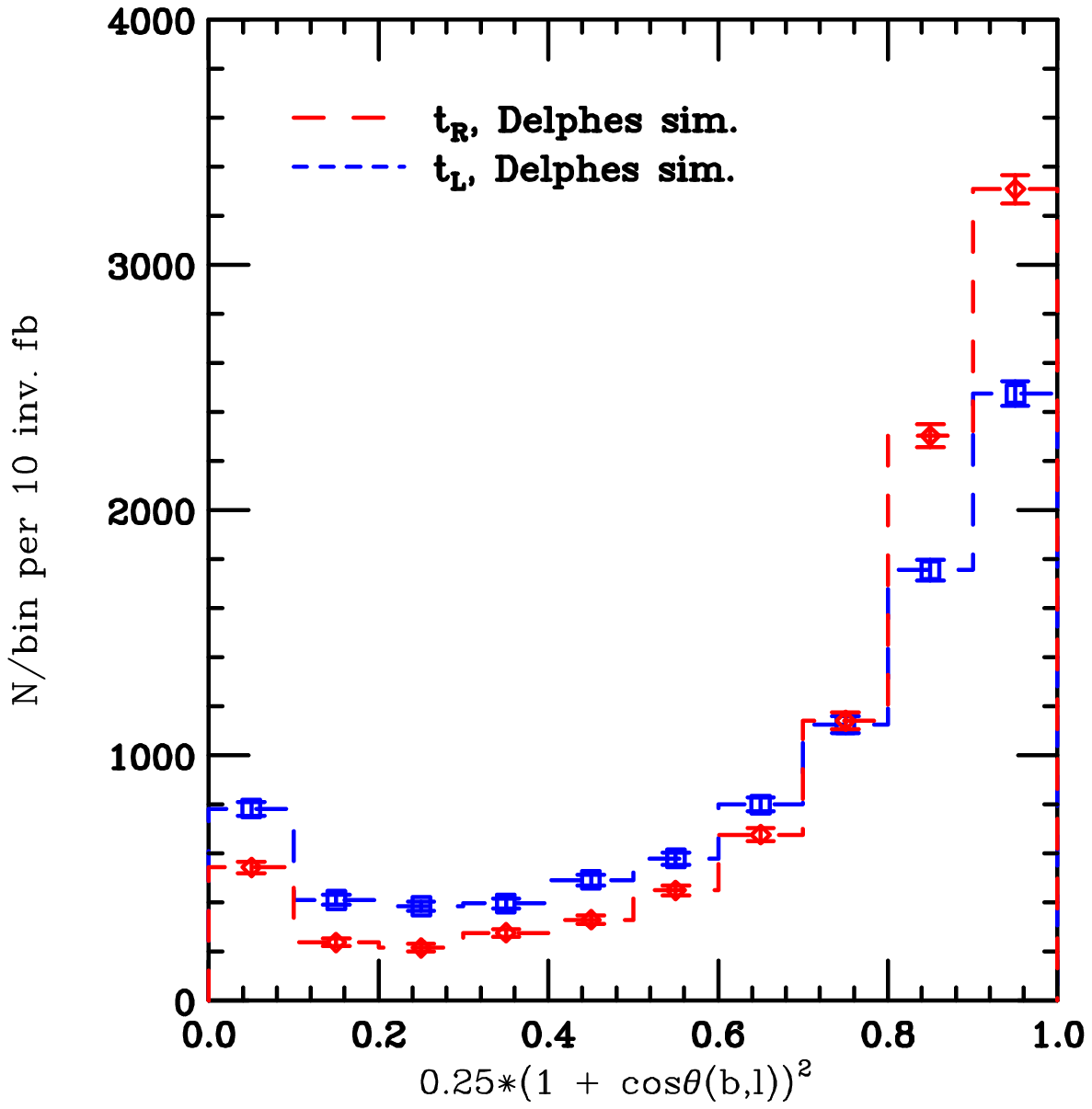}  
  \hspace{2.5cm}      
  \includegraphics[angle=0,scale=0.55]{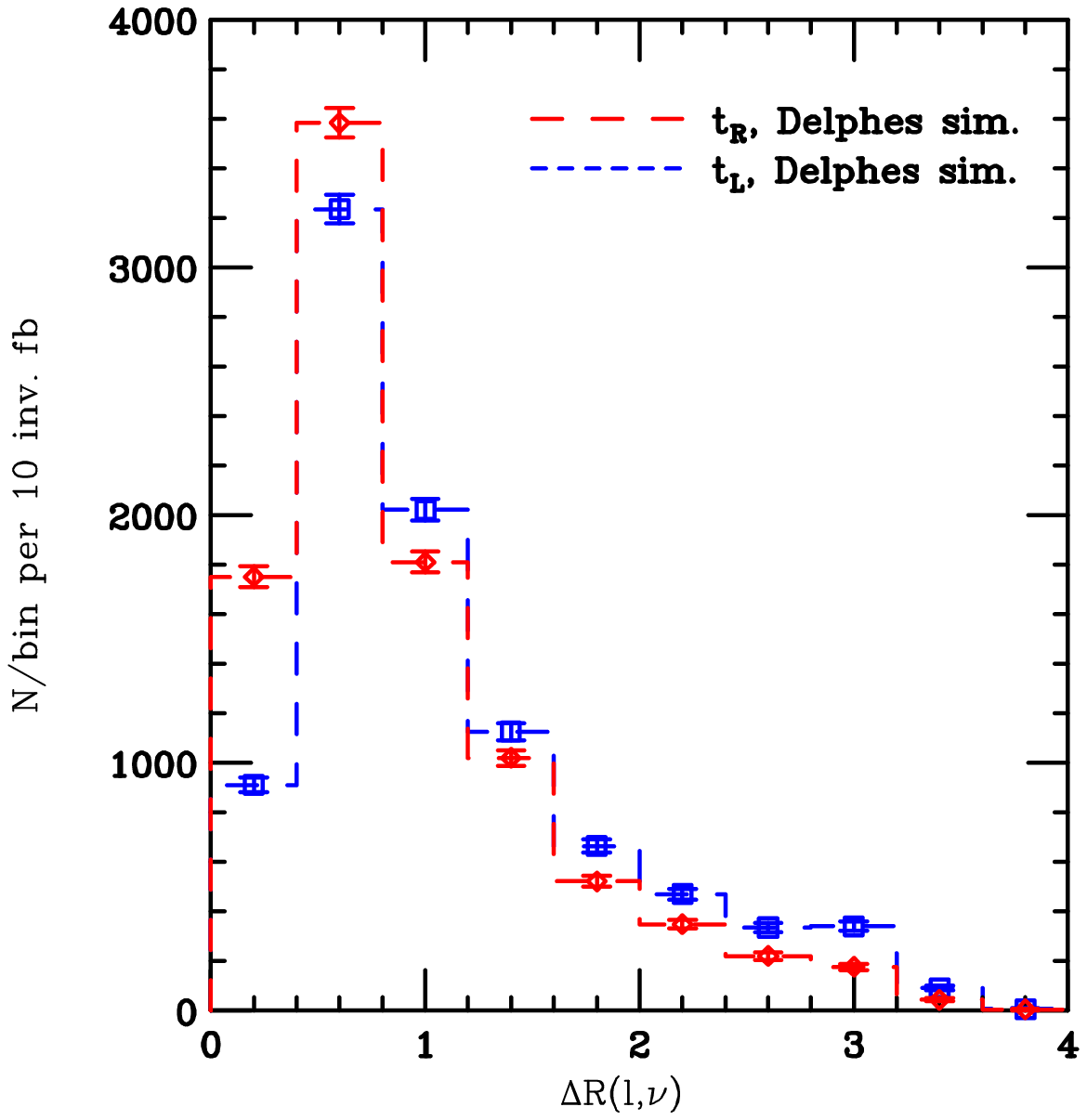}  
  \caption{The $f(\cos \theta_{b,\ell})$ (left panel) and  $\Delta R
    (\ell,\nu)$ (right panel) variables for a left- or right-handed $Z'$ bosons decaying to $u\bar{t}$ or $\bar{u}t$, obtained from the reconstructed events for an LHC run at 14
    TeV, with 10 fb$^{-1}$.} 
  \label{fig:angdelpheszp}
\end{figure}

\begin{figure}[!htb]
  \centering 
  \vspace{5.0cm}
 \includegraphics[scale=0.55, angle=0]{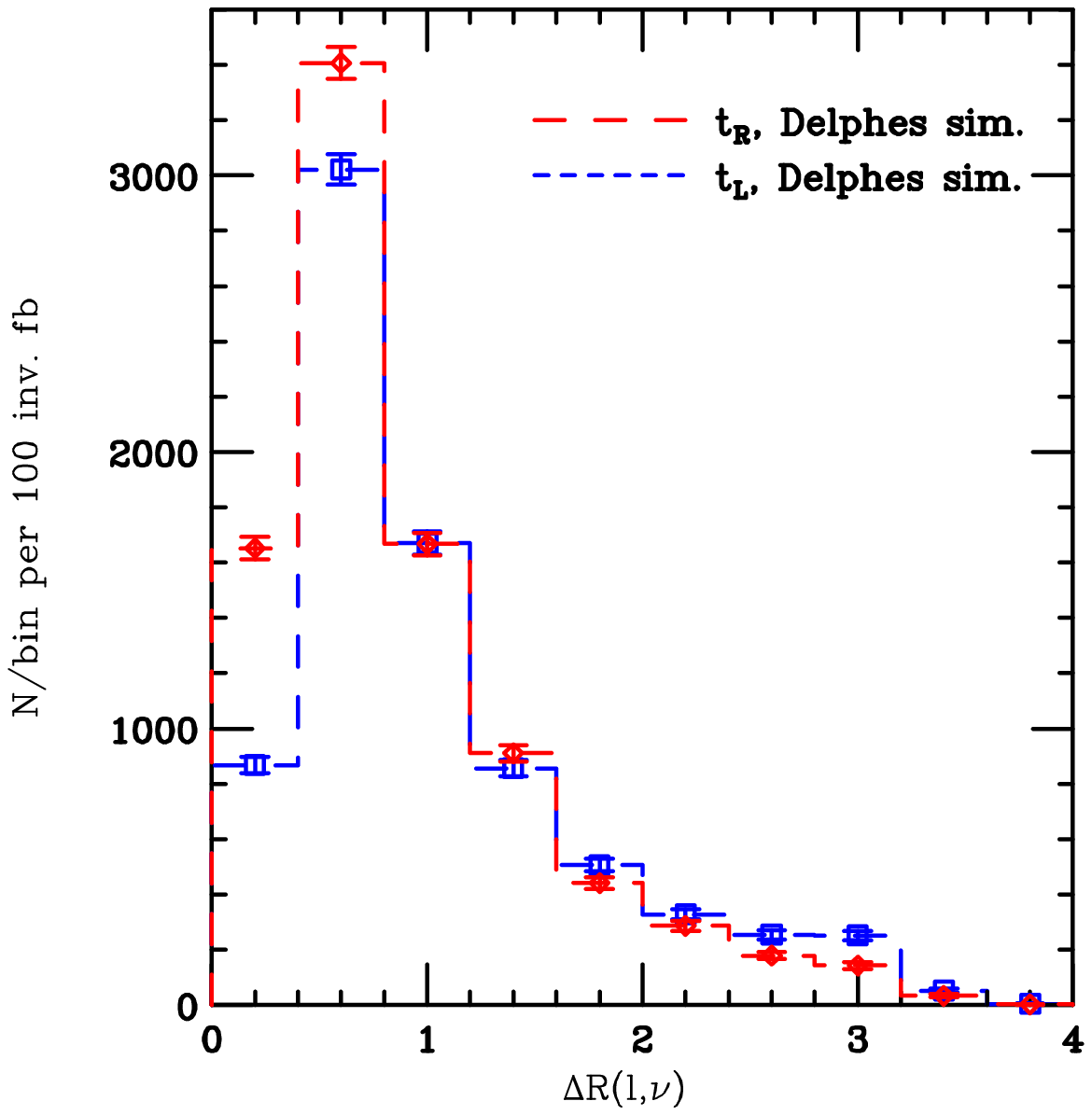}
  \hspace{2.5cm}
  \includegraphics[scale=0.55,angle=0]{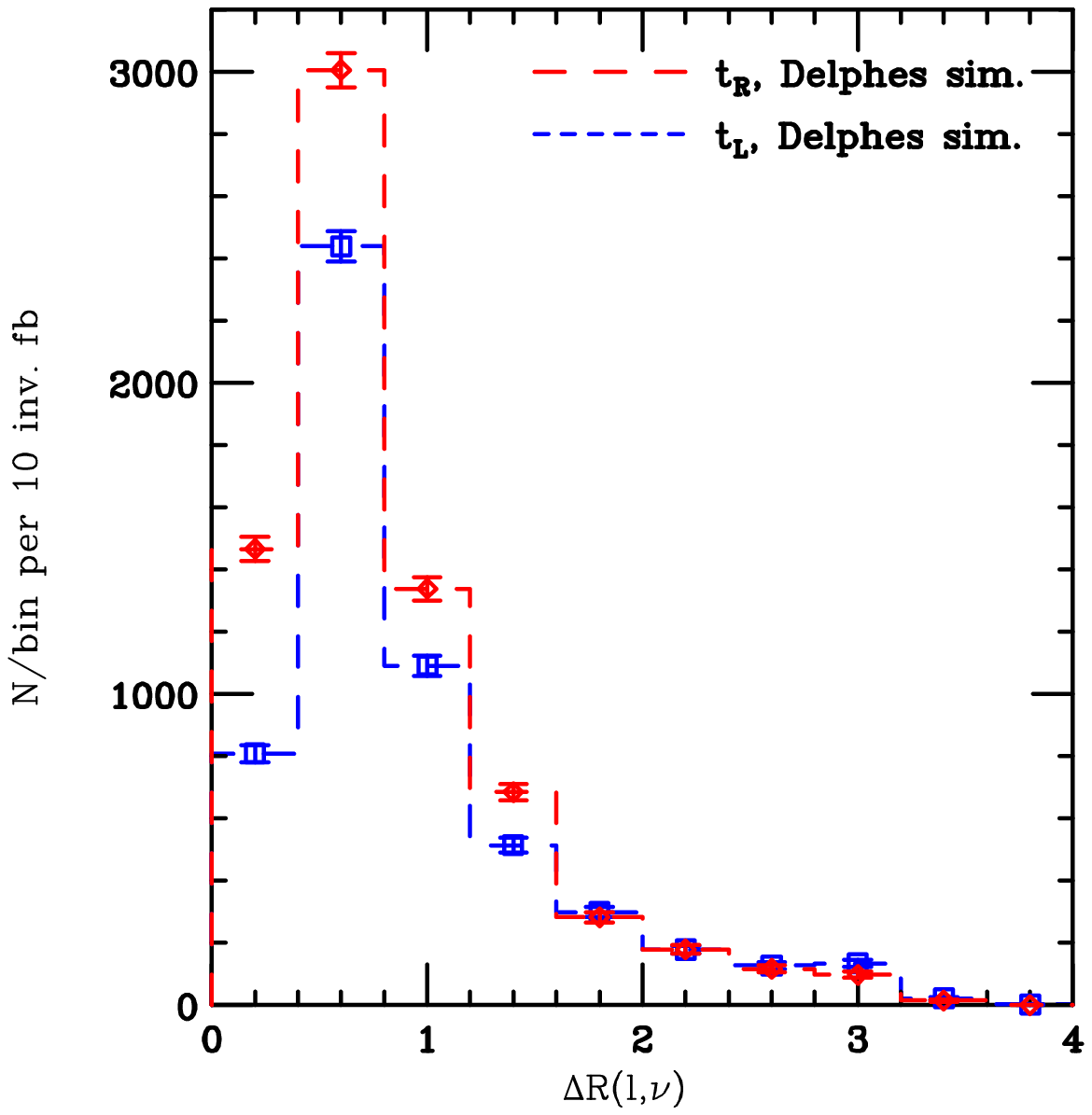}
  \caption{The $\Delta R (\ell,\nu)$ variable for a left- or right-handed
  $Z'$ bosons decaying to $u\bar{t}$ or $\bar{u}t$, obtained from the reconstructed events for an LHC run at 14
  TeV, with 10 fb$^{-1}$, with the set of cuts A (left) and B
  (right), as explained in the text.}
\label{fig:drellnuzprimecuts}
\end{figure}

\begin{figure}[!htb]
  \centering 
  \vspace{5.0cm}
 \includegraphics[scale=0.55, angle=0]{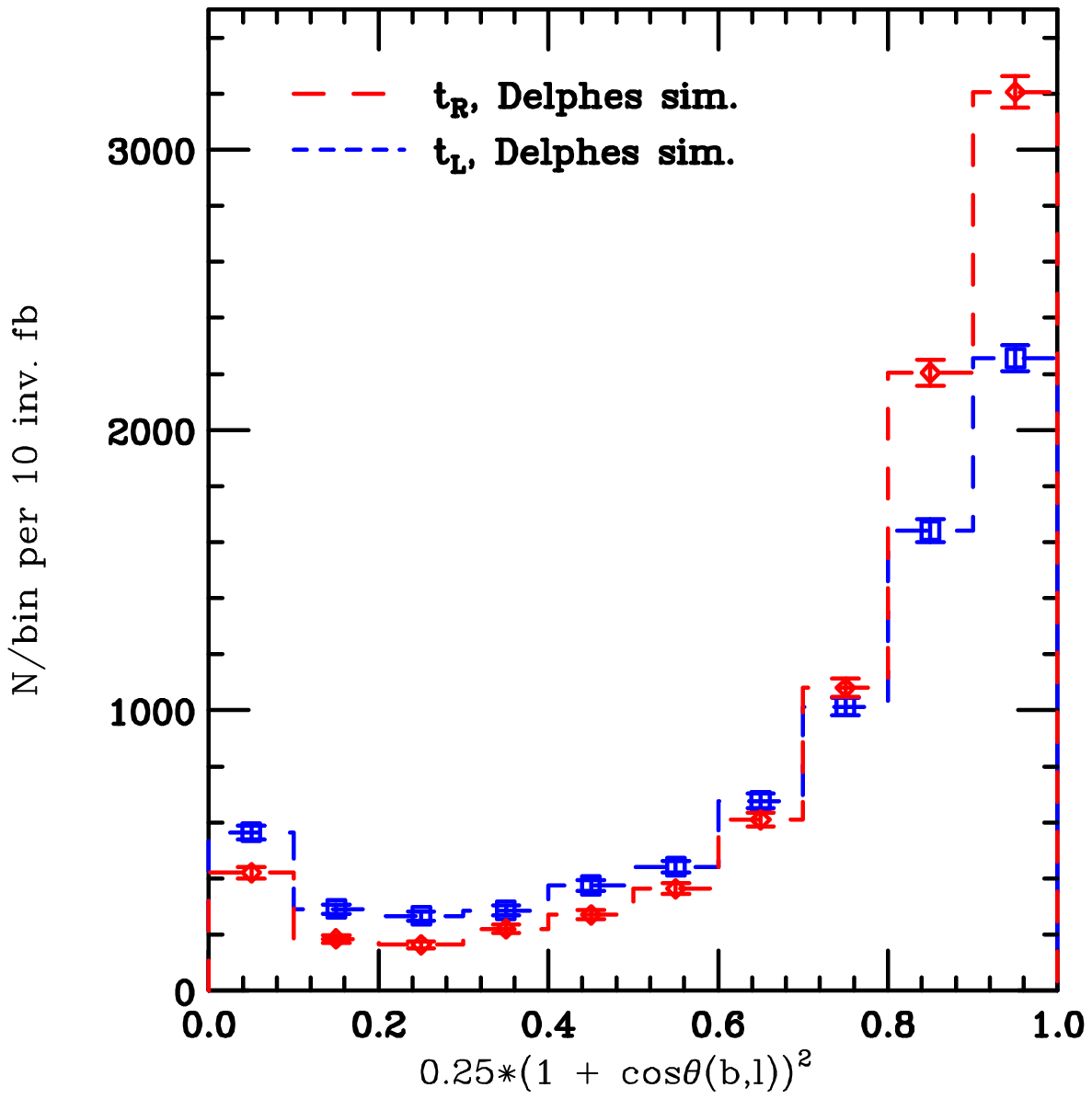}
  \hspace{2.5cm}
  \includegraphics[scale=0.55,angle=0]{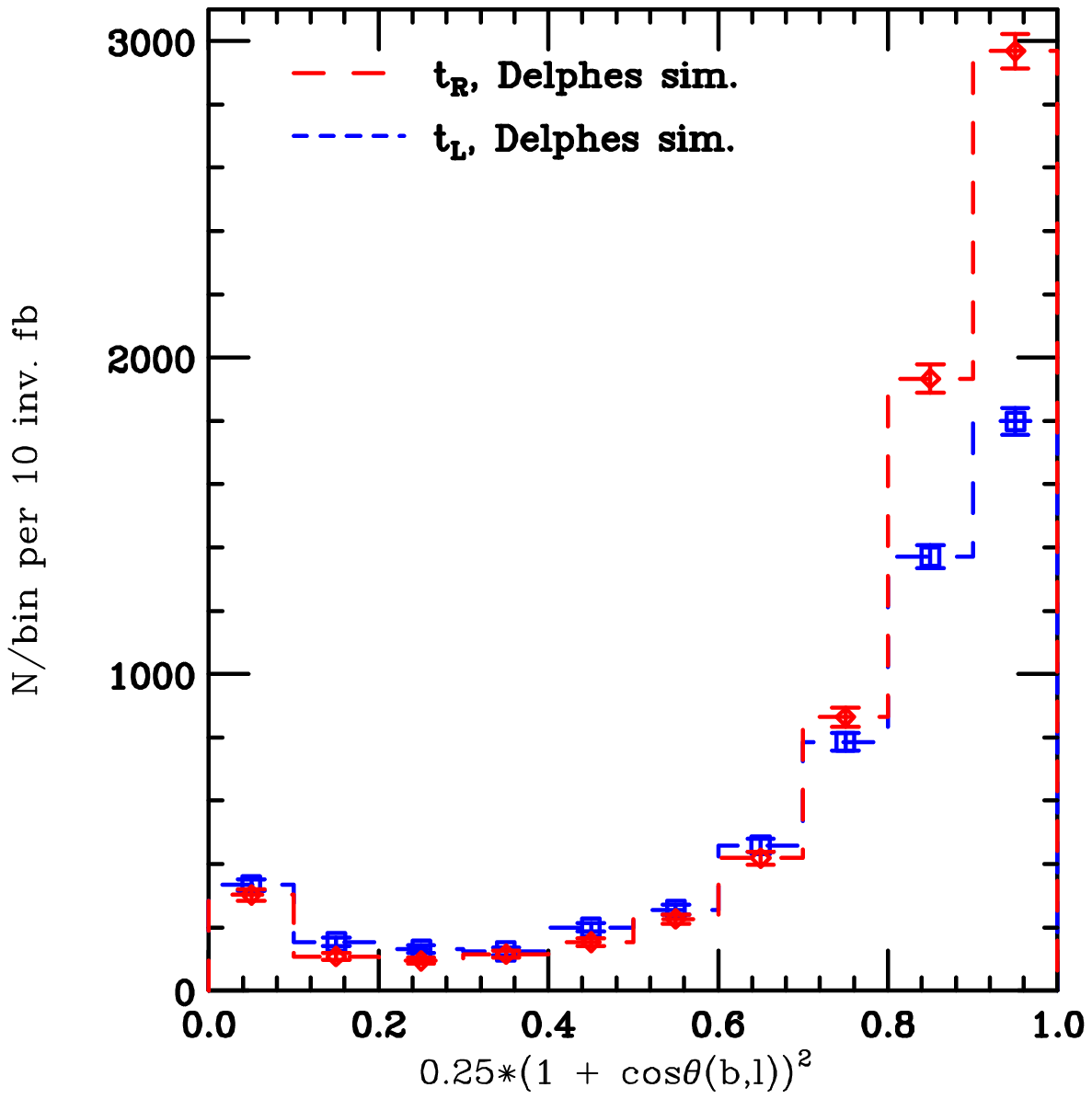}
  \caption{The $f(\cos \theta_{b,\ell})$ variable for a left- or right-handed
  $Z'$ bosons decaying to $u\bar{t}$ or $\bar{u}t$, obtained from the reconstructed events for an LHC run at 14
  TeV, with 10 fb$^{-1}$, with the set of cuts A (left) and B
  (right), as explained in the text.}
\label{fig:chtblzprimecuts}
\end{figure}

Figure~\ref{fig:angdelpheszpparton} shows the angular variables at
parton-level for the $Z'$ model described by the Lagrangian density of
Eq.~(\ref{eq:zprimelag}), for a mass of 1.5 TeV at a 14 TeV LHC. In
Fig.~\ref{fig:angdelpheszp} we show the reconstructed
distributions after Delphes simulation for the minimal set of cuts,
and Figs.~\ref{fig:drellnuzprimecuts} and~\ref{fig:chtblzprimecuts}
show the corresponding reconstructed distributions for the set of cuts
$A$ and $B$, defined in section~\ref{sec:zprimereco}.
Table~\ref{tb:chisqzprime2} shows the corresponding $\chi^2 /
N_{\mathrm{d.o.f.}}$ between the left- and right-handed
distributions corresponding to the figures. It is evident that these
angular variables can provide equivalent magnitudes of
discrimination between left- and right-handed top quarks as the energy
function variables that have been used throughout the main part of the
paper. 

\begin{table}[!t]
\begin{center}
\begin{tabular}{|c|c|c|} \hline
Variable & $\Delta
R(\ell,\nu)$ & $f(\cos \theta (b,\ell))$ \\\hline \hline
Min. & 45.7  &  29.1\\\hline 
A &  35.4 &  19.1\\\hline 
B &  34.0  & 12.6\\\hline 
\end{tabular}
\end{center}
\caption{The value of $\chi^2 / N_{\mathrm{d.o.f.}}$ between the left-
and right-handed distributions in the $Z'$ model with decays to $u\bar{t}$ or $\bar{u}t$, for the three
different sets of cuts. It is evident that the distributions are
distinguishable even for the higher cuts.}
\label{tb:chisqzprime2}
\end{table}

\section{Detector-level smearing}\label{app:delphesVparton}
\begin{figure}[!t]
  \centering 
  \vspace{3.5cm}
 \includegraphics[scale=0.35, angle=0]{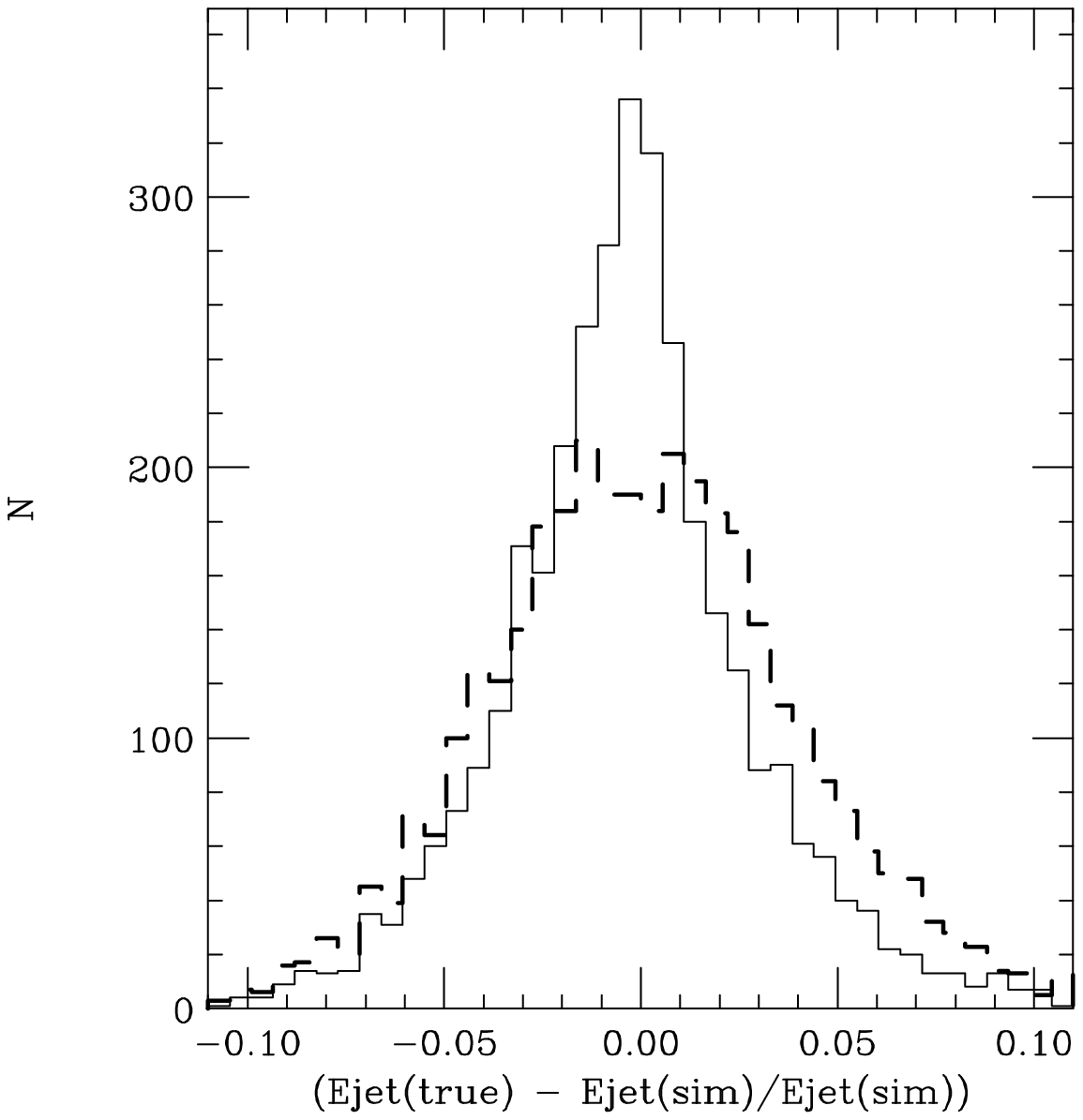}
  \hspace{1.0cm}
  \includegraphics[scale=0.35,angle=0]{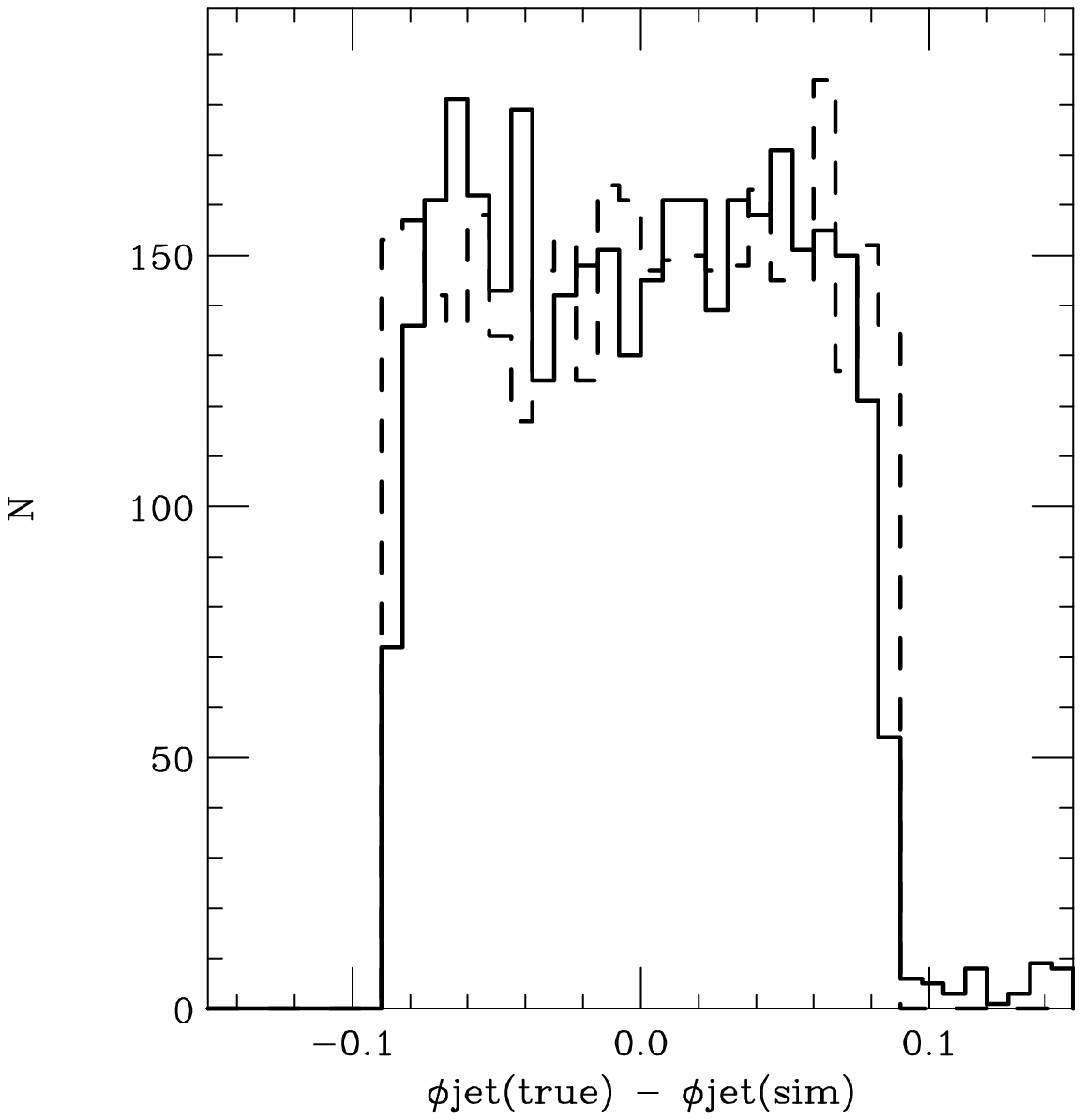}
    \hspace{1.0cm}
  \includegraphics[scale=0.35,angle=0]{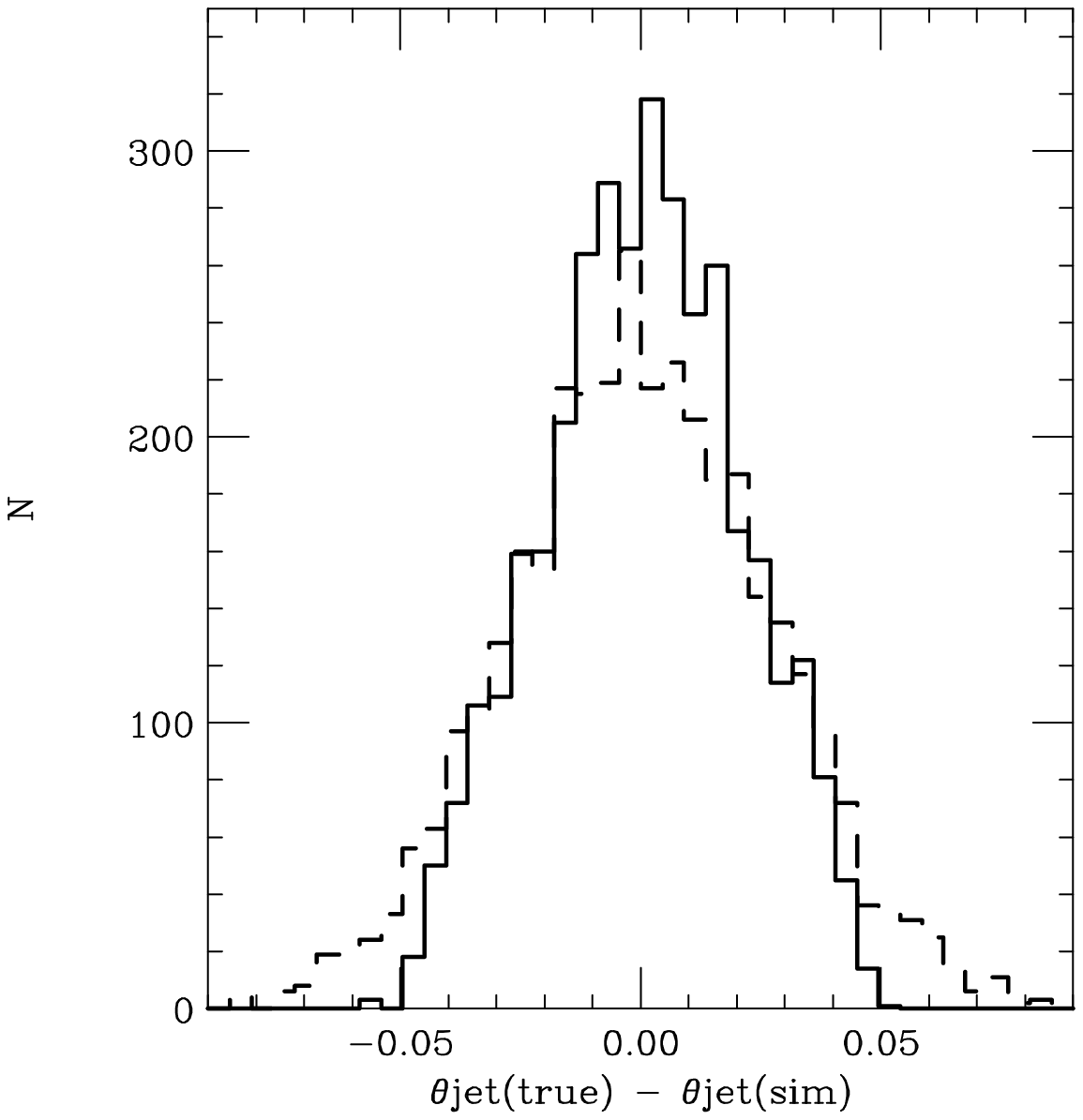}
  \caption[]{The probability density functions for smearing of jet
    energy, the angle on the transverse plane $\phi$ of jets, and the
    azimuthal angle $\theta$ of jets. The solid histograms show the
    actual Delphes simulation results whereas the dashed one show the
    functions that were actually used for the likelihood method of section~\ref{sec:ztautau}.}
\label{fig:jet-discrepancy}
\end{figure}

\begin{figure}[!t]
  \centering 
  \vspace{3.5cm}
   \includegraphics[scale=0.35, angle=0]{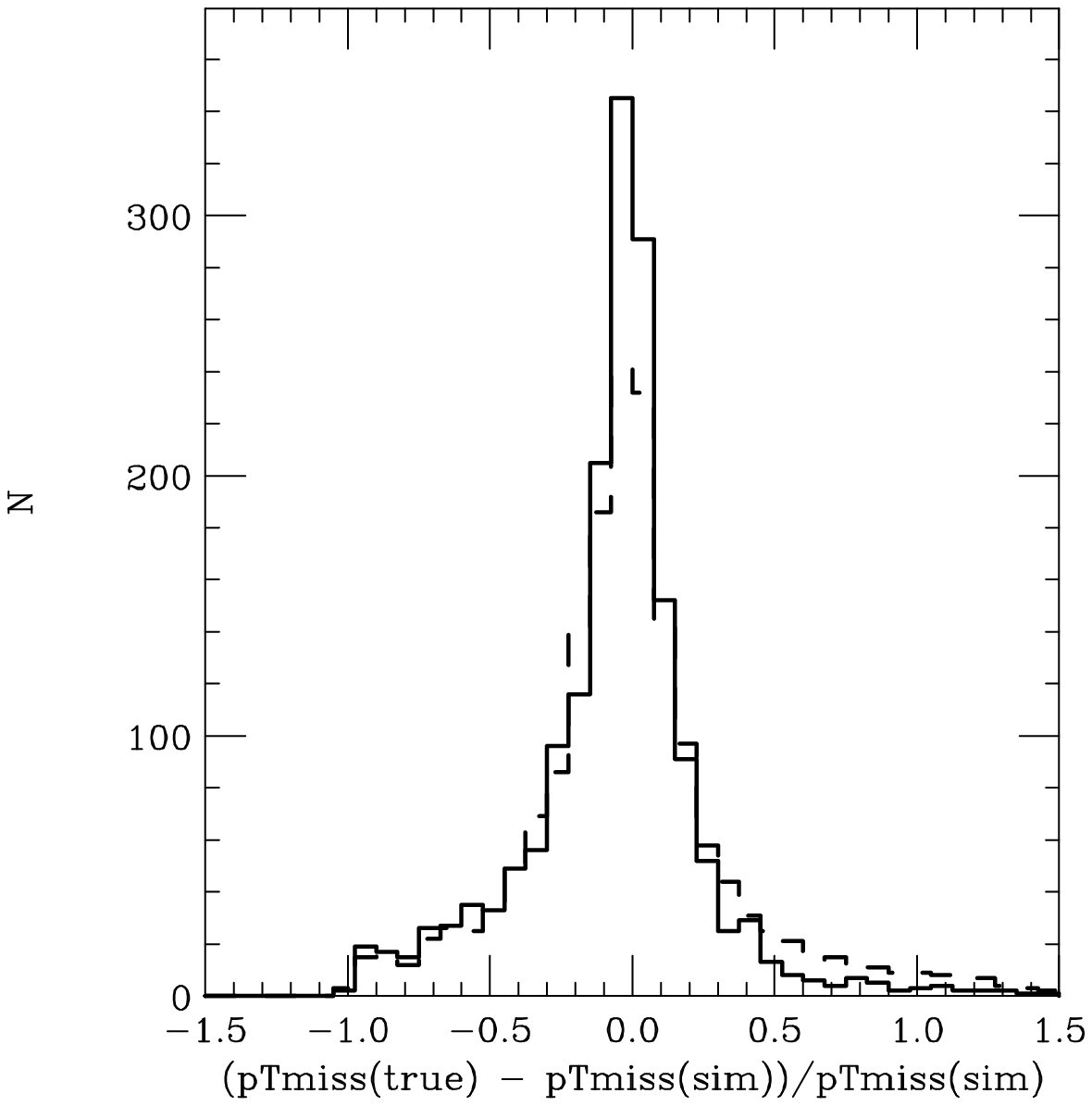}
  \hspace{1.0cm}
  \includegraphics[scale=0.35,angle=0]{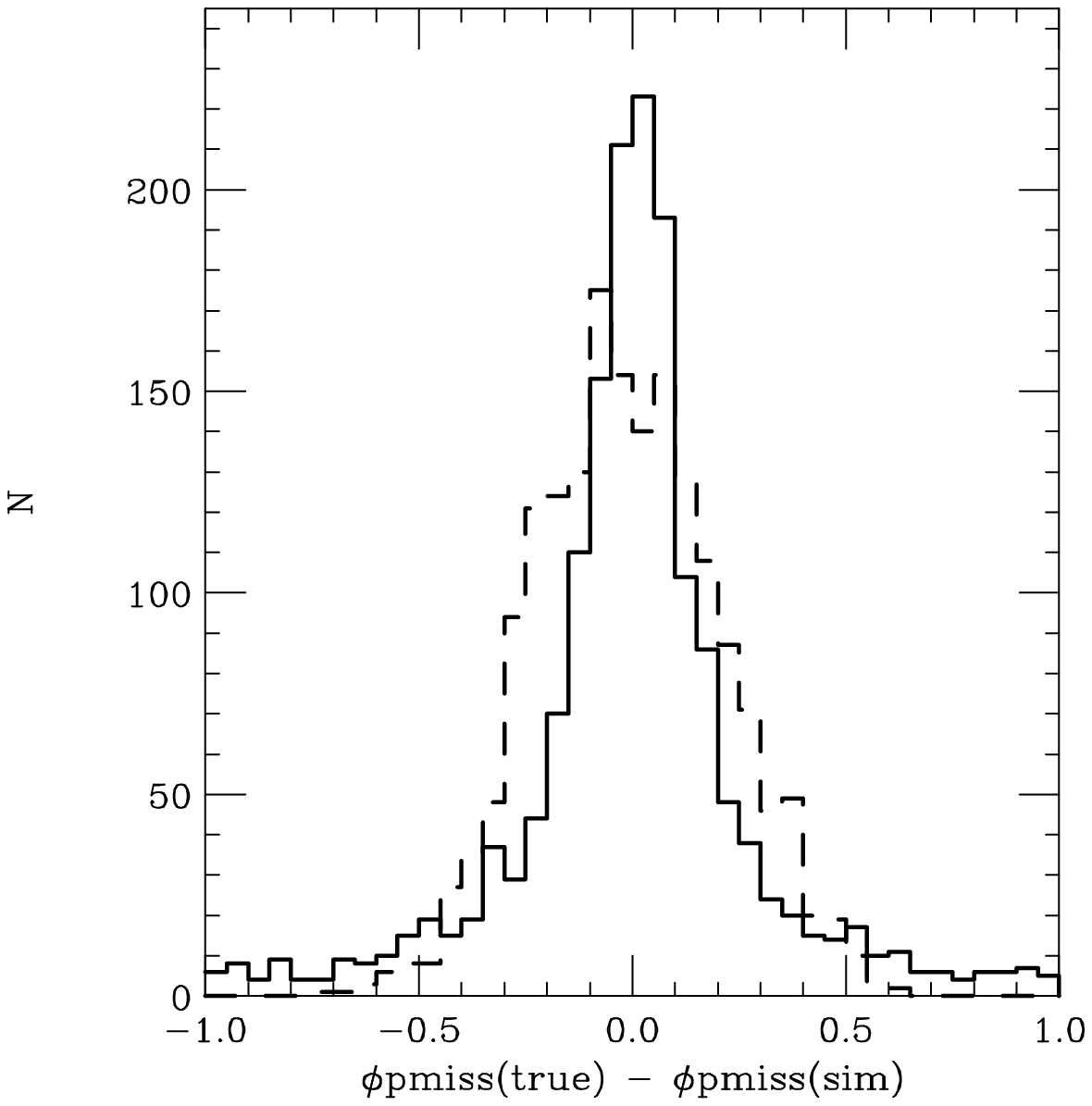}
    \hspace{1.0cm}
  \includegraphics[scale=0.35,angle=0]{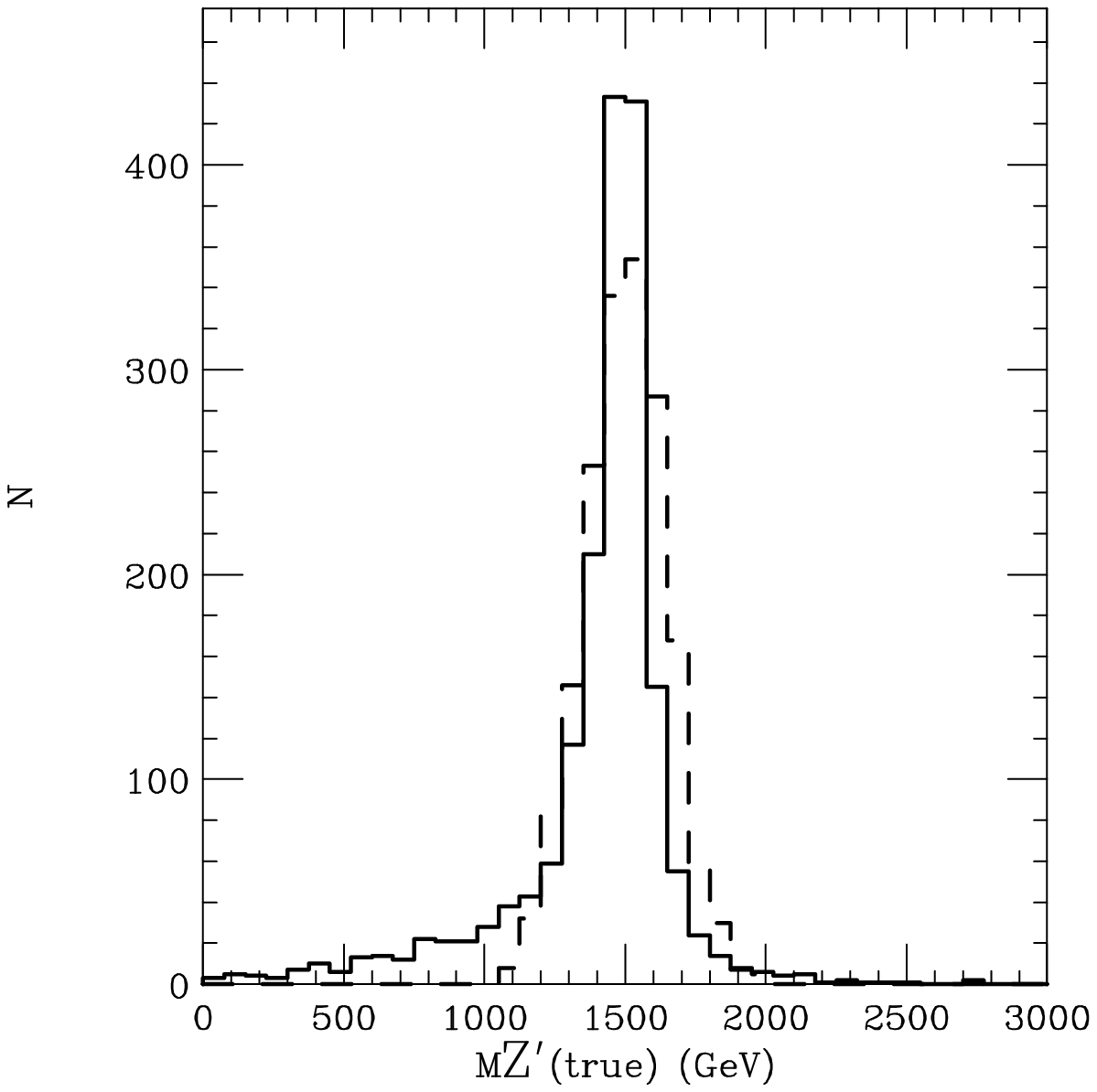}
  \caption[]{The probability density functions for smearing of missing
    transverse energy, the angle on the transverse plane, $\phi$, of
    the missing transverse momentum vector, and the actual $Z'$ mass
    distribution in the model described by the Lagrangian of
    Eq.~(\ref{eq:zprimetautau}) (measured through some other decay
    mode than the $\tau \tau$ mode). The solid histograms show the actual Delphes simulation results whereas the dashed one show the
    functions that were actually used for the likelihood method of section~\ref{sec:ztautau}.}
\label{fig:met-discrepancy}
\end{figure}

The probability density functions used for the likelihood method that
corrects for likelihood methods are shown in Fig.~\ref{fig:jet-discrepancy} and 
Fig.~\ref{fig:met-discrepancy} with dashed histograms.
The solid histograms show the actual discrepancy 
between parton-level and detector-level objects as obtained by
Delphes. In practice these distributions would be known features of the detector
performance. 

\section{Alternative leptoquark decay: $S_{XY} \rightarrow t_R
  \bar{\tau}_L, t_L \bar{\tau}_R$}\label{app:leptoalternative}
\begin{figure}[!t]
  \centering 
\vspace{5.0cm}  
   \includegraphics[angle=0,scale=0.55]{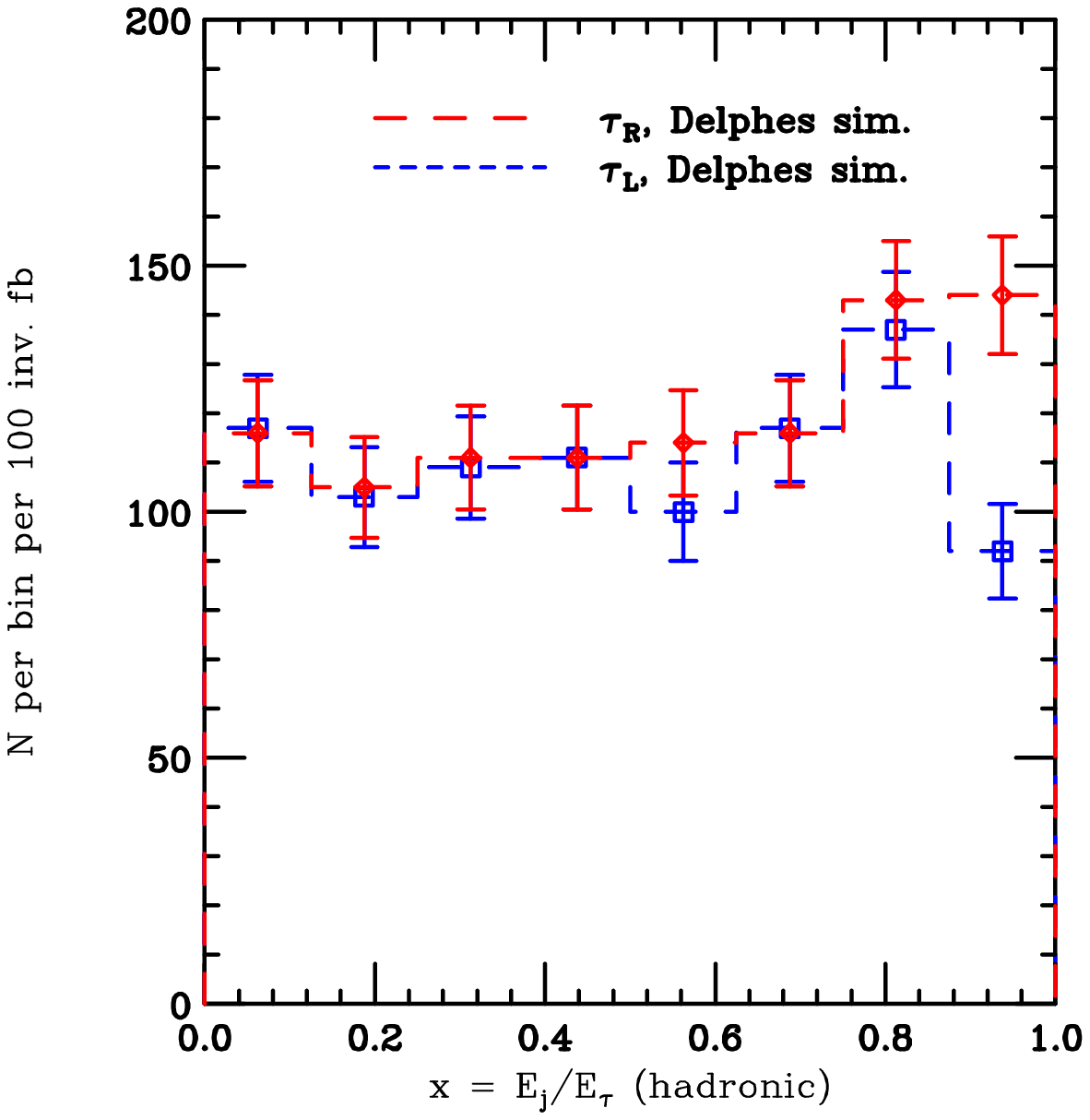}  
    \hspace{1.5cm}  
   \includegraphics[angle=0,scale=0.55]{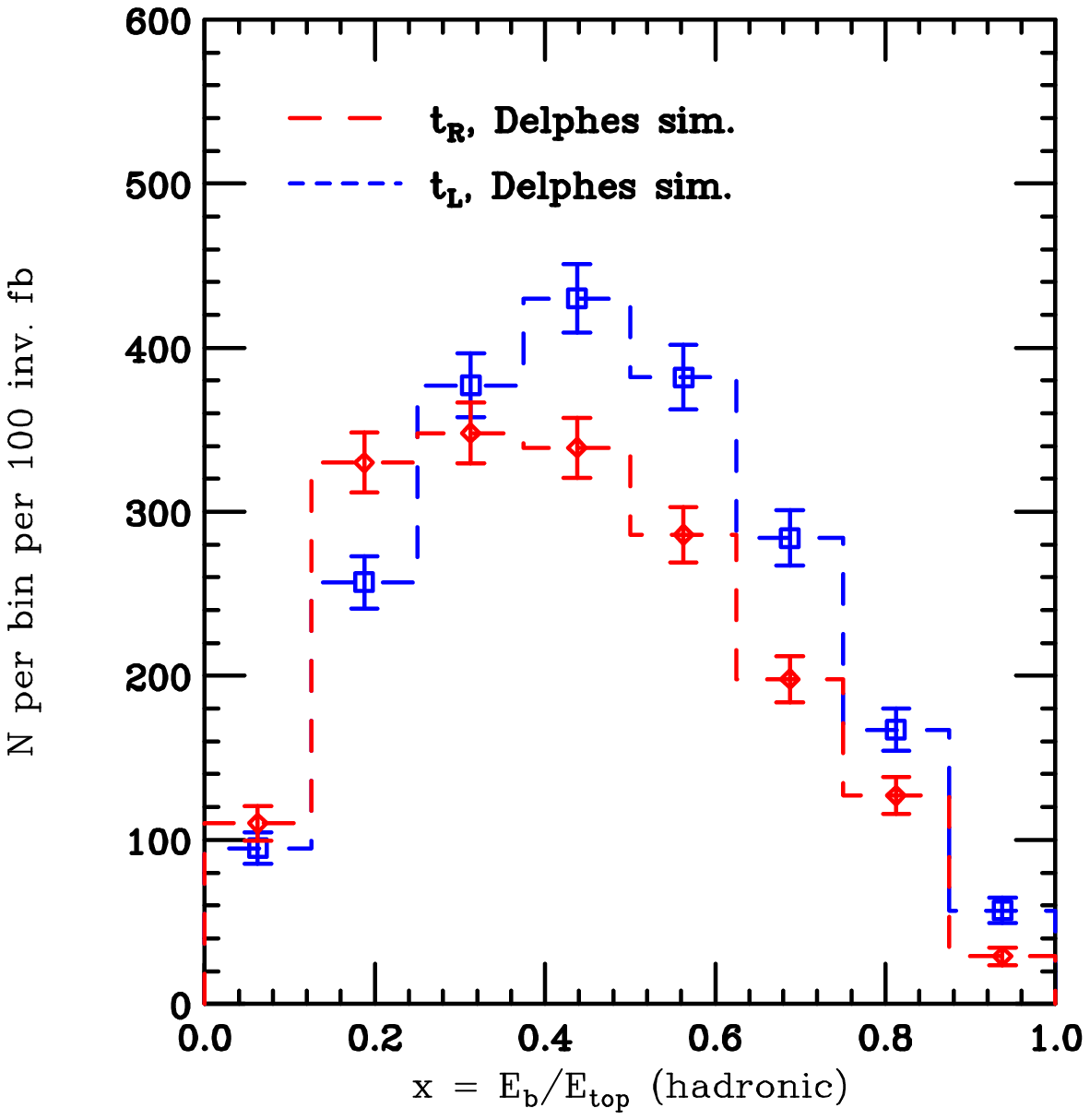}  

\caption{The $x_{\tau}$ (left panel) and $x_{\rm top}$ (right panel) variables for left- or
  right-handed tops from the $S_{XY}$ leptoquark, for an integrated luminosity of
  100 fb$^{-1}$ for a 14 TeV LHC.} 
\label{fig:xtaudelphesttau_had_s12}
\end{figure}

In section~\ref{sec:leptoquark} we considered the decay of a
leptoquark which we called $S_{XX}$, where $X \in \{L,R\}$, with electromagnetic charge $\pm 5/3$ to
either $t_R \tau_R$ or $t_L \tau_L$. Here we
consider the alternative combination of helicities, corresponding to a
leptoquark with charge $\pm 1/3$ which we call $S_{XY}$, where $X \neq Y$ and $X,Y \in \{L,R\}$.\footnote{In the
  notation of Ref.~\cite{Gripaios:2010hv}, we are actually considering the
decays of the upper component of the $S_{1/2}$ doublet,
$S^{(+)}_{1/2}$.} The results for $x_{\mathrm{top}}$ and $x_\tau$,
corresponding to the fully-hadronic leptoquark
analysis constructed in section~\ref{sec:leptoquark} are shown in Fig.~\ref{fig:xtaudelphesttau_had_s12}, for the minimal
set of cuts. Note that in this case one should understand that the left-handed $\tau$ should be paired up
with the right-handed top and vice versa (i.e. red in one plot with
blue in the other).  

The values of $\chi^2 / N_{\mathrm{d.o.f.}}$ were found to differ
compared to those of the $S_{XX}$ leptoquark, corresponding to $\sim 1.6$ for the $x_\tau$
variable and $\sim 8.2$ for the $x_{\mathrm{top}}$ variable. The
differences between the $S_{XY}$ and $S_{XX}$ leptoquarks arise due to
the difference in acceptances of the left- and right-handed
particles. These are a consequence the interplay of the efficiencies
associated with the $b$-tagging, $\tau$-tagging and reconstruction of
the top quarks. 

\section{Matrix element for polarised top decay}\label{app:polarizedtop}
\begin{figure}[!t]
  \centering 
  \hspace{4cm}
  \includegraphics[scale=0.80]{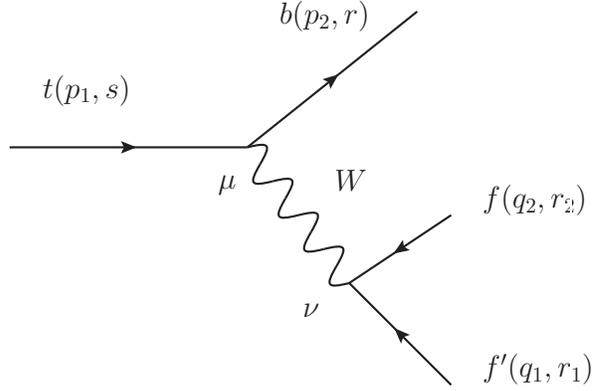}
  \caption{The Feynman diagram for polarised top decay. The
    parentheses indicate the 4-momentum and spin labels respectively, for each
    particle. }
\label{fig:topfeyn}
\end{figure}

It is useful to outline here the derivation of the polarised top decay differential
width. The coupling of the $W$ boson to the fermions is given by:
\beq
\mathcal{L} = \frac{g}{2\sqrt{2}} V_{ff'} \bar{f} \gamma_\mu (1 -
\gamma_5) f' W^\mu + \mathrm{h.c.}\;,
\eeq
where $g$ is the weak charge and $V_{ff'}$ is the CKM matrix element
corresponding to the fermions $f$ and $f'$. 
The matrix element, corresponding to the diagram shown in
Fig.~\ref{fig:topfeyn} is given by:
\begin{eqnarray}
\mathcal{M} = \frac{g^2}{8} V_{tb} V_{ff'} \bar{u}^r_b(p_2) \gamma_\mu
(1-\gamma_5) W^\mu u^s_t{p_1} \bar{u}_{f'}^{r_1} (q_1) \gamma_\nu
(1-\gamma_5) v^{r_2}_f(q_2) W^\nu G(q^2)\;,
\end{eqnarray}
where $u$ and $v$ are the positive and negative frequency spinors,
$W^\mu$ is the $W$ polarisation vector and
$G(q^2)$ is the $W$ propagator,
\beq
G(q^2) = \frac{1}{(q^2 - m_W^2) + i \Gamma_W m_W}\;.
\eeq
Squaring the matrix element and summing over the $b$ quark and fermion
spins, we obtain:
\begin{eqnarray}\label{eq:traces}
\sum_{r_1,r_2,r} \left| \mathcal{M} \right| &=& \Omega~ \mathrm{Tr} \left[
  \frac{1}{2} ( 1 + 2 s \gamma_5 \slashed{S} ) (\slashed{p}_1 + m_{\rm
    top})
  \gamma_k (1-\gamma_5) \slashed{p}_2 \gamma_\mu (1-\gamma_5) \right]
\nonumber \\
&\times& \mathrm{Tr}\left[ \slashed{q}_1 \gamma_\nu (1-\gamma_5)
  \slashed{q}_2 \gamma_\lambda (1- \gamma_5)\right] \nonumber \\
  &\times& W^\mu W^\nu W^{*\lambda} W^{*k}\;,
\end{eqnarray}
where $S = 1/m_{\rm top} ( |\vec{p}_1|, E_1 \vec{p}_1 /|\vec{p}_1|) $ is the
spin 4-vector for the top-quark and $\Omega$ is defined as:
\beq
\Omega \equiv \frac{g^4}{64} \left| V_{tb} \right|^2 \left|
  V_{ff'}\right|^2  \times \frac{1}{(q^2 - m_W^2)^2 + \Gamma_W^2 m_W^2}\;\;.
\eeq 
The traces can be calculated using the \texttt{FORM}
package~\cite{Vermaseren:2000nd}. The first trace in Eq.~(\ref{eq:traces}), corresponding to a top quark with
spin $s = \pm 1/2$ and the bottom
quarks, is given by:
\begin{eqnarray}
&4& ( p_1^\mu p_2^k + p_1^k p_2^\mu - g^{\mu k } p_1 \cdot p_2 -
\epsilon^{ij\mu k} p^i_1 p^j_2 ) \nonumber \\
+ &8& m_{\rm top} s ( -p_2^\mu S^k - S^\mu p_2^k + g^{\mu k } p_2 \cdot S -
\epsilon^{ij\mu k} p_2^i S^j )\;.
\end{eqnarray}
It is obvious that the second term in the above result vanishes if we sum
over $s$ or set $m_{\rm top} \rightarrow 0$. The second trace in Eq.~(\ref{eq:traces}), corresponding to the fermions $f$ and $f'$ is given
by:
\begin{eqnarray}
8(q^\nu_1 q_2^\lambda + q^\lambda_1 q^\nu_2 - g^{\nu\lambda} q_1 \cdot
q_2 + 8 \epsilon^{ij\mu k} q^i_1 q_2^j)\;.
\end{eqnarray}
Summing over the $W$ polarisations introduces $g^{k \lambda}$ and
$g^{\mu \nu}$, and gives, for the polarised top matrix
element squared,
\beq
\left| \mathcal{M} \right|^2(s) = 128 \Omega ( p_2 \cdot q_1 ) \left[( p_1 -
m_{\rm top} (2s) S ) \cdot q_2\right]\;.
\eeq

\section{Finite mass effects on top polarisation}\label{app:pd}
\subsection{Production polarisation}
\begin{figure}[!t]
  \centering 
  \vspace{3.0cm}
 \includegraphics[scale=0.40, angle=0]{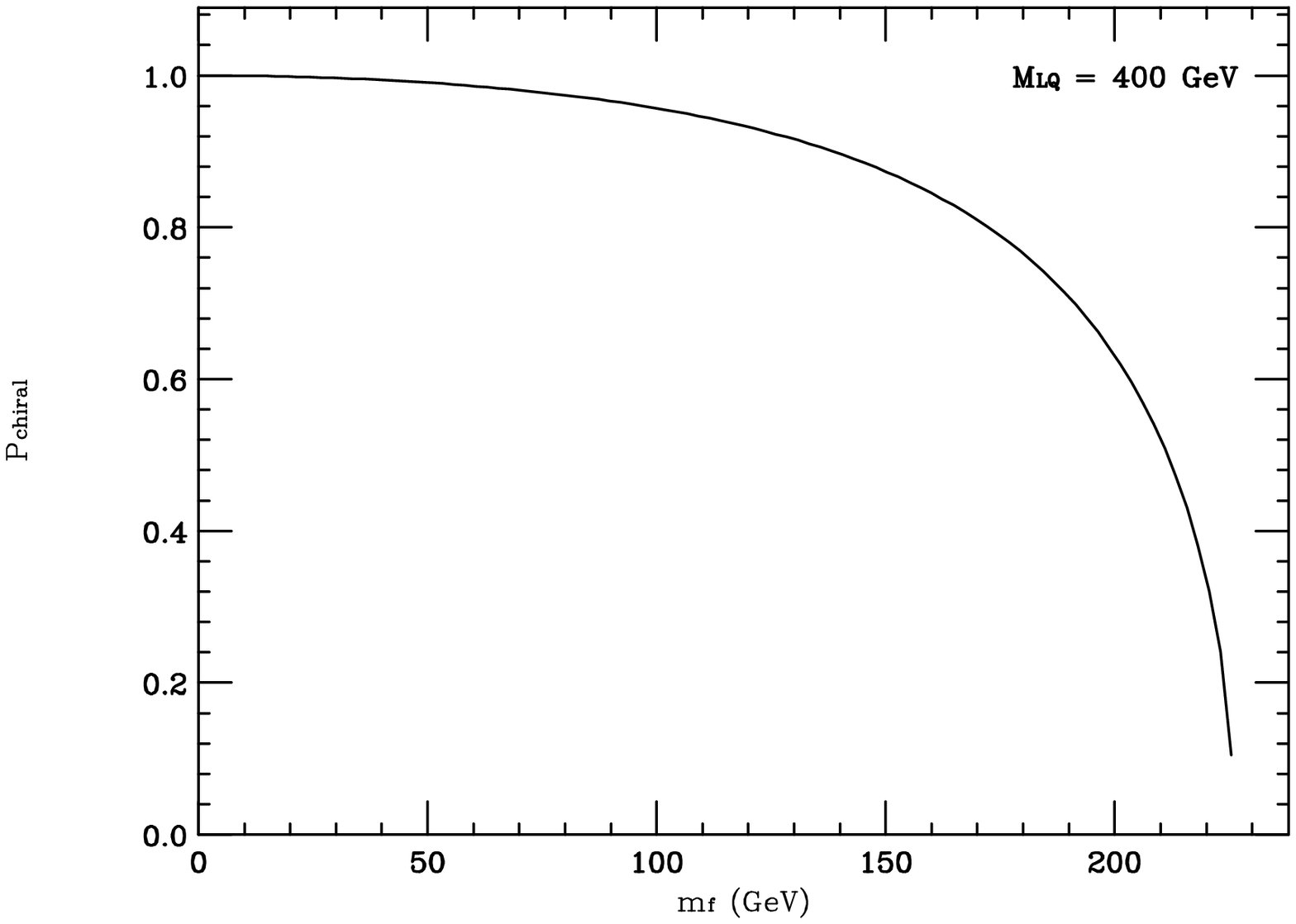}
  \hspace{2.0cm}
  \includegraphics[scale=0.40, angle=0]{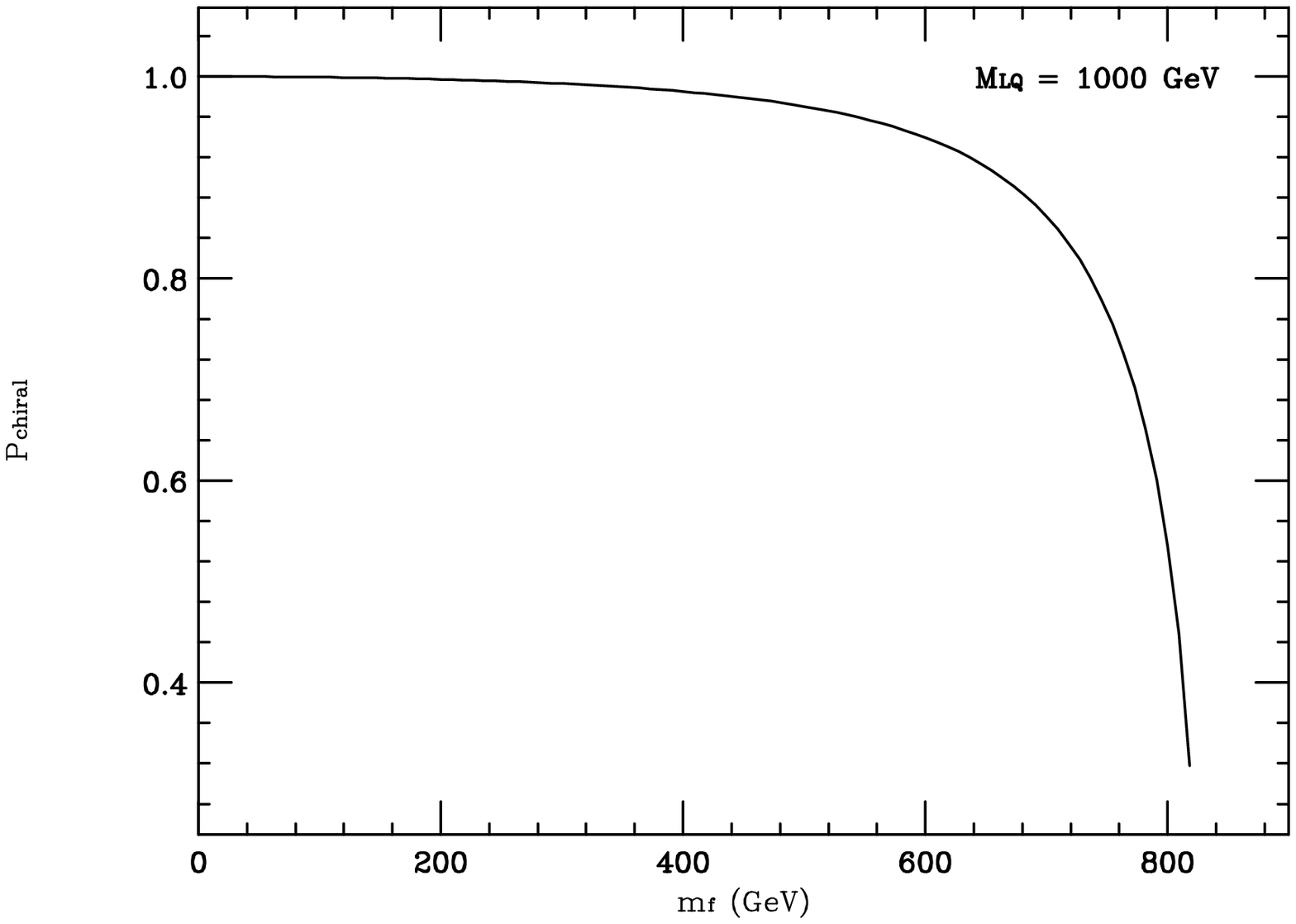}
  \caption{The effect of the sister fermion mass $m_f$ on the purely chiral
    top quark polarisation, $\left< P_P \right>_{\rm chiral}$, in the decay
    of a scalar leptoquark of mass $M_{LQ} = 400 \gev$ (left panel), or
    $M_{LQ} = 1000 \gev$ (right panel), to $tf$.}
\label{fig:mfeffect}
\end{figure}

We adapt the description given in Ref.~\cite{Shelton:2008nq} for the
case of stop decay to a top and a neutralino, to the decay of a scalar
leptoquark to a top and another fermion. The corresponding Lagrangian
for the decay of scalar leptoquark $S$ to $t_{L,R}$ and a fermion $f$ can be written as:
\beq
\mathcal{L} = g_L\bar{t}_LS f + g_R\bar{t}_R Sf + \mathrm{h.c.}\;,
\eeq
where $g_L$ and $g_R$ are the left- and right-handed couplings
respectively. The axis of spin quantisation is taken to lie along the
top direction of motion in the parent leptoquark rest frame. The
production amplitudes for positive and negative helicity top quarks
depend on two functions $F_\pm$:
\beq
F_\pm = \frac{ (E_t + m_{\rm top} \pm |p_t|) (E_f + m_f \pm |p_f|) } {\sqrt{4 ( E_t
  + m_{\rm top}) ( E_f + m_f ) }} \;,
\eeq 
where all quantities are given in the leptoquark rest frame. The
functions result from explicit evaluation of the matrix element for
the leptoquark decay using the spinor wavefunctions. These are given
by:
\begin{equation}
u(\vec{p},s) = 
\begin{pmatrix}
\sqrt{p\cdot\sigma} \chi_s  \\
\sqrt{p\cdot\bar{\sigma}} \chi_s   \\
\end{pmatrix},\;\;v(\vec{p},s) = 
\begin{pmatrix}
2s \sqrt{p\cdot\sigma} \chi_{-s}  \\
-2s \sqrt{p\cdot\bar{\sigma}} \chi_{-s}   \\
\end{pmatrix}\;,
\end{equation} 
where the spinors $\chi_s$ are eigenstates of the operator $1/2
\vec{\sigma} \cdot \hat{\vec{s}}$ with eigenvalue $s$, where $s = \pm
1/2$ corresponding to spin-up and spin-down respectively. The spinors are given,
for $\hat{\vec{s}}$ along the $z$-direction, by:
\begin{equation}
\chi_{1/2} (\hat{\vec{z}}) = 
\begin{pmatrix}
1\\
0\\
\end{pmatrix},\;\;\chi_{-1/2} (\hat{\vec{z}}) = 
\begin{pmatrix}
0\\
1\\
\end{pmatrix}\;.
\end{equation} 
For the decay of a scalar to left-handed tops, for example, the matrix element is
proportional to $\bar{v} (\vec{p},s) P_L u(\vec{k},r)$, where
$\vec{p}$ and $s$ are the top momentum and spin respectively, and
$\vec{k}$ and $r$ are the fermion momentum and spin respectively. The
$P_L$ operator projects the upper component of $u(\vec{k},r)$ and
hence the matrix element for the left-handed coupling is given by:
\beq
\mathcal{M}^L_{sr} \propto \begin{pmatrix}
-2s \sqrt{p\cdot\bar{\sigma}} \chi_{-s}^\dagger & ~2s
\sqrt{p\cdot\sigma} \chi_{-s}^\dagger   \\
\end{pmatrix}\begin{pmatrix}
 \sqrt{k\cdot\sigma} \chi_{r}  \\
0\\
\end{pmatrix}\;.
\eeq
Using the relations~\cite{Dreiner:2008tw}:
\begin{eqnarray}
\sqrt{p\cdot \sigma} = \frac{ (E_p + m) \mathbb{I}  - \vec{\sigma}\cdot
  \vec{p} } { \sqrt{2 (E_p + m) } }\;,\nonumber\\
\sqrt{p\cdot \bar{\sigma}} = \frac{ (E_p + m) \mathbb{I}  + \vec{\sigma}\cdot
  \vec{p} } { \sqrt{2 (E_p + m) } }\;,
\end{eqnarray}
we obtain the following:
\beq
\mathcal{M}^L_{sr} \propto -2s \chi^\dagger_{-s} \frac{ [ (E_t + m_{\rm top})
    \mathbb{I} + \vec{\sigma}\cdot\vec{p} ] [ (E_f + m_f )\mathbb{I}
    - \vec{\sigma}\cdot \vec{k} ]
} { \sqrt{4 (E_t + m_{\rm top}) ( E_f + m_f ) } } \chi_r\;,
\eeq
or, writing out the matrices explicitly:
\beq
\mathcal{M}^L_{sr} \propto -\frac{2s}{D} \chi^\dagger_{-s}  \begin{pmatrix}
E_t + m_{\rm top} + p_z & p_x - i p_y  \\
p_x + ip_y & E_t + m_{\rm top}- p_z  \\
\end{pmatrix} \begin{pmatrix}
E_f + m_f - k_z & -k_x + i k_y  \\
k_x + ik_y & E_f + m_f + k_z  \\
\end{pmatrix} \chi _r \;,
\eeq
where we have defined the denominator as $D = \sqrt{4 (E_t + m_{\rm
    top}) ( E_f + m_f )}$. In the rest frame of the scalar leptoquark,
we can take the $z$-axis to
lie along the top direction of motion, and hence we have $p_x = p_y =
k_z = k_y = 0$ and $ p_z = -k_z $ and the matrix element becomes:
\beq
\mathcal{M}^L_{sr} \propto -\frac{2s}{D} \chi^\dagger_{-s}  \begin{pmatrix}
E_t + m_{\rm top} + |p_z| & 0  \\
0 & E_t + m_{\rm top} - |p_z|  \\
\end{pmatrix} \begin{pmatrix}
E_f + m_f + |k_z| & 0 \\
0 & E_f + m_f - |k_z|  \\
\end{pmatrix} \chi _r \;,
\eeq
Performing the matrix multiplication, one immediately notices that the
functions $F_{\pm}$ appear as elements of the resulting matrix:
\beq
\mathcal{M}^L_{sr} \propto -2s \chi^\dagger_{-s}  \begin{pmatrix}
F_+ & 0  \\
0 & F_-  \\
\end{pmatrix} \chi _r \;,
\eeq
If we choose $s=+1/2$ then:
\beq
\mathcal{M}^L_{+r} \propto - ( 0, F_-) \chi _r\;,
\eeq
or, if we choose $s=-1/2$ then:
\beq
\mathcal{M}^L_{-r} \propto - ( F_+, 0) \chi _r\;.
\eeq
Upon squaring and summing over the fermion spin, $r$, we obtain:
\beq
\sum_{r} |\mathcal{M}^L_{+r}|^2 = |\mathcal{M}^L_{+-}|^2 \propto | g_L
F_+ |^2\;,
\eeq
and 
\beq
\sum_{r} |\mathcal{M}^L_{-r}|^2 = |\mathcal{M}^L_{-+}|^2 \propto | g_L
F_- |^2\;.
\eeq
Since only one component of $|\mathcal{M}^L_{sr}|$ contributes, we may
also write:
\beq
\mathcal{M}^L_{\pm} \equiv \mathcal{M}^L_{\pm\mp}\;,
\eeq
and hence:
\beq
\mathcal{M}^L_{\pm} \propto g_L F_\mp\;.
\eeq
The extension to $\mathcal{M}^R_{\pm}$ is trivial, for which the resulting matrix
elements are thus given by:
\begin{eqnarray}
\mathcal{M}^R_\pm \propto g_R F_ \pm\;.
\end{eqnarray}
For finite $m_{\rm top}$ this gives a non-vanishing amplitude for top quarks
of both helicities even in the limit of a purely chiral vertex. It can
be shown that the polarisation along the production axis at a parent rest frame is given by:
\beqn
\left< P_P \right> &\equiv& 
\frac{|{\cal M}_+^L + {\cal M}_+^R|^2 - |{\cal M}_-^L + {\cal M}_-^R|^2}
{|{\cal M}_+^L + {\cal M}_+^R|^2 + |{\cal M}_-^L + {\cal M}_-^R|^2} \nonumber \\
&=&
\frac{ (|g_R|^2 - |g_L|^2) M_{\rm LQ} |p_t|} { (|g_R|^2
  + |g_L|^2) (M_{\rm LQ} E_t - m_{\rm top}^2 ) + 2 g_R g_L m_f m_{\rm top} } \;,
\eeqn
where 
\beqn
&& |p_t| = \frac{\sqrt{(M_{\rm LQ}^2 + m_{\rm top}^2 - m_f^2)^2 - 4
    M^2_{\rm LQ} m_{\rm top}^2}}{2 M_{\rm LQ}} \,, \nonumber \\
&& E_t = \frac{M_{\rm LQ}^2 - m_f^2 + m_{\rm top}^2}{2 M_{\rm LQ}}\,,
\eeqn
and $M_{\rm LQ}$ is the parent leptoquark mass. It is interesting to
consider the effect of the mass of the sister fermion produced in the decay in the purely chiral
case. Setting either $g_R = 0$ or $g_L = 0$ defines:
\beq
\left< P_P \right>_{\rm chiral} \equiv \left< P_P
\right>_{(g_{R}~{\rm or}~g_{L}  = 0)}  = \pm \frac{ M_{\rm LQ} |p_t|} {
  M_{\rm LQ} E_t - m_{\rm top}^2  } \;.
\eeq
If we had $m_f = m_{\rm top} = 0$, then we would get $\left< P_P \right>_{\rm
  chiral} = \pm 1$. For a non-zero top quark mass, it turns out that for a wide range of values of
$m_f$, the deviation from the value of $\left< P_P \right>_{\rm
  chiral} = \pm 1$ is small. Figure~\ref{fig:mfeffect} shows the effect of the fermion mass for the
case $M_{LQ} = 400 \gev$ and $M_{LQ} = 1000 \gev$. It is clear that
when $m_f \ll m_{\rm top}$, then the
effect is small, and $\left< P_P \right>_{\rm chiral}$ is close to the
$m_f = 0$ value. The effect is $\mathcal{O}(\rm few~\%)$ even for $m_f
\sim 100 \gev$ in the $M_{LQ} = 400 \gev$ case. For $M_{LQ} = 1000
\gev$, the effect of the sister particle mass is of $\mathcal{O}(\rm few~\%)$ even for masses as high as $\sim 600 \gev$.\footnote{Note that it may be interesting
to investigate whether the helicity distributions of the top can provide independent
information on the mass of the accompanying sister particle,
especially if this is weakly-interacting (and hence invisible).}

One may also consider effects of relativistic rotation via the Wigner angle between the
production and detection axes of the top quark~\cite{Shelton:2008nq}. However, if the velocity of parent particle is small or the top is
highly boosted in the parent rest frame, the effect of the Wigner
rotation is negligible. Therefore we can ignore it for our $Z'$ examples, since
we have assumed $Z'$ being much heavier than the daughter particles. 
Even for the leptoquark example, the effect has been checked and found
not to be significant, since most of the events have $\beta_p \ll 1$ and $\beta_t \sim 1$, 
as we will see in the next subsection.   

\subsection{$\beta$ distributions from Monte Carlo}\label{app:beta}
\begin{figure}[!t]
  \centering 
  \vspace{0.5cm}
 \hspace{4.0cm}
  \includegraphics[scale=0.40, angle=90]{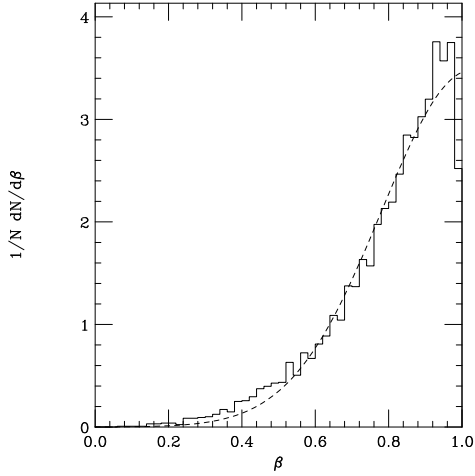}
  \caption{The fit to the top quark $\beta_t$ distribution for $400
    \gev$ leptoquarks is shown in solid black dashes. The
\Herwigpp histogram extracted from parton-level Monte Carlo events is
shown in solid black.}
\label{fig:beta400}
\end{figure}
\begin{figure}[!t]
  \centering 
  \vspace{0.5cm} 
  \includegraphics[scale=0.40, angle=270]{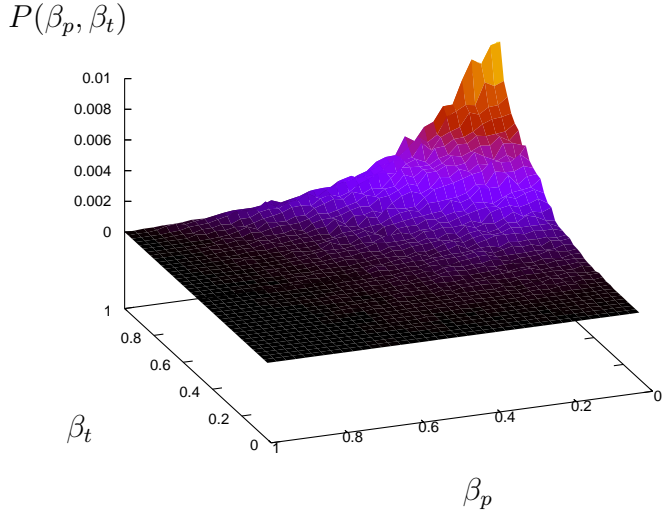}
  \put(-75,-174){$\beta_p$}
  \put(-225,-150){$\beta_t$}
  \put(-247,5){$P(\beta_p,\beta_t)$}
  \caption{The $\beta_p - \beta_t$ distribution for $400
    \gev$ leptoquarks, extracted from the \Herwigpp event generator,
    is shown.}
\label{fig:beta2d}
\end{figure}

We also show the form of the two-dimensional $\beta_p - \beta_t$ distribution for $400
    \gev$ leptoquarks, extracted from the \Herwigpp event generator in Fig.~\ref{fig:beta2d}.
For the $\beta_t$ distribution (i.e. integrated over $\beta_p$) in the
decay of a $400$~GeV leptoquark a fit can be made, shown in
Fig.~\ref{fig:beta400}. The fit has the form of a Gaussian:
\beq\label{eq:boostdist}
P(\beta_t) = a  \exp( -(\beta_t - b)^2 / c )\;\;,
\eeq
where the parameters $a$, $b$ and $c$ were given by the fit to be $a \simeq
3.48$, $b \simeq 1.03$, $c \simeq 0.13$. The distribution integrates to
$\sim 1$ in $\beta_t \in (0,1)$.
\bibliography{thirdgenhel}

\providecommand{\href}[2]{#2}\begingroup\raggedright\begin{thebibliography}{10}

\bibitem{Gripaios:2009dq}
B.~Gripaios, ``{Composite Leptoquarks at the LHC},'' {\em JHEP} {\bf 1002}
  (2010) 045, \href{http://arXiv.org/abs/0910.1789}{{\tt 0910.1789}}.

\bibitem{Jezabek:1988ja}
M.~Jezabek and J.~H. Kuhn, ``{Lepton Spectra from Heavy Quark Decay},'' {\em
  Nucl.Phys.} {\bf B320} (1989) 20.

\bibitem{Czarnecki:1990pe}
A.~Czarnecki, M.~Jezabek, and J.~H. Kuhn, ``{LEPTON SPECTRA FROM DECAYS OF
  POLARIZED TOP QUARKS},'' {\em Nucl.Phys.} {\bf B351} (1991) 70--80.

\bibitem{Shelton:2008nq}
J.~Shelton, ``{Polarized tops from new physics: signals and observables},''
  {\em Phys.Rev.} {\bf D79} (2009) 014032,
  \href{http://arXiv.org/abs/0811.0569}{{\tt 0811.0569}}.

\bibitem{DTP-91-42}
B.~K. Bullock, K.~Hagiwara, and A.~D. Martin, ``{Tau polarization as a signal
  of charged Higgs bosons},'' {\em Phys.Rev.Lett.} {\bf 67} (1991) 3055.

\bibitem{KEK-TH-332}
B.~K. Bullock, K.~Hagiwara, and A.~D. Martin, ``{Tau polarization and its
  correlations as a probe of new physics},'' {\em Nucl.Phys} {\bf B395} (1993)
  499.

\bibitem{Anderson:1992jz}
J.~D. Anderson, M.~H. Austern, and R.~N. Cahn, ``{Measurement of Z-prime
  couplings at future hadron colliders through decays to tau leptons},'' {\em
  Phys.Rev.} {\bf D46} (1992) 290--302.

\bibitem{Gieseke:2011na}
S.~Gieseke, D.~Grellscheid, K.~Hamilton, A.~Papaefstathiou, S.~Platzer, {\em et
  al.}, ``{Herwig++ 2.5 Release Note},''
  \href{http://arXiv.org/abs/1102.1672}{{\tt 1102.1672}}.

\bibitem{Abazov:2011rq}
{\bf D0 Collaboration} Collaboration, V.~M. Abazov {\em et al.},
  ``{Forward-backward asymmetry in top quark-antiquark production},''
\href{http://arXiv.org/abs/1107.4995}{{\tt 1107.4995}}.

\bibitem{Aaltonen:2011kc}
{\bf CDF Collaboration} Collaboration, T.~Aaltonen {\em et al.}, ``{Evidence
  for a Mass Dependent Forward-Backward Asymmetry in Top Quark Pair
  Production},'' {\em Phys.Rev.D} (2011)
  \href{http://arXiv.org/abs/1101.0034}{{\tt 1101.0034}}.

\bibitem{Duraisamy:2011pt}
M.~Duraisamy, A.~Rashed, and A.~Datta, ``{The Top Forward Backward Asymmetry
  with general Z' couplings},'' {\em Phys.Rev.} {\bf D84} (2011) 054018,
\href{http://arXiv.org/abs/1106.5982}{{\tt 1106.5982}}.

\bibitem{Buckley:2011vc}
M.~R. Buckley, D.~Hooper, J.~Kopp, and E.~Neil, ``{Light Z' Bosons at the
  Tevatron},'' {\em Phys.Rev.} {\bf D83} (2011) 115013,
\href{http://arXiv.org/abs/1103.6035}{{\tt 1103.6035}}.

\bibitem{Gripaios:2010hv}
B.~Gripaios, A.~Papaefstathiou, K.~Sakurai, and B.~Webber, ``{Searching for
  third-generation composite leptoquarks at the LHC},'' {\em JHEP} {\bf 1101}
  (2011) 156, \href{http://arXiv.org/abs/1010.3962}{{\tt 1010.3962}}.

\bibitem{Brandenburg:2002xr}
A.~Brandenburg, Z.~Si, and P.~Uwer, ``{QCD corrected spin analyzing power of
  jets in decays of polarized top quarks},'' {\em Phys.Lett.} {\bf B539} (2002)
  235--241,
\href{http://arXiv.org/abs/hep-ph/0205023}{{\tt hep-ph/0205023}}.

\bibitem{Jezabek:1994qs}
M.~Jezabek, ``{Top quark physics},'' {\em Nucl.Phys.Proc.Suppl.} {\bf 37B}
  (1994) 197,
\href{http://arXiv.org/abs/hep-ph/9406411}{{\tt hep-ph/9406411}}.

\bibitem{Bernreuther:2008ju}
W.~Bernreuther, ``{Top quark physics at the LHC},'' {\em J.Phys.G} {\bf G35}
  (2008) 083001,
\href{http://arXiv.org/abs/0805.1333}{{\tt 0805.1333}}.

\bibitem{Godbole:2010kr}
R.~M. Godbole, K.~Rao, S.~D. Rindani, and R.~K. Singh, ``{On measurement of top
  polarization as a probe of $t \bar t$ production mechanisms at the LHC},''
  {\em JHEP} {\bf 1011} (2010) 144,
\href{http://arXiv.org/abs/1010.1458}{{\tt 1010.1458}}.

\bibitem{Godbole:2011vw}
R.~M. Godbole, L.~Hartgring, and I.~N. C.~D. White, ``{Top polarisation studies
  in $H^-t$ and $Wt$ production},'' \href{http://arXiv.org/abs/1111.0759}{{\tt
  1111.0759}}.

\bibitem{Aad:2011wc}
{\bf ATLAS Collaboration} Collaboration, G.~Aad {\em et al.}, ``{Search for New
  Phenomena in ttbar Events With Large Missing Transverse Momentum in
  Proton-Proton Collisions at sqrt(s) = 7 TeV with the ATLAS Detector},''
  \href{http://arXiv.org/abs/1109.4725}{{\tt 1109.4725}}.

\bibitem{Chatrchyan:2011ay}
{\bf CMS Collaboration} Collaboration, S.~Chatrchyan {\em et al.}, ``{Search
  for a Vector-like Quark with Charge 2/3 in t + Z Events from pp Collisions at
  sqrt(s) = 7 TeV},'' \href{http://arXiv.org/abs/1109.4985}{{\tt 1109.4985}}.

\bibitem{Ovyn:2009tx}
S.~Ovyn, X.~Rouby, and V.~Lemaitre, ``{DELPHES, a framework for fast simulation
  of a generic collider experiment},''
  \href{http://arXiv.org/abs/0903.2225}{{\tt 0903.2225}}.

\bibitem{Bellan:2011vm}
P.~Bellan, ``{Measurements of Inclusive b-quark Production at 7 TeV with the
  CMS Experiment},''
\href{http://arXiv.org/abs/1109.2003}{{\tt 1109.2003}}.

\bibitem{Salam:2009jx}
G.~P. Salam, ``{Towards Jetography},'' {\em Eur.Phys.J.} {\bf C67} (2010)
  637--686, \href{http://arXiv.org/abs/0906.1833}{{\tt 0906.1833}}.

\bibitem{numericalrec}
W.~H. {Press}, S.~A. {Teukolsky}, W.~T. {Vetterling}, and B.~P. {Flannery},
  {\em Numerical Recipes: The Art of Scientific Computing}.
\newblock Cambridge University Press, UK, third edition~ed., 2007.

\bibitem{Elagin:2010aw}
A.~Elagin, P.~Murat, A.~Pranko, and A.~Safonov, ``{A New Mass Reconstruction
  Technique for Resonances Decaying to di-tau},'' {\em Nucl.Instrum.Meth.} {\bf
  A654} (2011) 481--489, \href{http://arXiv.org/abs/1012.4686}{{\tt
  1012.4686}}.

\bibitem{Gripaios:2011jm}
B.~Gripaios, K.~Sakurai, and B.~Webber, ``{Polynomials, Riemann surfaces, and
  reconstructing missing-energy events},'' {\em JHEP} {\bf 1109} (2011) 140,
  \href{http://arXiv.org/abs/1103.3438}{{\tt 1103.3438}}.

\bibitem{Plehn:2011sj}
T.~Plehn, M.~Spannowsky, and M.~Takeuchi, ``{How to Improve Top Tagging},''
  \href{http://arXiv.org/abs/1111.5034}{{\tt 1111.5034}}.

\bibitem{Vermaseren:2000nd}
J.~Vermaseren, ``{New features of FORM},''
  \href{http://arXiv.org/abs/math-ph/0010025}{{\tt math-ph/0010025}}.

\bibitem{Dreiner:2008tw}
H.~K. Dreiner, H.~E. Haber, and S.~P. Martin, ``{Two-component spinor
  techniques and Feynman rules for quantum field theory and supersymmetry},''
  {\em Phys.Rept.} {\bf 494} (2010) 1--196,
  \href{http://arXiv.org/abs/0812.1594}{{\tt 0812.1594}}.

\end{thebibliography}\endgroup
\bibliographystyle{utphys}

\end{document}